\documentclass[reprint,superscriptaddress,nofootinbib,
amsmath,amssymb,aps,prd,floatfix
]{revtex4-2}

\usepackage{amsmath}
\usepackage{amssymb}
\usepackage{graphicx}
\usepackage{subcaption}
\usepackage{gensymb}
\usepackage{multirow}
\usepackage{color}
\usepackage{fontawesome}  
\usepackage{xspace}
\usepackage{lineno}

\newcommand*\patchAmsMathEnvironmentForLineno[1]{%
  \expandafter\let\csname old#1\expandafter\endcsname\csname #1\endcsname
  \expandafter\let\csname oldend#1\expandafter\endcsname\csname end#1\endcsname
  \renewenvironment{#1}
     {\linenomath\csname old#1\endcsname}
     {\csname oldend#1\endcsname\endlinenomath}}
\newcommand*\patchBothAmsMathEnvironmentsForLineno[1]{%
  \patchAmsMathEnvironmentForLineno{#1}%
  \patchAmsMathEnvironmentForLineno{#1*}}%
\AtBeginDocument{%
  \patchBothAmsMathEnvironmentsForLineno{equation}%
  \patchBothAmsMathEnvironmentsForLineno{align}%
  \patchBothAmsMathEnvironmentsForLineno{flalign}%
  \patchBothAmsMathEnvironmentsForLineno{alignat}%
  \patchBothAmsMathEnvironmentsForLineno{gather}%
  \patchBothAmsMathEnvironmentsForLineno{multline}%
}

\usepackage{ragged2e}
\usepackage{caption}
\newcommand{\vect}[1]{\boldsymbol{\mathbf{#1}}}
\newcommand{\dd}{{\rm d}}

\definecolor{deepgreen}{rgb}{0.2,0.8,0.2}

\definecolor{deepblue}{rgb}{0.2,0.4,0.8}

\definecolor{deepred}{rgb}{0.8,0.2,0.2}

\definecolor{linkcolor}{rgb}{0.7752941176470588, 0.22078431372549023, 0.2262745098039215}

\newcommand{\nbicon}{{\color{linkcolor}\faFileCodeO}\xspace}
\newcommand{\nblink}[1]{\href{https://github.com/kenvantilburg/quaDM_lens/blob/main/#1}{\nbicon}}

\newcommand\githubicon[1]{\href{#1}{\faGithub}\xspace}

\newcommand{\micronblink}[1]{\href{https://github.com/abby-moran/Quasar_microlensing_public/blob/main/#1}{\nbicon}}

\usepackage[colorlinks=true,linkcolor=deepblue,citecolor=deepblue,urlcolor=deepblue]{hyperref}

\begin{document}

%\linenumbers

\title{Measuring Extragalactic Microlens Masses and Motions in Strongly Lensed Quasar Systems with Intensity Interferometry}

\author{Abigail Moran}
\email{abigail.moran@columbia.edu}
\affiliation{Department of Astronomy, Columbia University, New York, NY 10027, USA}

\author{Ken Van Tilburg}
\email{kenvt@stanford.edu}
\affiliation{Leinweber Institute for Theoretical Physics, Department of Physics, Stanford University, Stanford, CA 94305, USA}

\date{\today}

\begin{abstract}
The direct measurement of masses and transverse motions of individual stars and compact objects in distant galaxies remains an outstanding challenge, leaving extragalactic stellar populations constrained only through model-dependent population fits.
We show that time-resolved intensity interferometry of strongly lensed quasars, combined with flux-ratio photometry, can break the source-size, microlens-mass, and transverse-velocity degeneracies of conventional light-curve microlensing.
Using the quadruply lensed quasar B1422+231 as a fiducial system and a next-generation interferometer ($\sim$800\,m$^2$ collecting area, spectral resolving power $\mathcal{R}\sim2\times10^4$, and few-picosecond timing), we forecast the precision on the quasar source size and on the mass, position, and transverse motion of an individual stellar-mass microlens in the lens galaxy.
Over a single epoch, such an intensity interferometer can determine the accretion disk size to $4\%$ in favorable geometries ($10\%$ for a typical detectable configuration) and the microlens mass to order unity for $m_l \gtrsim 0.1\,M_\odot$.
An eight-year, 33-epoch campaign reaches $\lesssim1\%$ on the disk size, localizes the microlens' impact parameter to $\sim0.1$--$0.5\,\mu\mathrm{as}$, and resolves its proper motion.
These observations would enable direct measurements of individual stellar masses and transverse motions in galaxies at moderate redshift.
After subtraction of the modeled stellar microlensing, they would yield residual uncertainties as low as $10^{-2}\,\mu\mathrm{as}\,\mathrm{yr}^{-2}$ on the angular acceleration of lensed quasar images, which can help separate stellar backgrounds from dark matter substructure lensing, the subject of a companion paper~\cite{DMpaper}.
\end{abstract}

\maketitle

\section{Introduction}\label{sec:intro}

Direct, object-by-object measurements of stellar and compact-object masses beyond the Local Group remain largely out of reach.
Stellar masses can be derived from binary orbital parameters, but in practice, this approach is limited to stars within our Galaxy~\cite{binary_masses} and Andromeda~\cite{binary_masses_and}.
The extragalactic stellar mass function is instead constrained indirectly by population-synthesis fits to photometric data (see e.g.~\cite{mass_metallicity, mag_color_fit, Evolutionary_mass}), introducing uncertainties in the stellar mass function that propagate into studies of galactic structure and evolution~\cite{uncertainty_prop}.

Measuring individual relative stellar transverse velocities poses an equally difficult challenge. The high-precision astrometry of \textit{Gaia} has enabled such measurements for many Milky Way stars (e.g.~\cite{Gaia_ind_vel, Gaia_Rev}), and decades of observations with \textit{HST} have resolved proper motions in the Large Magellanic Cloud~\cite{LMC_vels}. Current astrometric observatories generally cannot track individual stars in more distant galaxies~\cite{Gaia_galaxy_vel}.

Direct measurements of transverse stellar motions would constrain the internal kinematics and the bulk transverse peculiar velocity of a galaxy, complementing the radial velocity inferred from galactic redshift measurements. Together with distance information, proper motions can be turned into peculiar velocities gravitationally induced from large-scale structure, and thus constrain the matter power spectrum~\cite{Kochanek_vel, galaxy_pair_vel, LSS_vel, growth_rate_vel}.

Strongly lensed quasars provide access to these otherwise unresolved stars and remnants. The lensing deflection by a foreground galaxy produces multiple quasar images whose small-scale distortions and variability depend on compact objects in the lens plane~\cite{chang_refsdal}. Conventional analyses use unresolved photometric light curves and magnification maps generated by inverse ray shooting~\cite{Wambsganss99}, but the source size, microlens mass scale, and effective transverse velocity are degenerate after discarding (or integrating over) angularly resolved information. These limitations are further compounded by intrinsic quasar variability and extinction. We show that intensity interferometry (II) supplies complementary spatial-frequency information that can break these degeneracies and directly constrain individual microlens masses and motions.

Intensity interferometry measures second-order coherence of light, or equivalently the square of the visibility modulus of an (astronomical) image. The technique was developed in the 1950s~\cite{Brown_Twiss} and used to measure stellar angular diameters~\cite{Brown_Davis}, but its sensitivity was historically limited by collecting area, photodetection efficiency, timing resolution, and the absence of spectral multiplexing. Existing optical implementations on Cherenkov arrays, including VERITAS~\cite{abeysekara2020demonstration,2011arXiv1110.4702N}, MAGIC~\cite{abe2024performance}, and HESS~\cite{zmija2024first}, therefore target relatively bright stars; ASTRI and CTA are expected to improve this reach~\cite{Zampieri:2024imn,Saha:2025hwd}.

A dedicated modern design can gain substantially from each instrumental factor. Relative to a $1\,\mathrm{ns}$, single-channel, 5-hour measurement, $30\,\mathrm{ps}$ photon timing precision increases the SNR by $(1\,\mathrm{ns}/30\,\mathrm{ps})^{1/2}\simeq5.8$, spectral multiplexing with spectral resolution $\mathcal{R}\sim5000$ adds $\mathcal{R}^{1/2}\simeq71$, and a 50-hour exposure adds $(50/5)^{1/2}\simeq3.2$. Their product is $\sim1.3\times10^3$, corresponding to a depth gain of $\Delta m\simeq7.8$ magnitudes at fixed threshold SNR. Current concepts pursue either spectral multiplexing on optical-quality telescopes (QUASAR~\cite{Walter:2025lrq} and I2C~\cite{Leopold:2024juv}) or large arrays of smaller apertures (LFAST~\cite{LFAST} and MAST~\cite{2024arXiv240816822S,MAST_Stellar}). Because an intensity interferometer's signal-to-noise ratio scales linearly with its collecting area and is proportional to the square root of independent spectral channels and inverse timing resolution, an architecture combining several of these improvements can target sources much fainter than the bright stars that have been historically studied with intensity interferometry~\cite{pirsa_PIRSA:24100096}. We envision observing capabilities that combine all of these techniques to perform II on sources as faint as (the brightest) quasars.

Recent work has explored II for high-angular-resolution imaging and precision astrometry~\cite{KVT2023,Galanis23,Rai25,II_cosmo,II_sne}. For lensed quasars, microarcsecond angular sensitivity probes the morphology of each macro-image and its microlensing distortions. Ref.~\cite{microlens_swarm} considered the statistics of an unresolved swarm of microimages; here we instead forecast inference of individual microlens parameters from resolved squared visibilities, flux ratios, and their time evolution. Photometric microlensing has measured accretion-disk sizes and temperature profiles~\cite{quasar_microlens_star,Kochanek,Poindexter2008,Morgan2010,Mosquera,int_var_sigma}, constrained stellar mass functions and compact-object fractions~\cite{Schechter_Wambsganss,Mediavilla2009}, and identified candidate dark compact objects~\cite{quasar_microlens_dark,microlensing_BH}; see Ref.~\cite{Quasar_microlens_rev} for a review. Spatial information from II makes it possible to separate source and lens parameters that photometry alone cannot determine.

We simulate II and photometric observations of the quadruply lensed quasar B1422+231, perturbed by a single point-like microlens, and forecast constraints on the source radius, microlens mass and position, transverse motion, and light-centroid acceleration. Section~\ref{sec:setup} defines the source, lens, and observing models; Sec.~\ref{sec:analysis} describes the single-epoch and time-domain forecasts; and Sec.~\ref{sec:results} presents the results. We conclude with a discussion of future applications in Sec.~\ref{sec:discussion}. Throughout this work, we introduce notation which is summarized in App.~\ref{app:Notation} in Table \ref{tab:notation}. Appendix~\ref{app:pipeline} documents the numerical pipeline, publicly available on GitHub at \githubicon{https://github.com/abby-moran/Quasar_microlensing_public}
\href{https://github.com/abby-moran/Quasar_microlensing_public}
{\texttt{github.com/abby-moran/Quasar\_microlensing\_public}}. Additionally, the \nbicon in each figure caption links to the notebook that generated the corresponding figure.

\section{Setup and System} \label{sec:setup}

We will take the quadruply-imaged quasar B1422+231 as our benchmark system for studying stellar microlensing. It is an ideal target because it is one of the brightest lensed quasars (and thus one of the first within reach of future intensity interferometers) and has been well characterized over its long observational record~\cite{Patnaik1992,Mosquera}. We first define the intrinsic quasar emission, then the macro- and microlensing model, and finally the mock II and photometric measurements.

\subsection{Quasar Emission Morphology}\label{subsec:QEM}

We simulate the unlensed quasar in source-plane angular coordinates $\vect{\beta}$. The accretion disk follows the geometrically thin, optically thick model of Ref.~\cite{quasar_mod}, with local blackbody emission. A single angular scale, $\theta_\mathrm{src}$, sets the radial temperature profile.  
Here $\theta_\mathrm{src}$ is a circularized angular scale, defined such that an area $\pi \theta_\mathrm{src}^2$ corresponds to the emitting region size. At $|\vect{\beta}|=\theta_\mathrm{src}$, the temperature is $T_{500}\equiv hc/(\lambda_{500}k_\mathrm{B})=28{,}776\,\mathrm{K}$ for $\lambda_{500} \equiv 500\,\mathrm{nm}$. Thus
\begin{equation}\label{Toftheta}
T(\vect{\beta})= T_{500}\left(\frac{|\vect{\beta}|}{\theta_\mathrm{src}}\right)^{-\frac{3}{4}}.
\end{equation}

At observing wavelength $\lambda$, the unlensed specific surface brightness per unit wavelength is
\begin{equation}\label{Boftheta}
B_\lambda(\vect{\beta}) = \frac{2 hc^2}{\lambda^5}\left(\exp{\frac{hc}{\lambda k_\text{B} T(\vect{\beta})}}-1\right)^{-1},
\end{equation}
in units of $\mathrm{erg\,s^{-1}\,cm^{-2}\,cm^{-1}\,sr^{-1}}$, where $h$ is Planck's constant, $c$ is the speed of light, and $k_\mathrm{B}$ is Boltzmann's constant. 
A redshifted blackbody retains the same spectral form, with the source temperature shifted to the observer-frame temperature $T_\mathrm{obs}=T/(1+z)$. We can therefore calculate the observed spectrum of a given quasar by evaluating Eq.~\ref{Boftheta} using $T_\mathrm{obs}$. The additional redshift dependence is incorporated through the temperature normalization defined in Eq.~\ref{Toftheta}. This produces an image parameterized by $\theta_\mathrm{src}$ on a grid of source-plane coordinates $\vect{\beta}$.

Lensing increases the apparent linear scale as
We convert $\theta_\mathrm{src, lensed}$ to an angular circularized radius $\theta_\text{src}$ via
\begin{equation}\label{eq:theta500}
    \theta_\mathrm{src, lensed} = \sqrt{\mu} \theta_\mathrm{src},
\end{equation}
because the image area scales as $\mu$. We define the magnification of image (i) relative to image (ii) as the ratio of their integrated surface brightnesses over the observed solid angle $\Omega$,
\begin{equation}\label{eq:mag_def}
\mu_{\mathrm{i,ii}}=\frac{\int_\Omega B_{ \mathrm{(i)}}}{\int_\Omega B_\mathrm{(ii)}},
\end{equation}
where the first subscript labels the numerator and the second the reference image in the denominator. Taking image (ii) to be the unlensed source recovers the absolute magnification $\mu$ used above. 
We also apply Eq.~\ref{eq:mag_def} to cases where the two surface-brightness distributions (i) and (ii) describe the \emph{same} quasar image but differ in their model parameters.

We determine $\theta_\mathrm{src, lensed}$ by normalizing the surface-brightness profile to the observed flux of image~A. Specifically, we integrate $B(\vect{\beta})$ over the source plane to obtain the total observed flux density. We adopt observational data for image A from the CASTLES survey\footnote{\url{https://lweb.cfa.harvard.edu/castles/Individual/B1422.html}}, which reports a total magnitude of $m = 15.88$ mag in the F791W band (effective wavelength $\lambda=7969~\mathrm{\AA}$).
Following Ref.~\cite{Galanis23}, we convert this magnitude to a flux density and solve $F_\lambda = \frac{16\pi}{3}\,\Gamma(\tfrac{8}{3})\,\zeta(\tfrac{8}{3})\, h^{-5/3} c^{-2/3} k_\text{B}^{8/3}\, T_{500}^{8/3}\, \lambda^{-7/3}\, \theta_{\mathrm{src, lensed}}^2$ for $\theta_{\mathrm{src, lensed}} = 5.13 \times 10^{-13}~\mathrm{rad}$. As a cross-check, we repeat the same flux-normalization procedure using \textit{Gaia} DR3 G-band photometry, $G = 16.32\,\mathrm{mag}$~\cite{GAIADR3}, obtaining $\theta_{\mathrm{src, lensed}} = 5.58\times 10^{-13}~\mathrm{rad}$. The two values agree reasonably well; the modest discrepancy likely reflects the different bands and the decades-long gap between the measurements, over which dust and time-variable lensing can alter the apparent flux.

\subsection{Macrolensing}

We model the foreground galaxy as a smooth, homogeneous mass distribution. Following the conventions of Ref.~\cite{Lensing_rev}, source-plane coordinates $\vect{\beta}$ and image-plane coordinates $\vect{\theta}$ are related by the lens equation,
\begin{equation}\label{eq:basic_lensing}
    \vect{\beta} = \vect{\theta} - \vect{\alpha}(\vect{\theta}),
\end{equation}
where $\vect{\alpha}$ is the deflection angle.

Approximating the local potential near an image by constant convergence $\kappa$ and shear $\gamma$, the distortion is described by the Jacobian
\begin{equation}\label{eq:jac_macro}
    \vect{A} =
    \begin{pmatrix}
    1 - \kappa - \gamma_1 & -\gamma_2 \\
    -\gamma_2 & 1 - \kappa + \gamma_1
    \end{pmatrix},
\end{equation}
where $\gamma_1$ and $\gamma_2$ are the two shear components.
The eigenvectors of $\vect{A}$ define the orthogonal principal axes of the distortion, and the eigenvalues give the inverse linear magnifications along those axes. The total magnification is $\mu=|\det\vect{A}|^{-1}$. The companion paper denotes the inverse matrix $\vect{A}^{-1}$ by $B^\mathrm{I}$~\cite{DMpaper}.
The corresponding deflection angle is given by
\begin{equation}
    \vect{\alpha}(\vect{\theta}) = (\vect{I} - \vect{A}) \cdot \vect{\theta} =
    \begin{pmatrix}
    \kappa + \gamma_1 & \gamma_2 \\
    \gamma_2 & \kappa - \gamma_1
    \end{pmatrix}
    \cdot \vect{\theta}.
    \label{eq:deflection_angle}
\end{equation}

For B1422+231, we take a lens redshift $z_l = 0.34$~\cite{ODowd} and adopt the best-fit parameters $(\kappa, \gamma_1, \gamma_2) = (0.385, 0.254, 0.410)$ for image~A, which give a magnification $\mu=6.86$. They come from an independent singular-isothermal-ellipsoid plus external-shear fit to the four image positions, following the methods of Ref.~\cite{Cosmograil_2012}, as described in the companion paper~\cite[App.~B2]{DMpaper}. 

Using Eq.~\ref{eq:theta500}, the CASTLES and \emph{Gaia} normalizations imply $\theta_\mathrm{src}=6.93\times10^{15}\,\mathrm{cm}$ and $7.54\times10^{15}\,\mathrm{cm}$, respectively. We use the CASTLES-based value below.

The unlensed source-plane brightness distribution and the macrolensed image are shown in the upper-left and upper-center panels of Fig.~\ref{fig:basic_lensing}, respectively.
The center panel also marks the macrolensing-shear axis, defined by the eigenvector of $\vect{A}$ with the smaller eigenvalue (the direction of greatest stretching); the angle $\phi_0$ sets its orientation relative to the real-space $x$-axis (right ascension).

\subsection{Microlensing}\label{sec:microlensing}
Compact objects in the lensing galaxy add microlensing deflections on top of this smooth distribution. Each lens of mass $m_l$ produces a deflection
\begin{equation}
    \vect{\alpha}_\star = \frac{4G m_l}{c^2}
    \frac{D_{ls}}{D_s D_l}
    \frac{\vect{\theta} - \vect{\theta}_\mathrm{sep}}{|\vect{\theta} - \vect{\theta}_\mathrm{sep}|^2},
\end{equation}
where $D_{ls}$, $D_s$, and $D_l$ are the angular-diameter distances between lens and source, observer and source, and observer and lens, respectively, and $\vect{\theta}_\mathrm{sep}$ is the angular position of the microlens. 
We use $z_s=3.62$~\cite{ODowd} and a flat $\Lambda$CDM cosmology with $H_0=70\,\mathrm{km\,s^{-1}\,Mpc^{-1}}$ and $\Omega_m=0.3$ when calculating these values.

We also define the Einstein radius of a point mass lens as
\begin{equation}\label{eq:thetaE}
    \theta_{E,m_l} = \sqrt{\frac{4Gm_l}{c^2}\frac{D_{ls}}{D_s D_l
    }} \approx 2.6 \, \mu as \,\left(\frac{m_l}{M_\odot}\right)^{\frac{1}{2}},
\end{equation}
which sets the characteristic angular scale of the microlens~\cite{microlensing_0}. The approximation above assumes the lensing geometry of B1422+231, using the corresponding values of $D_{ls}$, $D_s$, and $D_l$. Microlensing is most pronounced when the source-lens separation $\theta_\mathrm{sep} \equiv |\vect{\theta}_\mathrm{sep}|$ falls within an Einstein radius, $\theta_\mathrm{sep} \lesssim \theta_{E,m_l}$.

We assume that the microlenses reside in the lens galaxy. Near image~A, the stellar surface mass density is
\[
\Sigma_*=\kappa_*\Sigma_\mathrm{crit}
=\kappa_*\frac{c^2}{4\pi G}\frac{D_s}{D_lD_{ls}} \approx 0.017 \text{g cm}^{-2},
\]
where $\Sigma_\mathrm{crit}$ is the critical surface density~\cite{Lensing_rev} and $\kappa_*=0.04$ is the stellar convergence~\cite{kappa_vals}. The expected number of detectable lenses is
\[
N_\mathrm{det}=\frac{\Sigma_*}{\langle M_*\rangle}A_\mathrm{det}.
\]
Taking $\langle M_*\rangle=0.4\,M_\odot$~\cite{mean_mass} and the detectable area $A_\mathrm{det}/D_l^2= 60\ \theta_{E,0.4}^2$ inferred in Sec.~\ref{sec:results}, we obtain $N_\mathrm{det}=0.8$.

We estimate the event duration using 
\begin{align}
|\vect{v}|=\sqrt{v_\mathrm{bulk}^2+2\sigma_\mathrm{int}^2} \approx 472 \, \mathrm{km \, s^{-1}}, \label{eq:v_tot}
\end{align}
with $v_\mathrm{bulk}\simeq400\,\mathrm{km\,s^{-1}}$~\cite{bulk_vel} and $\sigma_\mathrm{int}=178\,\mathrm{km\,s^{-1}}$~\cite{vdisp}. Encounter rates and durations use the apparent image-fixed-frame velocities defined in Sec.~\ref{subsec:traj}. These velocities lie near the lower end of values in the literature; the companion paper adopts larger characteristic velocities~\cite{DMpaper}, citing Ref. \citep{Kundic:1997}. Drawing $10^5$ trajectories with random approach angles and closest-approach points inside $r_\mathrm{det}$ gives a mean crossing time $T_\mathrm{cross}=50\,\mathrm{yr}$ and a mean waiting time $T_\mathrm{wait}=T_\mathrm{cross}/N_\mathrm{det}=60\,\mathrm{yr}$. Because the duty cycle is high but the expected occupancy is below unity, we model one dominant microlens in image~A.

We selected image~A specifically because its low stellar density justifies the single-microlens simplification. Images B and C of B1422+231 have comparable stellar densities~\cite{kappa_vals}. Many systems (as well as image D of B1422+231) have substantially higher stellar surface densities in their image regions, where microlensing is routine. 
Of the 87 lensed quasars compiled by Ref.~\cite{Mosquera}, 26 are brighter than magnitude $17$ and thus amenable to an analysis like the one presented here. Typical lensed images have stellar densities of $\kappa_\star\sim0.1$ \citep{avg_density}, suggesting that the $\kappa_\star\lesssim 0.05$ required for our analysis is not atypical among observationally accessible systems. This sample is also growing as surveys such as \textit{Gaia} identify new lensed quasar systems: the current census of spectroscopically confirmed lensed quasars stands at 364 systems, of which 87 are quads or triples~\citep{gaia_quasars}.

At the other end of the spectrum, quasars lensed by galaxy \emph{clusters} provide another potential target for II observations and are typically even more star-poor. Only eight such systems are currently known~\cite{2003cluster, 2012cluster, 2013cluster, 2018Cluster, 2026cluster, 2026dual_cluster}, and their stellar densities can be extremely low; for example, SDSS J1029+2623 is estimated to have $\kappa_*=0.003$~\cite{DMpaper}.
At these low stellar densities, the probability of having even a single detectable microlens is so small that the single-lens approximation becomes very good---in fact, to the point that most images will not have a detectable microlens. The limited number of known cluster-lensed quasars makes it difficult to assess their suitability as a broader II target population. As future surveys such as Rubin, \textit{Euclid}, and \textit{Roman} expand the sample of cluster-lensed quasars~\cite{2026cluster, future_clusters}, these systems may become promising II targets and provide an alternative to galaxy-scale lenses.

We also define the image light centroid as
\begin{equation}\label{eq:cent_det}
    \vect{\theta}_c=\frac{\int_\Omega B_i \vect{\theta}_i}{\int_\Omega B_i} = \frac{\int_\Omega B_i \vect{\theta}_i}{I_\lambda}.
\end{equation}
In the second equality, we defined $I_\lambda$ as the integrated flux of an image over area $\Omega$ at a given wavelength.
After subtracting the microlensing contribution, precise centroid measurements can constrain additional line-of-sight perturbers, including dark matter halos~\cite{DMpaper}.

\subsection{Intensity Interferometry}
An intensity interferometer measures the intensity correlation between one (or more) pair(s) of telescopes separated by a baseline $\vect{d}$:
\begin{equation}
   C(\tau) = \frac{\langle I_1(t) I_2(t+\tau)\rangle }{\langle I_1(t)\rangle \langle I_2(t+\tau)\rangle}-1,
\end{equation}
where $I_i$ is the integrated flux for a given wavelength at telescope $i$ and time $t$, and $\tau$ is the relative time delay between detectors. For chaotic (thermal) light, this correlation (at optimal time delay and timing resolution) approaches the square visibility modulus,
\begin{align}\label{C_int}
    C\simeq \left| \mathcal{V}(\vect{d}, \bar{k})\right|^2,  
\end{align} 
where $\bar{k} = 2\pi / \bar{\lambda}$ is the mean photon wavenumber in a spectral channel with width $\Delta k$ and the visibility is the normalized Fourier transform of the source's sky intensity distribution: 
\begin{align}
    \mathcal{V}(\vect{d},\bar{k}) \simeq  \int_{\Delta k} \dd k \int_{\Delta \Omega} \dd \Omega \, I_{\bar{k}}^{-1} B_\lambda(k,\boldsymbol{\theta}) \, e^{i \boldsymbol{u} \cdot \boldsymbol{\theta}}
\end{align}
where $d\Omega = \dd^2\theta$ is the angular area and $\vect{u} \equiv (u,v) = k\vect{d}$ is the spatial frequency (angular wavevector) probed by baseline $\vect{d}$. We define $I_{\bar{k}} =\int_\Omega B_{\bar{\lambda}}$.
The surface brightness $B_\lambda$ follows Eq.~\ref{Boftheta}; for a uniform disk, the integral gives an Airy pattern centered at $(u,v)=(0,0)$.

In practice, we evaluate the visibility in a spectral channel centered on $\bar{k}$ by discretizing the source image over a finite angular region of extent $\theta_\mathrm{max}$ and computing its discrete Fourier transform (DFT). For a more detailed description of this procedure, see App.~\ref{app:pipeline}.

For a given interferometer, each measurement has an uncertainty given by
\begin{equation}
    \sigma_C = \sqrt{\frac{1}{\sqrt{4\pi}\,\sigma_t\,t_{\rm obs}}}\,
    \frac{\hbar c \bar{k}}{\eta_1 \eta_2 A_1 A_2}\,
    \frac{1}{\sqrt{\langle I_1 \rangle \langle I_2 \rangle}},
\end{equation}
where $\sigma_t$ is the detector time resolution, $t_{\rm obs}$ the integration time, $\eta_i$ the detector efficiencies, and $A_i$ the collecting areas~\cite{Galanis23}. We adopt $\sigma_{C, \text{base}}=4.66$, computed from the observed flux density of image~A of B1422+231 ($\approx 1.3\times10^{-26}~\mathrm{erg\,s^{-1}\,cm^{-2}\,Hz^{-1}}$, i.e.\ $m\approx15.8$, at $\bar{\lambda}=570$\,nm) for $t_{\rm obs} = 8~\mathrm{hr}$ per epoch and specifications comparable to Phase~III of~\cite{Galanis23}: a total collecting area of $785~\text{m}^2$ (e.g.~ten 10\,m-diameter apertures) with photodetection efficiencies $\eta_1=\eta_2=0.5$, a spectral resolving power of $\mathcal{R}=20{,}000$ (about four times finer than a DESI-like spectrograph~\cite{DESI:2016igz}), and a timing resolution $\sigma_t=3\,\mathrm{ps}$.
These somewhat idealized specifications have all been individually demonstrated, but have not been combined into a single observatory.
The design study in Ref.~\cite{Walter:2025lrq} argues that comparable capabilities may be achievable on a decadal timescale.

The fiducial specifications are broadly aligned with the goals of dedicated intensity interferometry concepts (spectral multiplexing combined with many small apertures or a few large-area ones) discussed in Sec.~\ref{sec:intro}. We assume a central wavelength $\bar{\lambda}=570\,\mathrm{nm}$, near the peak of the observed optical spectrum of B1422+231~\cite{Patnaik1992}.

\begin{figure*}[h]
    \includegraphics[width=0.98\linewidth]{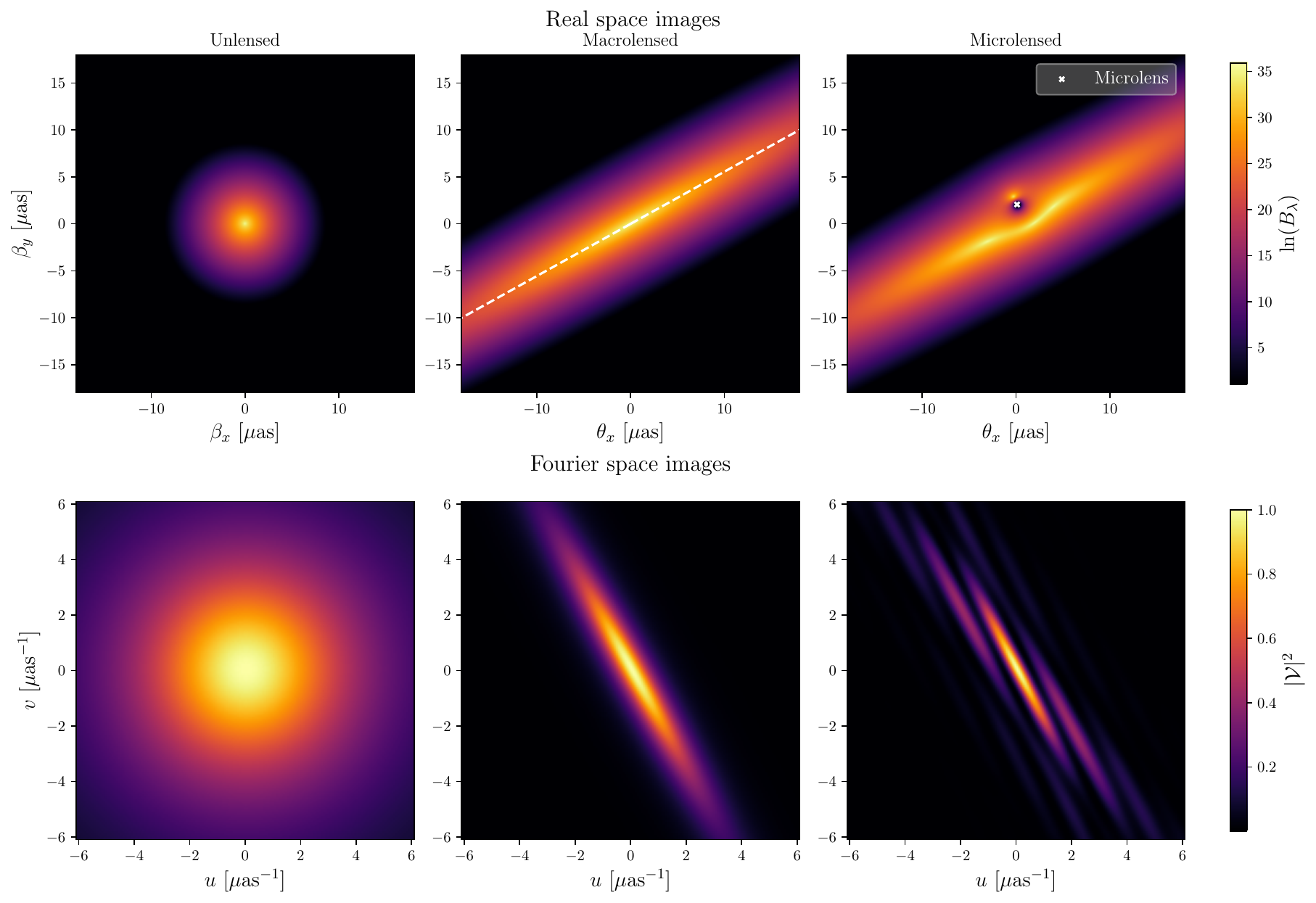}
    \caption{
    \micronblink{Plots.ipynb}
    Simulated real- and Fourier-space representations of image A of quasar B1422+231.
    Top row: Real-space surface brightness maps ($B_\lambda$; Eq.~\ref{Boftheta}); the color encodes the natural logarithm of the spectral radiance, $\ln B_\lambda$ (arbitrary units), and the axes are angular positions in $\mu\mathrm{as}$ ($\vect{\beta}$ for the source, $\vect{\theta}$ for the images). From left to right we show the unlensed source in $\vect{\beta}$-space, the macrolensed image in $\vect{\theta}$-space, and the image including both macrolensing and an example microlensing perturbation. The dotted line in the top-center panel marks the macrolensing shear axis defined by $\phi_0$.
    Second row: The corresponding Fourier ($\vect{u}$-) space representations, where the color shows the squared visibility modulus (correlation) $|V|^2 = C$ (dimensionless, normalized to $C=1$ at the origin) and the axes are the wavenumbers $(u,v)$ in $\mu\mathrm{as}^{-1}$.
    In the rightmost column, the microlensing perturbation is produced by a $0.4\,M_\odot$ lens located at $\vect{\theta}_\mathrm{sep} = [0.10, 2.06] \mu\mathrm{as}$, marked by the white ``x''. The images are generated over a sky region of $\theta_\mathrm{max}=4\times 10^{-10}$ radians with $8000$ pixels per dimension; the panels are zoomed by a factor of 50 to highlight small-scale structure.
    }
    \label{fig:basic_lensing}
\end{figure*}

\subsection{Photometry}

We combine II with unresolved flux-ratio photometry. For B1422+231, both \textit{HST} and \textit{Gaia} have measured the individual macro-images; all four images are resolved in \textit{Gaia} DR3, with sub-milliarcsecond astrometry and per-image photometry compiled in Ref.~\citep{gaia_quasars}. We use the magnification $\mu$ of Eq.~\ref{eq:mag} to quantify the flux of a microlensed image relative to the macrolensed-only image, holding the macrolensing parameters fixed in both. For reference, a $0.4\,M_\odot$ lens at $\theta_\mathrm{sep}=0.5\theta_{E,0.4}$ doubles the flux of the macrolensed-only image.

Measuring this change is well within reach of modern instruments: for B1422+231, we estimate $\sigma_\mu = 0.101$, corresponding to a total magnitude error of $0.11$~mag on a single measurement of image~A. We combine three contributions in quadrature: instrumental photometric noise, residual intrinsic quasar variability, and residual extinction. The photometric noise we take as $0.01$~mag, typical of current facilities such as \textit{Gaia}. 

The dominant contribution is residual intrinsic quasar variability, which we take to be $0.1$~mag. Although quasars can vary at much higher levels, most intrinsic variability is common to all lensed images and cancels in flux-ratio measurements. Only variability between the reference-baseline epoch and the observation propagates into $\sigma_\mu$.
This residual contribution in the optical we take to be $0.1$ mag for B1422+231 \citep{optical_int_crr}. 

The final contribution to $\sigma_\mu$ is differential-extinction uncertainty, which we take to be $0.03\,\mathrm{mag}$. When comparing lensed images, a uniform extinction component largely cancels, leaving differential extinction between images as the relevant systematic uncertainty. No direct estimate of differential extinction is currently available for B1422+231, so we instead adopt the value measured for Q2237+0305~\citep{Dust_Q2237}. The similar radio-to-optical macro-image flux ratios of B1422+231 and Q2237+0305~\cite{radio_1422,Dust_Q2237} suggest that different dust extinction is not a dominant contribution in either system. We therefore take this value as a representative systematic uncertainty in our forecast, resulting in $\sigma_\mu = \sqrt{0.01^2 + 0.1^2 + 0.03^2} \text{ mag} = 0.11 \text{ mag}.$

\section{Analysis Methods}\label{sec:analysis}

We first forecast the information in a one epoch of II and photometric data and then combine multiple epochs to infer the parameters of a typical microlensing trajectory. Section~\ref{subsec:sing} defines the single-epoch observables and Fisher analysis; Sec.~\ref{subsec:traj} extends this calculation to a full observing campaign. The results of these analyses are in Sec.~\ref{sec:results}.

\subsection{Single-Epoch Analysis}\label{subsec:sing}
In this subsection, we describe the information that can be extracted from a single observation of a microlensed quasar image. One epoch constrains four parameters, namely $\vect{p} = (\theta_\mathrm{src},\, m_l,\, \vect{\theta}_\mathrm{sep})$: the source (quasar) angular radius $\theta_\mathrm{src}$ (as defined in Sec.~\ref{subsec:QEM}), the lens mass $m_l$, and the two coordinates of the lens position $\vect{\theta}_\mathrm{sep}$ in the image plane. We hold the macrolensing parameters fixed at their best-fit values~\cite{Cosmograil_2012} and do not propagate uncertainties in the macrolensing model. The forecasts are therefore conditional on the adopted local distortion matrix. Sec~\ref{sec:discussion} explains how the II data could instead constrain that matrix directly.

To generate a lensed image with both macro- and microlensing components, we begin with a square grid of size $N_\text{pix} \times N_\text{pix}$ in image-space coordinates $\vect{\theta}$, where $N_\text{pix}$ is the number of pixels used. We then ray-trace this grid according to the lens equation (Eq.~\ref{eq:basic_lensing}) which maps each image-plane position $\vect{\theta}$ to a source-plane position $\vect{\beta}$. A source-plane point can map to several image-plane positions, producing  multiple macro-images.
The image temperature $T(\vect{\theta})$ equals the source-plane temperature $T(\vect{\beta})$ at the point to which $\vect{\theta}$ maps (Eq.~\ref{Toftheta}), which in turn fixes the imaged surface brightness (spectral radiance) $B(\vect{\theta})$ via Eq.~\ref{Boftheta}.

We show image A of B1422+231 without lensing, with macrolensing only, and with both macro- and microlensing in the top panels of Fig.~\ref{fig:basic_lensing}. The accretion disk is assumed face-on in the source plane (top-left panel), so its emission is azimuthally symmetric. Macrolensing magnifies and shears this region along a preferred axis, marked by the dashed white line in the top-center panel; from the eigenvectors of $\vect{A}$ we find this shear axis is rotated by $\phi_0=-2.07$\,rad relative to the real-space $x$-axis. In the top right panel, we show how a $0.4\,M_\odot$ microlens (location indicated by a white ``x'') would distort and fragment this macro-image into four separate microimages of varying brightness. The microlensed morphology encodes the lens mass and position; App.~\ref{app:microlens} illustrates these dependences.

Integrating $B(\vect{\theta})$ over the image with and without the microlens gives the microlensing magnification (at fixed macrolensing; Eq.~\ref{eq:mag}), which serves as our mock photometric observation with fractional uncertainty $\sigma_{\mu}=0.101$.

In the optical, state-of-the-art ground-based observatories reach resolutions of tens of milliarcseconds (mas) with adaptive optics, e.g.\ MagAO~\cite{MagAO}, and conventional (amplitude) interferometry improves on this by up to two orders of magnitude, e.g.\ CHARA at $0.4$\,mas~\cite{CHARA}. Neither reaches the $\mu\mathrm{as}$-level resolution required to resolve the real-space features in the top panels of Fig.~\ref{fig:basic_lensing}.

A long-baseline intensity interferometer, by contrast, samples the visibility modulus of Eq.~\ref{C_int} at angular wavenumber $\vect{u} = k\vect{d}$, with magnitude
\begin{equation}
\label{eq:baseline}
    |\vect{u}| = \frac{2\pi}{\lambda} |\vect{d}| \approx 0.53\,\mu\mathrm{as}^{-1} \left(\frac{570\,\mathrm{nm}}{\lambda}\right)\left(\frac{d}{10\,\mathrm{km}}\right),
\end{equation}
sufficient to resolve the features in Fig.~\ref{fig:basic_lensing}.
Extracting the full information in an image requires measuring the correlation $C \propto |\mathcal{V}|^2$ over many baseline lengths and orientations and across a range of spectral wavenumbers $k$.

At the origin $(u,v)=(0,0)$, $C=1$ by definition and decreases rapidly with increasing $|\vect{u}|$. Despite the typical decrease of the visibility modulus $|\mathcal{V}|$ with increasing $|\vect{u}|$, most of the information about microlensing-induced departures from the macrolensed-only image is encoded at $|\vect{u}| \sim \theta_{\mathrm{E},m_l}^{-1}$ (see the Fourier-space panels of Fig.~\ref{fig:basic_lensing}), so restricting to small $|\vect{u}|$ would discard much of the constraining power.

\begin{figure*}
    \centering
    \includegraphics[width=0.8\linewidth]{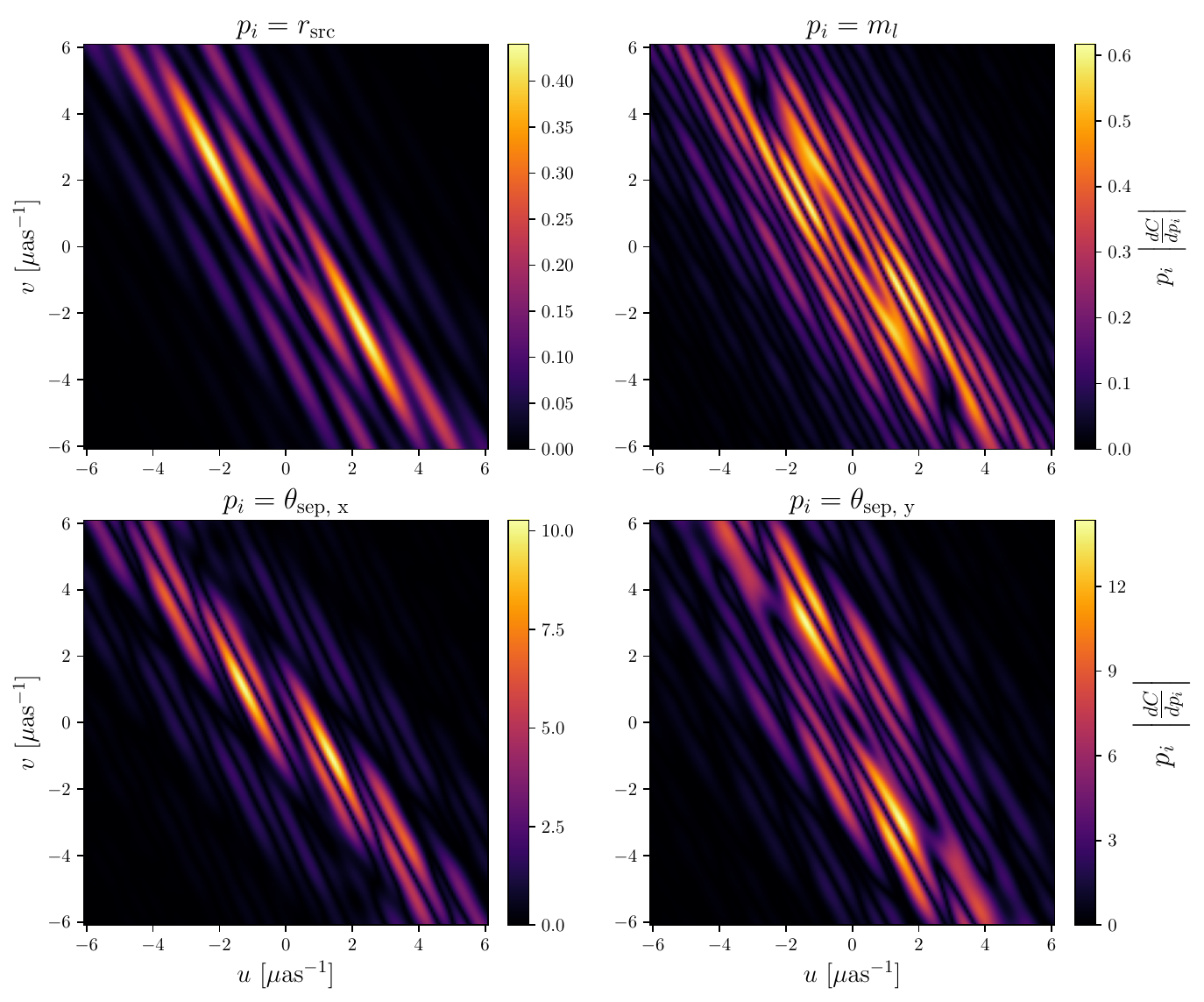}
    \caption{\micronblink{Plots.ipynb} Partial logarithmic derivatives of the correlation $C$ with respect to each parameter, for an image lensed by a compact object with $m_l=0.4\,M_{\odot}$ at $\vect{\theta}_\mathrm{sep}=[0.10, 2.1]~\mu\mathrm{as}$, shown in Fourier space. The color shows the parameter-scaled sensitivity $p_i\,\left|\partial C/\partial p_i\right|$ (dimensionless, so that panels for different parameters are directly comparable), and the axes are the Fourier-space wavenumbers $(u,v)$ in $\mu\mathrm{as}^{-1}$. This configuration is a representative example.}
    \label{fig:derivs}
\end{figure*}

To select a region that best balances amplitude and information content, we identify the optimal Fourier-space sampling location as the point that maximizes the Fisher information for the source radius parameter $\theta_\mathrm{src}$. For a Gaussian likelihood, the Fisher information density at each $(u,v)$ is proportional to $\big({\partial C}/{\partial \theta_\mathrm{src}}\big)^2$. The optimal measurement location is therefore
\begin{align}
(u,v)_\mathrm{opt}=\arg \max \left| \frac{\partial C(u,v)}{\partial \theta_\mathrm{src}} \right|. \label{eq:uv_opt}
\end{align}
We calculate the discrete forward derivative of the macrolensed image in Fourier space with respect to $\theta_\mathrm{src}$ and find four solutions: $(u,v)_\mathrm{opt}=\{ \pm[-1.83, 3.50], \ \pm[1.68, -2.28]\} \,\mu\mathrm{as}^{-1}$, where $C=0.47$.
These four points maximize the local source-size sensitivity, but the Fisher information occupies a finite region, and microlensing structure also appears at modes with smaller correlation amplitude. Figure~\ref{fig:derivs} shows the Fisher-information distribution in Fourier space for each parameter in a fiducial microlensing scenario.
We therefore sample the region $C>0.1$, which retains measurable signal while covering the informative microlensing structure.

\begin{figure}
    \centering
    \includegraphics[width=0.96\linewidth]{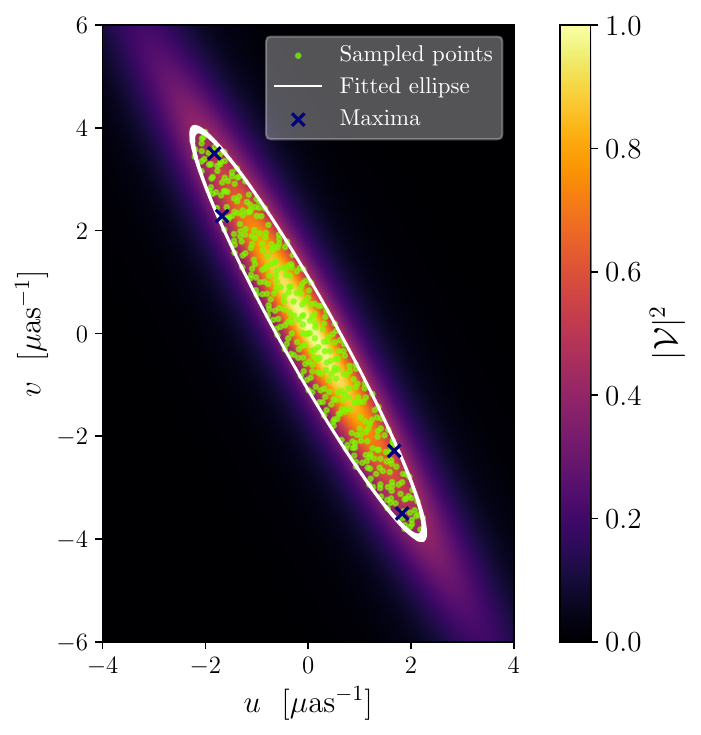}
    \caption{\micronblink{Plots.ipynb} Image A of B1422+231 shown in $\vect{u} = (u,v)$ Fourier space, where the color shows the square modulus of the visibility $|\mathcal{V}|^2 = C$ and the axes are the wavenumbers $(u,v)$ in $\mu\mathrm{as}^{-1}$.
    The white ellipse is the $C=0.1$ contour, which bounds the sampling region over which we generate mock interferometric measurements. We draw $1000$ points uniformly across this region to construct the synthetic observations, but for visual clarity show only $400$ of these sampled locations in green. The white dotted line again shows the macrolensing shear axis; note that in Fourier space the image is elongated along the direction perpendicular to this axis. The blue points marked with an `X' are the maxima of  $\frac{\partial C}{\partial \theta_\mathrm{src}}$ in the macrolensing-only model, which guided our choice of sampling region. }
    \label{fig:sampling}
\end{figure}

The boundary of this region, $C = 0.1$, is an ellipse.
The Fourier transform inverts spatial scales, such that an elongation in real space produces a compression in the corresponding Fourier direction and an elongation perpendicular to it. The major axis of the visibility pattern in Fourier space thus lies at $90\degree$ to the macrolensing shear axis (Fig.~\ref{fig:basic_lensing}).
As shown in Fig.~\ref{fig:sampling}, the sampling extends to a maximum angular wavenumber $|\vect{u}| \approx 4.5\,\mu\mathrm{as}^{-1}$, corresponding via Eq.~\ref{eq:baseline} to an effective maximum baseline $|\vect{d}|=85\,\mathrm{km}$. Baselines of this length pose no fundamental difficulty for intensity interferometry: unlike amplitude interferometry, it requires only fast photon timing and electronic delay calibration between stations, not optical path control, so the baseline is limited by clock synchronization and correlation bandwidth rather than by optics~\cite{KVT2023,Galanis23}.

Distinct spatial frequencies are sampled by varying the projected baseline, the component of $\vect{d}$ perpendicular to the line of sight. As Earth rotates, a fixed physical baseline traces an arc through $(u,v)$ space, sweeping through many Fourier modes. We approximate this for an idealized array by drawing points uniformly within the sampling region. 
In practice the coverage of a real array is sparse and highly non-uniform, set by the telescope layout, the source declination, and the observable hour angles. However, with many collecting elements, and hence many simultaneous projected baselines, it can nonetheless be made to span this region of the $(u,v)$ plane reasonably well~\cite{Dravins2013CTA}. Our uniform sampling should therefore be understood as an idealization, and a realistic layout would degrade the sensitivities quoted below somewhat.

A real interferometer measures visibilities at discrete $(u,v)$ points rather than over a continuum of modes.
We approximate this binning by drawing $10^3$ representative points, each standing in for $10$ underlying measurements, for $10^4$ effective measurements in total, and rescale the uncertainty as $\sigma_C = \sigma_{C,\text{base}}/\sqrt{10}$ to reflect this averaging (see Fig.~\ref{fig:sampling}).
The sampling coordinates are generated only once using the macrolensed-only image and then held fixed for all microlensed realizations when generating the mock observations.

We forecast the precision attainable on $\vect{p}$ from the curvature of the likelihood about its true value. The Cram\'er--Rao bound states that no unbiased estimator can do better than the inverse Fisher matrix, $\mathrm{Cov}(\vect{p}) \succeq F^{-1}$ (see e.g.~\cite{Fisher_overview}). We take the bound to be saturated, which is exact when the likelihood is Gaussian in $\vect{p}$ near $\vect{p}_0$ and approximate otherwise.
We construct a likelihood that compares mock observations generated at the true parameters $\vect{p}_{0}$ with predictions evaluated at trial parameters $\vect{p}$:
\begin{equation}\label{eq:LL_full}
    \log \mathcal{L}=\log \mathcal{L}_{\mathrm{II}}+\log \mathcal{L}_{\mathrm{mag}},
\end{equation}
where
\begin{equation}\label{eq:LL_II}
\log \mathcal{L}_{\mathrm{II}} =
-\sum_i \frac{\left(C_{i}(\vect{p}) - C_{i}(\vect{p}_{0})\right)^2}{2\sigma_C^2},
\end{equation}
and
\begin{equation}\label{eq:llmag}
\log \mathcal{L}_{\mathrm{mag}} = -\frac{1}{2}\left(\frac{\mu_{\vect{p}, \vect{p}_0} - \mu_{\vect{p}_0, \vect{p}_0}}{\sigma_\mu}\right)^2 = -\frac{1}{2}\left(\frac{\mu_{\vect{p}, \vect{p}_0} - 1}{ \sigma_\mu}\right)^2.
\end{equation}
Here $C_i(\vect{p}_0)$ is the correlation of the mock data (generated at the true parameters) and $C_i(\vect{p})$ is the model prediction at trial parameters $\vect{p}$. Likewise, $\mu_{\vect{p}, \vect{p}_0}$ is Eq.~\ref{eq:mag_def} applied to the same macro-image at the two parameter points: the magnification of the trial-parameter image relative to the true-parameter one. The second equality in Eq.~\ref{eq:llmag} uses $\mu_{\vect{p}_0,\vect{p}_0}=1$, since an image has unit magnification relative to itself.

We build a Fisher matrix $F$ for each likelihood term, with $(i,j)$ component
\begin{equation}\label{eq:Fisher}
F_{ij}=\mathbb{E}\!\left[
\frac{\partial\log\mathcal{L}}{\partial p_i}
\frac{\partial\log\mathcal{L}}{\partial p_j}
    \right].
\end{equation}
See App.~\ref{app:pipeline} for a more detailed explanation of the Fisher matrix construction pipeline. From $F$, we derive parameter uncertainties for a single set of observations.

Although the two terms are computed separately, the magnification measurement alone yields no meaningful constraint: it fixes only a single direction in the four-dimensional parameter space, along which simultaneous changes in the parameters leave the magnification unchanged. The II term breaks this degeneracy, so the joint analysis is far more constraining.

We then propagate these parameter uncertainties into a light-centroid precision. Assuming a Gaussian likelihood, we perform a first-order error propagation from $\vect{p}$ to $\vect{\theta}_c$ via the Jacobian
\begin{equation}\label{eq:Jcent}
J_{ij}=\frac{\partial \theta_{c,i}}{\partial p_j}.
\end{equation}
using a finite difference approximation, as we do for the magnification term in Eq.~\ref{eq:LL_full}. The full covariance of $\vect{\theta}_c$ is given by
\begin{equation}\label{eq:cov_thetac}
    \Sigma_{\vect{\theta}_c} = J F^{-1} J^T.
\end{equation}
From this covariance matrix, we define two centroid-precision metrics. As a $2\times2$ covariance, $\Sigma_{\vect{\theta}_c}$ defines a one-$\sigma$ uncertainty ellipse in centroid-position space, whose semi-axes are the square roots of its eigenvalues. The first metric, $\sigma_{|\vect{\theta}_c,1|}$, characterizes the overall localization precision (the geometric mean of the standard deviations given by the two semi-axes),
\begin{equation}\label{eq:theta_C1}
\sigma_{|\vect{\theta}_c,1|} =
\left[\det\left(\Sigma_{\vect{\theta}_c}\right)\right]^{1/4}.
\end{equation}
The second metric, $\sigma_{|\vect{\theta}_c,2|}$, is the standard deviation along the best-constrained principal direction:
\begin{equation}\label{eq:theta_C2}
    \sigma_{|\vect{\theta}_{c},2|} = \sqrt{\min{\left(\upsilon \ \middle|\ \Sigma_{\vect{\theta_c}}\vect{z}=\upsilon \vect{z}\right)}}
\end{equation}
for eigenvalue $\upsilon$ and eigenvector $\vect{z}$.

We consider lens masses $m_l=[0.1, 0.2, 0.4, 0.8]\,M_\odot$ and, for each mass, place the lens at every point on a dense two-dimensional grid of positions $\vect{\theta_{\text{sep}}}$ and generate mock II and magnification observations. For each configuration, we compute the Fisher matrix, the $\chi^2=-2\log\mathcal{L}$ relative to the macrolensing-only model, and the light centroid with its Jacobian $J$. In this single-epoch analysis, the microlensing model introduces three additional free parameters ($m_l$, $\theta_{\text{sep, }x}$, $\theta_{\text{sep, }y}$) relative to the macrolensing-only model, for which $r_\mathrm{src}$ is the only free parameter. We therefore define microlensing configurations with $\chi^2>6$, to be detectable.

\subsection{Time-Domain Analysis}\label{subsec:traj}
We combine epochs along a simulated lens trajectory to improve the static-parameter constraints and measure transverse proper motion. The fiducial campaign observes four times a year (every 91 days) for eight years, giving 33 epochs when both endpoints are included.

We model a lens trajectory relative to a stationary quasar macro-image, parametrized by the impact parameter $\vect{\xi}$, the lens--image separation at closest approach: its magnitude is $|\vect{\xi}| = \min_t |\vect{\theta}_\mathrm{sep}(t)|$, attained at time $t=0$, and its direction makes an angle $\psi$ with the macrolensing shear axis. 
We draw $\psi$ uniformly over $[0, 2\pi]$ and give every lens the same relative transverse speed, $|\vect{v}| = 472\,\mathrm{km\,s^{-1}}$ for B1422+231, combining bulk and internal motion as in Eq.~\ref{eq:v_tot}.

We assign the full relative source--lens transverse motion to the lens and work in an image-fixed frame in which the quasar macro-image is stationary by construction. Physically, relative source--lens motion displaces the quasar's macro-image, and small displacements are amplified in the image plane by the inverse magnification tensor $A^{-1}_{ij}$; boosting to the image-fixed frame transfers this amplification to the apparent motion of the lens relative to the image. The lens position on our $\vect{\theta}_\mathrm{sep}$ grids therefore evolves as $\vect{\theta}_\mathrm{sep}(t) = \vect{\xi} + \tilde{\vect{\nu}}\, t$, with apparent proper motion
\begin{align}
\tilde{\nu}_i = A^{-1}_{ij}\nu_{j}, \label{eq:proper_motion_lens}
\end{align}
where $\nu_j = v_{\perp,j}/[(1+z_l)\,D_l]$ is the unlensed relative proper motion and $\tilde{\nu}_i$ its apparent counterpart in the image-fixed frame \citep{Kochanek_vel}.
We expect a microlens in the B1422+231 lens galaxy to have unlensed proper motion $|\vect{\nu}|=0.069\,\mu\mathrm{as}\,\mathrm{yr}^{-1}$. The apparent rate ranges from $|\vect{\tilde{\nu}}|=0.52\,\mu\mathrm{as}\,\mathrm{yr}^{-1}$ for relative motion along the direction of maximal image stretching (the shear axis) down to $|\vect{\tilde{\nu}}|=0.063\,\mu\mathrm{as}\,\mathrm{yr}^{-1}$ for motion perpendicular to it, with an angle-averaged RMS of $0.37\,\mu\mathrm{as}\,\mathrm{yr}^{-1}$.

To forecast the measurement uncertainties at any point along a trajectory, we linearly interpolate the Fisher matrix of Eq.~\ref{eq:Fisher} (element-wise), the $\chi^2$ values, and the centroid Jacobian of Eq.~\ref{eq:Jcent} for each lens mass, using \texttt{scipy}'s \texttt{RegularGridInterpolator}~\cite{scipy}; we also interpolate the derived uncertainties as a function of position for illustration (Figs.~\ref{fig:rad_04}~and~\ref{fig:lens_grid}). We do not extrapolate beyond the $\vect{\theta}_\mathrm{sep}$ grid used in Sec.~\ref{subsec:sing}. From these position-dependent functions, we then assemble the Fisher matrix $F$, the $\chi^2$ values, and the centroid Jacobian for each epoch along the trajectory.

At each epoch, Eq.~\ref{eq:cov_thetac} gives the centroid covariance. We propagate these covariances to the light-centroid acceleration. We build a Fisher matrix $F_{\vect{\theta}_c}$ for the centroid's kinematic parameters $(\theta_{c,x0},\, \theta_{c,y0},\, v_x,\, v_y,\, a_x,\, a_y)$,
\begin{equation}\label{eq:fisher_theta_c}
F_{\vect{\theta}_c}=\sum_i G_i^T\Sigma_{\vect{\theta}_c,t_i}^{-1}G_i.
\end{equation}
where the sum runs over epochs $t_i$, $\Sigma_{\vect{\theta}_c,t_i}$ is the centroid covariance at epoch $t_i$ (Eq.~\ref{eq:cov_thetac}), and $G_i$ is the Jacobian of the centroid position with respect to these kinematic parameters given by
\begin{equation}\label{eq:cent_dt}
G_i =\begin{bmatrix}
    1 & 0 & t_i & 0 & t_i^2/2 & 0 \\
    0 & 1 & 0 & t_i & 0 & t_i^2/2
\end{bmatrix}.
\end{equation}
Inverting $F_{\vect{\theta}_c}$ gives the kinematic-parameter covariance. The acceleration uncertainties are the diagonal standard deviations of its acceleration sub-block:
\begin{equation}\label{eq:centroid_acc}
\sigma\left[\frac{\partial^2 \vect{\theta}_c}{\partial t^2}\right] = \sqrt{\left(F_{\vect{\theta}_c}^{-1}\right)_{ii}}, \quad i\in\{a_x,a_y\}.
\end{equation}

From the per-epoch 4D Fisher matrices for $\vect{p}$, we next build a six-dimensional Fisher matrix that adds the closest-approach impact parameter $\vect{\xi}$ and the apparent proper motion $\tilde{\vect{\nu}}$ (Eq.~\ref{eq:proper_motion_lens}) as parameters; physical velocities quoted below are recovered as $\vect{v}_\perp = (1+z_l)\,D_l\,\vect{A}\,\tilde{\vect{\nu}}$. We define the 6D parameter vector $\vect{\Theta} = [\theta_\mathrm{src}, m_l, \xi_x, \xi_y, \tilde{\nu}_x, \tilde{\nu}_y]$. Along the trajectory, $\vect{p}$ is a function of $\vect{\Theta}$ and time (the lens position evolves as $\vect{\theta}_\mathrm{sep}(t)=\vect{\xi}+\tilde{\vect{\nu}}\,t$), so to quadratic order in the offsets from the truth,
\begin{align}
-2\ln \mathcal{L} 
&=\sum_t \sum_{ij} F^{\vect{p}}_{ij}(t)\, \Delta\vect{p}_i(\vect{\Theta}, t)\,\Delta\vect{p}_j(\vect{\Theta}, t) \\
&= \sum_{mn} F^{\vect{\Theta}}_{mn}\, \Delta\vect{\Theta}_m\,\Delta\vect{\Theta}_n.
\end{align}
Equating the two forms gives
\begin{equation} \label{eq:ftheta}
F^{\vect{\Theta}} = \sum_k M(t_k)\, F^{\vect{p}}(t_k)\, M^T(t_k),
\end{equation}
where $F^{\vect{p}}(t_k)$ is the per-epoch 4D Fisher matrix and $M(t_k) = (\partial \vect{p}/\partial \vect{\Theta})^T$ is the transpose of the Jacobian of $\vect{p}$ with respect to $\vect{\Theta}$,
\[M= \begin{pmatrix}
    1 & 0& 0 &0 \\
    0 & 1& 0 &0 \\
    0 & 0& 1 &0 \\
    0 & 0& 0 &1 \\
    0 & 0& t &0 \\
    0 & 0& 0 &t \\
\end{pmatrix}.\]
Inverting $F^{\vect{\Theta}}$ then yields the 6D covariance, properly combining information across epochs.

For some microlensing configurations, a single epoch may not produce a decisive detection. We therefore combine the information from repeated observations by summing the $\chi^2$ values over the epochs of the eight-year campaign and use the resulting total to determine whether the microlensing model is favored over the macrolensing-only model. In this time-domain analysis, the threshold for detectability (critical $\chi^2$ value) is $\chi^2>10$, compared to $\chi^2 > 6$ in the single-epoch analysis due to the four additional parameters (both components of $\vect{\xi}$ and $\vect{nu}$). 
This lets us probe $|\vect{\xi}|$ values beyond those detectable in a single image; we accordingly choose the $\vect{\theta}_\mathrm{sep}$ grid of Sec.~\ref{subsec:sing} broad enough that any trajectory reaching cumulative preference for microlensing stays within the interpolation bounds.

For each lens mass and value of $|\vect{\xi}|$, we generate at least $10^5$ trajectories, drawing $\psi$ independently each time to marginalize over approach angle (in the unlensed frame); this yields the distribution of parameter uncertainties expected at that $|\vect{\xi}|$ and lens mass.
If a trajectory leaves the interpolation domain, we redraw $\psi$ and recompute it. Since frequent redraws would bias the results, we chose the $\vect{\theta}_\mathrm{sep}$ grid broad enough that this is rare.

We repeat this procedure for increasing $|\vect{\xi}|$, marginalizing over approach angle at each step, and continue as long as the median combined $\chi^2$ exceeds $10$ (evidence for microlensing across the full trajectory). Beyond that threshold, the macrolensing-only model is favored and we report no parameter uncertainties. We use the same range of $|\vect{\xi}|$ for every lens mass.

\section{Results} \label{sec:results}

\subsection{Single-Epoch Results}
We present single-epoch forecasts for microlenses of mass $m_l=[0.1, 0.2, 0.4, 0.8]M_\odot$. Main-text figures use $m_l=0.4\,M_\odot$; App.~\ref{app:microlens} shows plots for the other masses.
We generate the macrolensing-only image of B1422+231 and sample its correlation values (Fig.~\ref{fig:sampling}) to obtain the baseline macrolensing-only model for $C_0$ (the square modulus of the visibility), which carries no microlensing magnification by construction. Appendix~\ref{app:pipeline} details how we mitigate numerical noise when generating the real- and Fourier-space images.

\begin{figure*}
    \centering
    \includegraphics[width=0.8\linewidth]{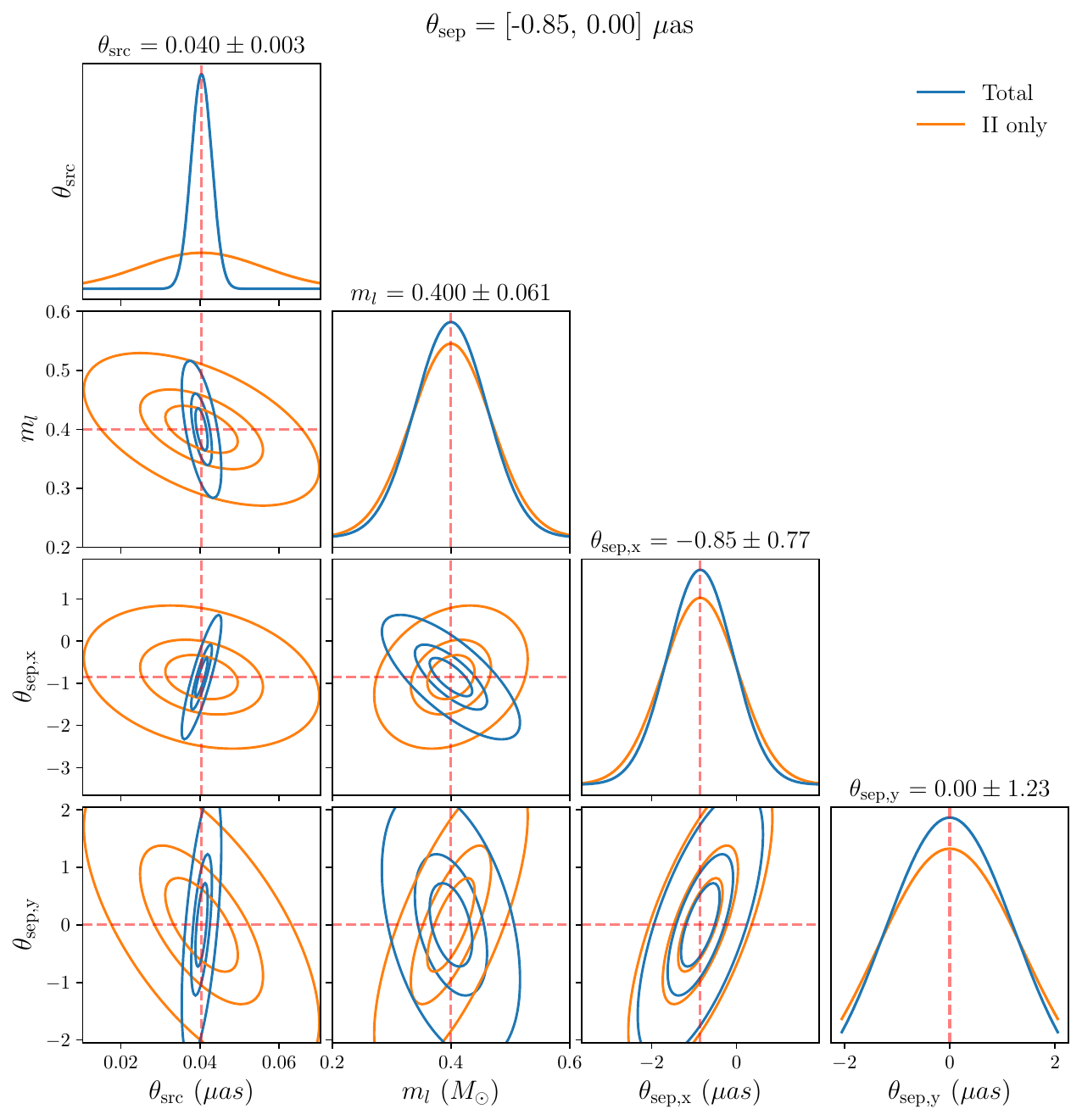}
    \caption{\micronblink{interp_grid04.ipynb} Corner plot for a single simulated observation with lens mass $m_l=0.4\,M_\odot$ and $\vect{\theta}_\mathrm{sep}=[-0.85, 0.00]\ \theta_{E,0.4}$. The analysis of this configuration illustrates the impact of including the magnification likelihood term in the ``Total'' constraint. Contours correspond to the 16th, 50th, and 84th percentiles, and the red lines indicate the true parameter values.}
    \label{fig:corner_ex}
\end{figure*}

Following Sec.~\ref{sec:analysis}, we generate mock data for II and magnification measurements over a dense grid of microlens positions for each lens mass. We evaluate only one quadrant of microlens positions (relative to the macrolensing shear axis) and reflect the results across both axes, exploiting the symmetry of the configuration, to cover all $\vect{\theta}_\mathrm{sep}$ within the grid.
Figure~\ref{fig:corner_ex} compares the II-only and total (II $+$ magnification) constraints for a fiducial configuration. As anticipated, the magnification term constrains a single direction in the four-dimensional parameter space, tightening the overall constraint; additional corner plots appear in Appendix~\ref{app:microlens}.

\begin{figure*}
    \centering
    \begin{subfigure}[b]{0.49\textwidth}
            \includegraphics[width=\textwidth]{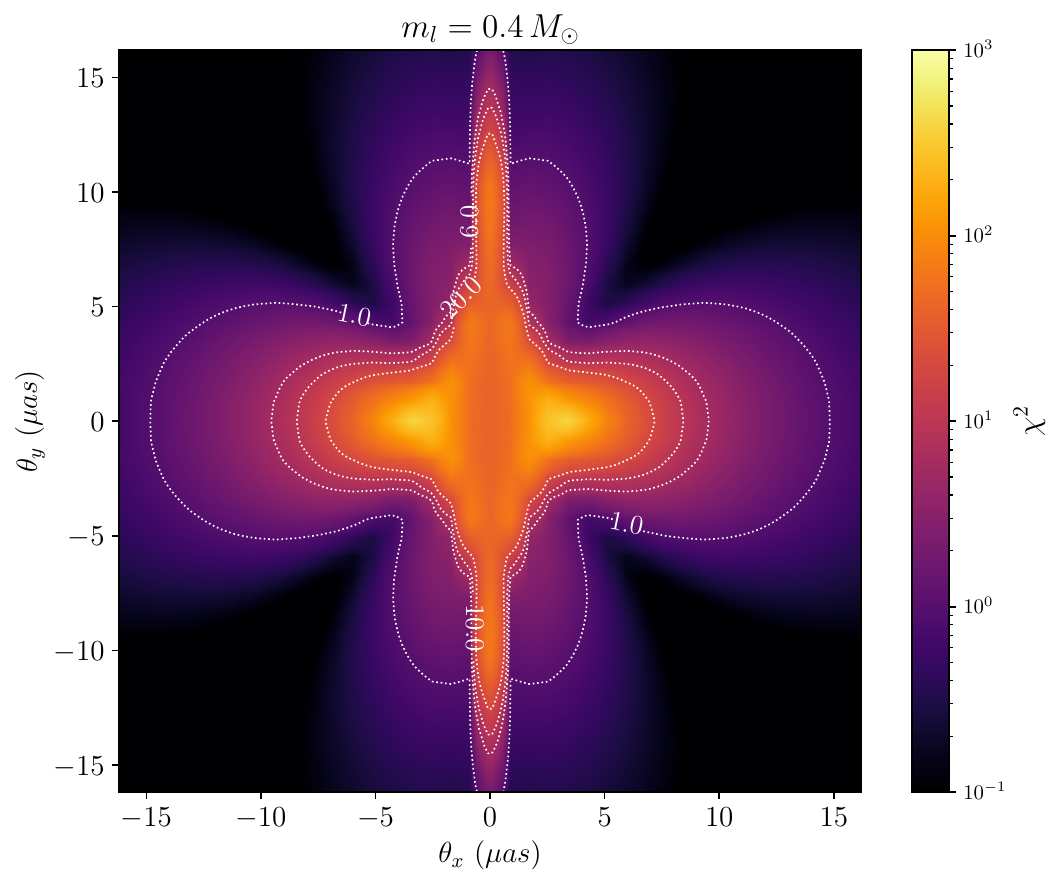}
            \caption{Single-epoch $\chi^2$}
            \label{fig:chi2_2}
        \end{subfigure}
        \begin{subfigure}[b]{0.49\textwidth}
            \includegraphics[width=\textwidth]{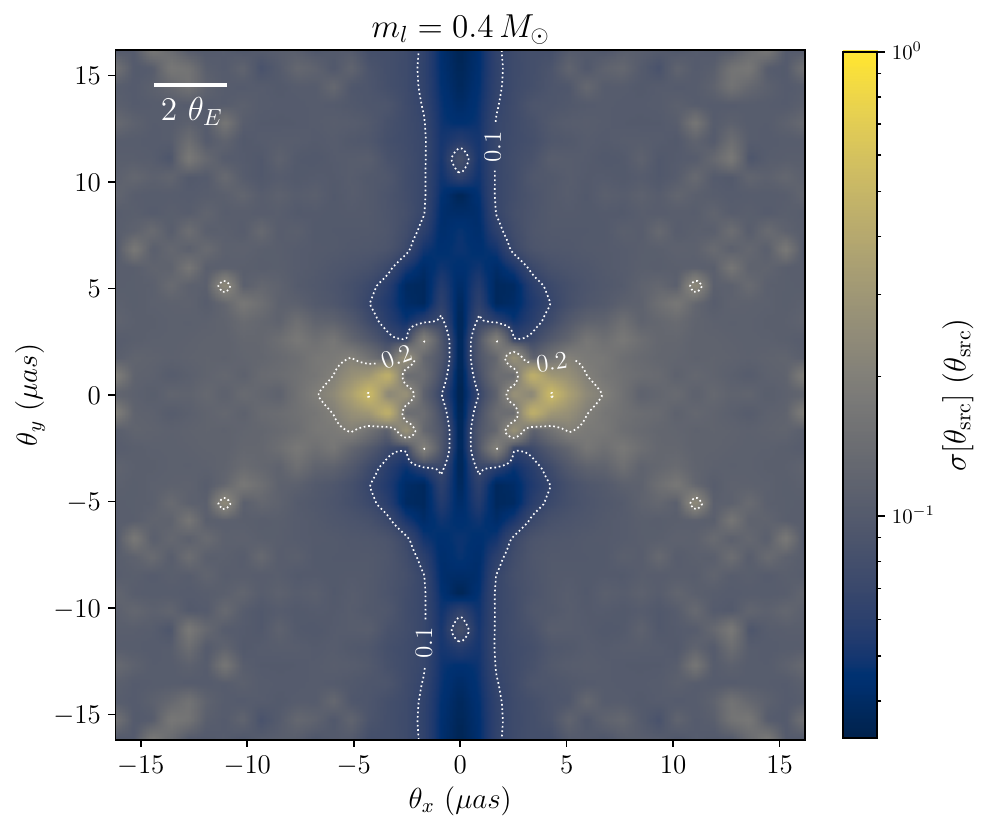}
            \caption{Single-epoch uncertainty of $\theta_\mathrm{src}$ in units of $\theta_\mathrm{src}$}
            \label{fig:rad_04}
        \end{subfigure}
        \caption{\micronblink{interp_grid04.ipynb} Single-epoch quantities for $m_l=0.4\,M_\odot$ in image-plane coordinates $\vect{\theta}=(\theta_x,\theta_y)$ (in $\mu\mathrm{as}$), rotated so that the macrolensing shear axis is vertical. The left colorbar is the detection statistic $\chi^2$ (dimensionless, logarithmic scale) and the right colorbar is the uncertainty associated with $\theta_\mathrm{src}$ in units of $\theta_\mathrm{src}$. The $\chi^2$ map shows that microlensing is detectable in a single snapshot over much of the parameter space, generally decreasing with radius. The source radius is likewise resolvable throughout the region, especially for lenses along the shear axis. The right panel also marks twice the Einstein radius, $2\theta_{E,0.4}$, for reference. }
\end{figure*}

For each microlensing configuration, we compute the $\chi^2$ relative to the macrolensing-only case; Fig.~\ref{fig:chi2_2} shows the interpolated map for $m_l=0.4\,M_\odot$. This figure, and all that follow, are rotated by $\phi_0$ so that the macrolensing shear axis (the white dashed line in Fig.~\ref{fig:basic_lensing}) is vertical.
For all four lens masses, the microlensing model is preferred ($\chi^2 > 6$, since the microlensing model includes three more parameters than the macrolensing-only model) in a single observation when the lens lies within $\sim 3\,\theta_{E,m_l}$ of the source for any approach direction; the detectable region is strongly elongated, however, extending to $\sim8\,\theta_{E,m_l}$ along the macrolensing shear axis and $\sim5\,\theta_{E,m_l}$ perpendicular to it. We find a total detectable area of $\sim 60 \theta_{E, m_l}^2$ by integrating the area over the region satisfying the detectability criterion.

\begin{figure*}[htbp]
    \centering
    \def\arraystretch{1}
    \setlength\tabcolsep{0pt}
    \begin{tabular}{cc}
        \begin{subfigure}[b]{0.48\textwidth}
        \includegraphics[height=0.8\textwidth]{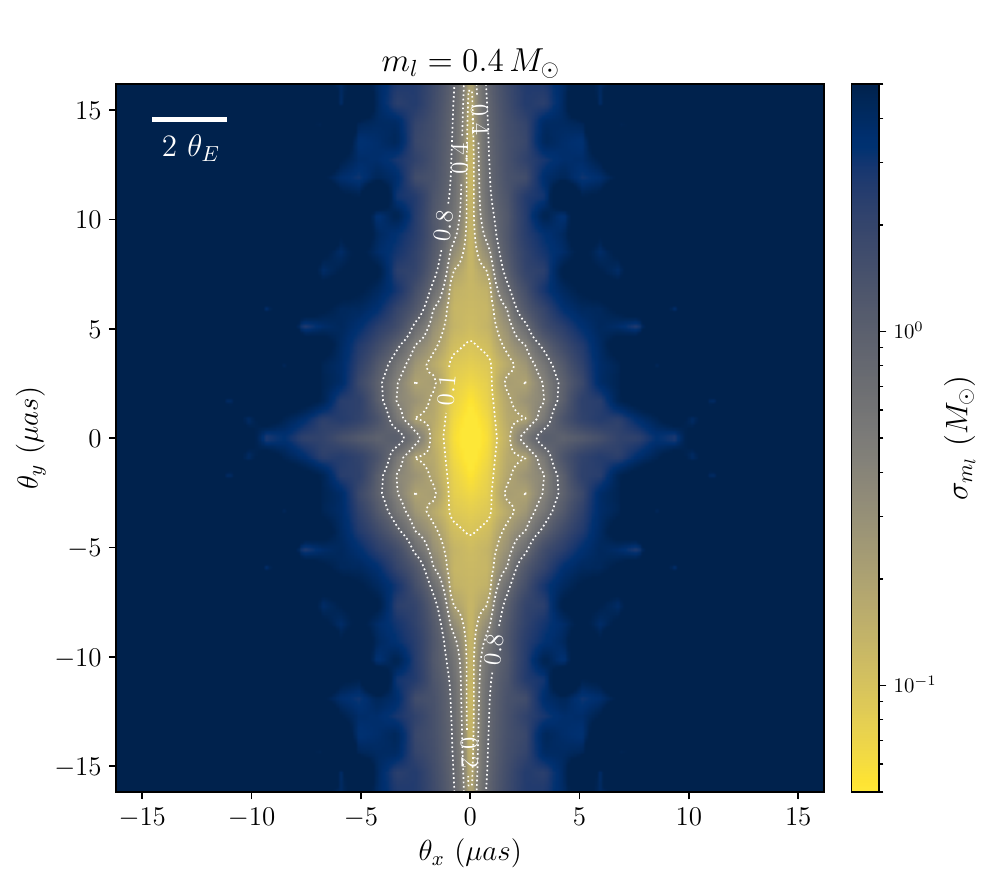}
        \end{subfigure}
        &
        \begin{subfigure}[b]{0.48\textwidth}
            \includegraphics[height=0.8\textwidth]{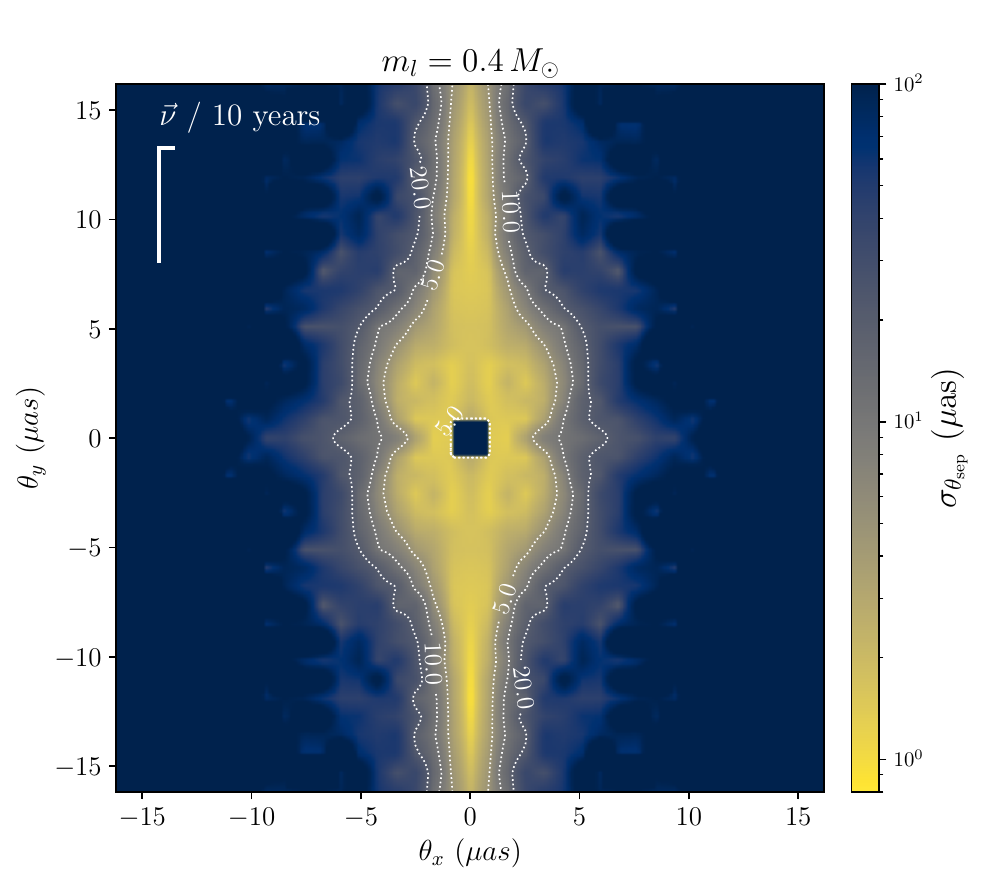}
        \end{subfigure} \\
        \begin{subfigure}[b]{0.48\textwidth}
        \includegraphics[height=0.8\textwidth]{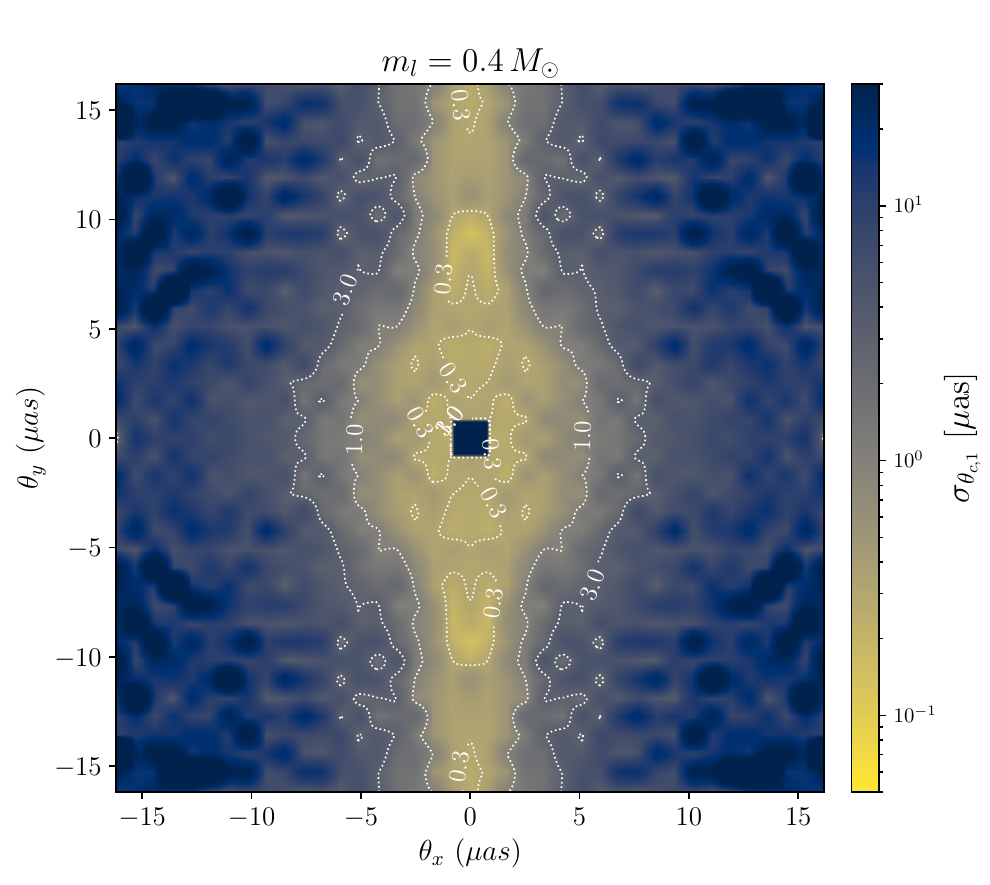}
        \end{subfigure}
        &
        \begin{subfigure}[b]{0.48\textwidth}
        \includegraphics[height=0.8\textwidth]{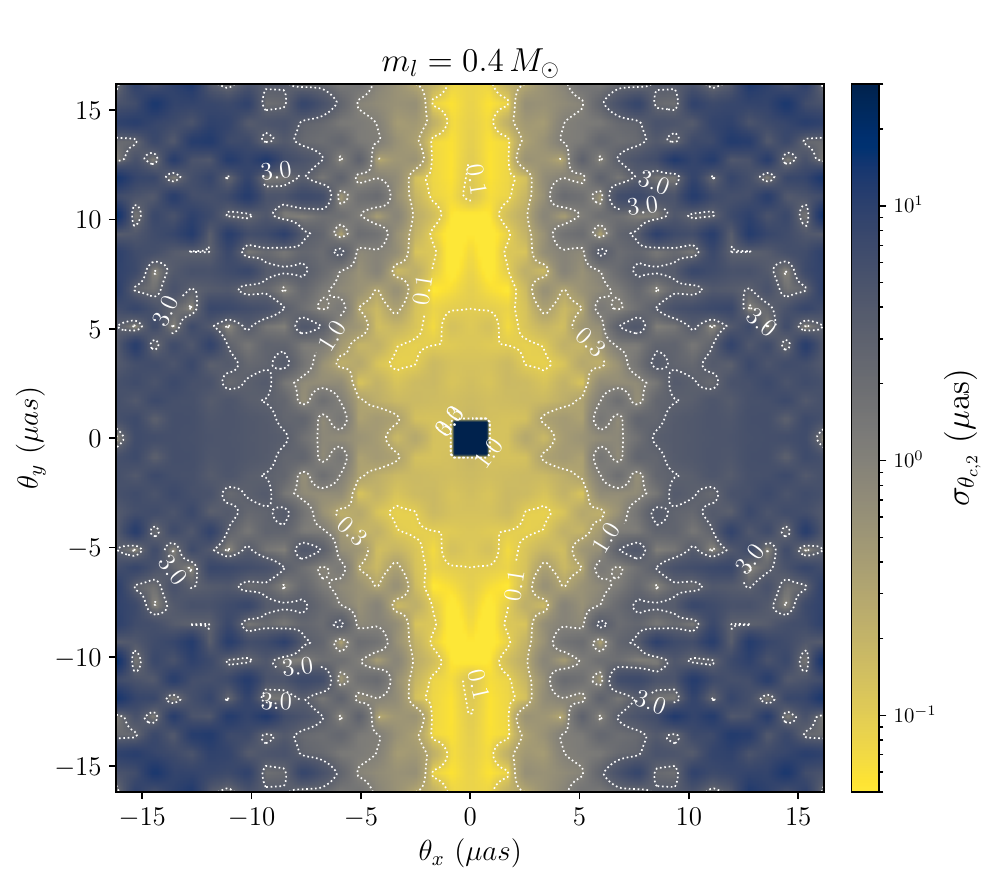}
        \end{subfigure} \\
    \end{tabular}
    \caption{\micronblink{interp_grid04.ipynb} Interpolated single-epoch precisions (including II and magnification information) as a function of lens position $\vect{\theta}_\mathrm{sep} = (\theta_x, \theta_y)$ for a lens of mass $m_l=0.4\,M_\odot$. In the top row we show the parameter uncertainty associated with lens mass, $\sigma[m_l]$ (left; colorbar in $M_\odot$), and lens position, $\sigma[\theta_\mathrm{sep}]$ (right; colorbar in $\mu\mathrm{as}$). In the second row we show interpolated single-epoch light-centroid precision, $\sigma[\theta_{c,1}]$ defined in Eq.~\ref{eq:theta_C1} (left) and $\sigma[\theta_{c,2}]$ defined in Eq.~\ref{eq:theta_C2} (right), both with colorbars in $\mu\mathrm{as}$. All four colorbars use a logarithmic scale, and the axes are the image-plane angular coordinates $(\theta_x,\theta_y)$ in $\mu\mathrm{as}$. Results combine II and magnification constraints. We mark twice the Einstein radius $2\theta_{E,0.4}$; the white scale bars on the position panel show the apparent proper-motion scale $|\tilde{\vect{\nu}}|\times10\,\mathrm{yr}$ along (vertical) and perpendicular to (horizontal) the shear axis (Eq.~\ref{eq:proper_motion_lens}).
    }
    \label{fig:lens_grid}
\end{figure*}

The source radius is best resolved for lenses along the macrolensing shear axis (Fig.~\ref{fig:rad_04}), to precisions of $\sim 4\%$; for lenses in a band along the perpendicular ($x$) axis the precision drops to $\sim60\%$, its floor being set by the magnification constraint, which retains sensitivity to $\theta_\mathrm{src}$ at any lens position. The same behavior holds for $m_l=0.1$, $0.2$, and $0.8\,M_\odot$ at the same $\vect{\theta}_\mathrm{sep}$ (in units of $\theta_{E,m_l}$).

We interpolate the remaining parameter uncertainties and every element of the Fisher matrices, allowing us to reconstruct the full information content of an observation for a lens anywhere in the grid. Figure~\ref{fig:lens_grid} shows the resulting uncertainty maps for $m_l=0.4$, and the same results are shown in Fig.~\ref{fig:other_mass_grid} for all four fiducial lens masses, 
with identical color scales for each quantity. The lens-mass precision is worse than that of the source radius, but a band of positions along the $y$ (shear) axis allows the mass to be resolved, widening for larger $m_l$.
Lens-position constraints are weaker still: the precision reaches $\sim5 \ \mu \text{as}$ directly above and below the source but falls below unity over most of the parameter space.

The bottom two panels in Fig.~\ref{fig:lens_grid} show the two metrics of light-centroid precision. Both are limited by the lens-position uncertainty, which propagates into the centroid precision; where $\theta_\mathrm{sep}$ is resolvable, the best-constrained centroid direction reaches $\sigma_{|\vect{\theta}_c,2|} \approx 0.04$--$0.1\,\mu\mathrm{as}$ (for lenses along the shear-axis band), while the geometric-mean localization is $\sigma_{|\vect{\theta}_c,1|} \approx 0.2$--$0.5\,\mu\mathrm{as}$. We also consider the eigenvectors associated with the minimum eigenvalue (Eq.~\ref{eq:theta_C2}), which identify the direction of greatest centroid precision. We find that regardless of lens position this direction is typically aligned with the x-axis, perpendicular to the macrolensing shear axis. The mottled numerical noise at large separations in Fig.~\ref{fig:lens_grid} comes from inverting the poorly conditioned Fisher matrix, a consequence of the large dynamic range between the source radius and $\vect{\theta}_\mathrm{sep}$.
These interpolated uncertainty maps are illustrative and are not used in the time-domain analysis: there we interpolate the \emph{pre-inversion} quantities---the Fisher matrix $F$ of Eq.~\ref{eq:Fisher} (element by element), the centroid Jacobian $J$ of Eq.~\ref{eq:Jcent}, and the $\chi^2$. Each of these vary smoothly across the grid since they are free of noise due to matrix inversion. We then use these interpolated quantities to form $\Sigma_{\vect{\theta}_c}=JF^{-1}J^T$ (Eq.~\ref{eq:cov_thetac}) epoch by epoch before accumulating it into Eqs.~\ref{eq:fisher_theta_c} and \ref{eq:ftheta}. The single ill-conditioned inversion is thus never interpolated through, and the sum over the 33 epochs of the campaign further suppresses the imprint of this numerical noise.

The divergence of the source--lens separation uncertainty, and hence of $|\vect{\theta}_c|$, at the smallest separations is a signature of an exact degeneracy of our observables. Because the macrolensing deflection is linear in $\vect{\theta}$ and the source brightness depends only on $|\vect{\beta}|$, the configurations $+\vect{\theta}_\mathrm{sep}$ and $-\vect{\theta}_\mathrm{sep}$ produce images related by a point reflection about the source center. Intensity interferometry is sensitive only to $C \simeq |\mathcal{V}|^2$ (Eq.~\ref{C_int}), and a real intensity distribution satisfies $\mathcal{V}(-\vect{u}) = \mathcal{V}^*(\vect{u})$, so the two configurations yield \emph{identical} correlations at every $\vect{u}$; their total magnifications are likewise equal. The likelihood of Eq.~\ref{eq:LL_full} is therefore exactly even in $\vect{\theta}_\mathrm{sep}$. At the origin, $\partial C/\partial\vect{\theta}_\mathrm{sep}$ vanishes identically, and the corresponding block of the Fisher matrix is exactly singular. The leading dependence on separation is quartic, $\Delta\chi^2\propto|\vect{\theta}_\mathrm{sep}|^4$, so the separation uncertainty remains bounded around the origin. This higher-order information is by construction invisible to a Fisher forecast, so the innermost errors are overestimated. Consistently, $\sigma[m_l]$ and $\sigma[\theta_\mathrm{src}]$, which are even under the reflection, remain finite there.

The same symmetry leaves one discrete degeneracy in the trajectory analysis: reflecting the lens path through the source, $(\vect{\xi},\tilde{\vect{\nu}}) \mapsto (-\vect{\xi},-\tilde{\vect{\nu}})$, changes no observable at any epoch. The lens may pass on either side of the source, moving either way along its track. Both $\vect{\xi}$ and $\tilde{\vect{\nu}}$ are thus determined only up to a common sign. Their magnitudes are not ambiguous to intensity-interferometric data.

\begin{figure*}
    \centering
    \includegraphics[height=0.44\linewidth]{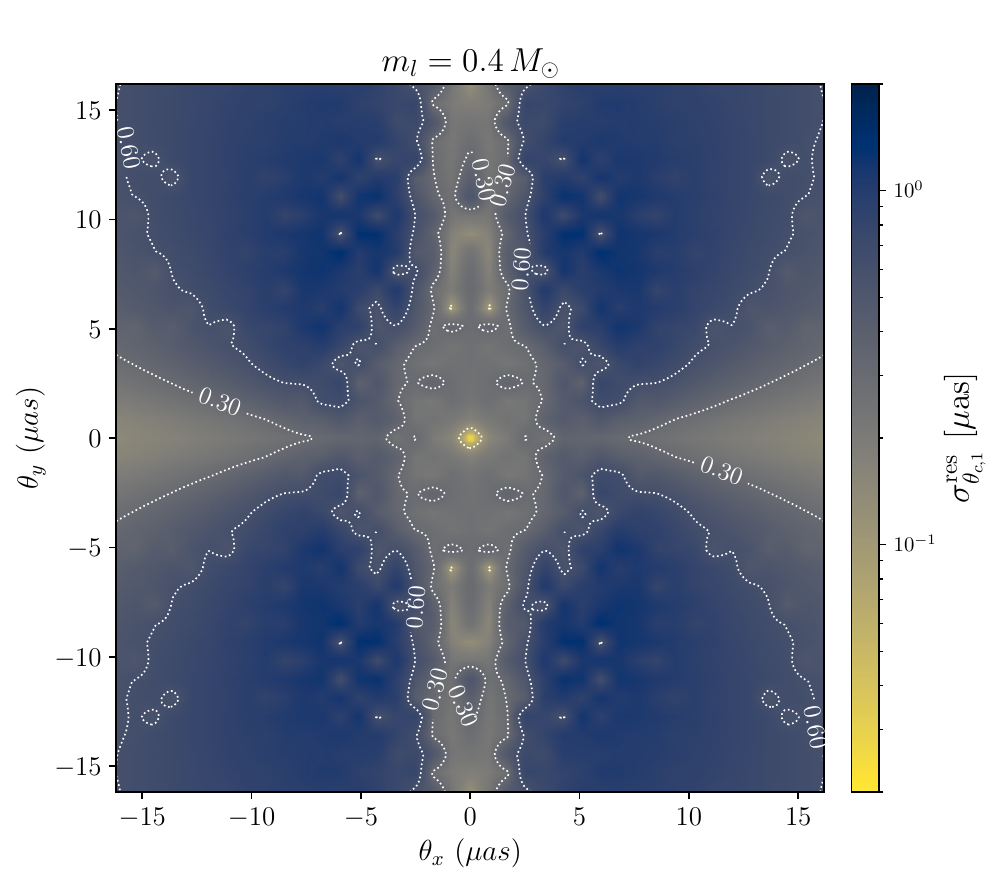}
    \includegraphics[height=0.44\linewidth]{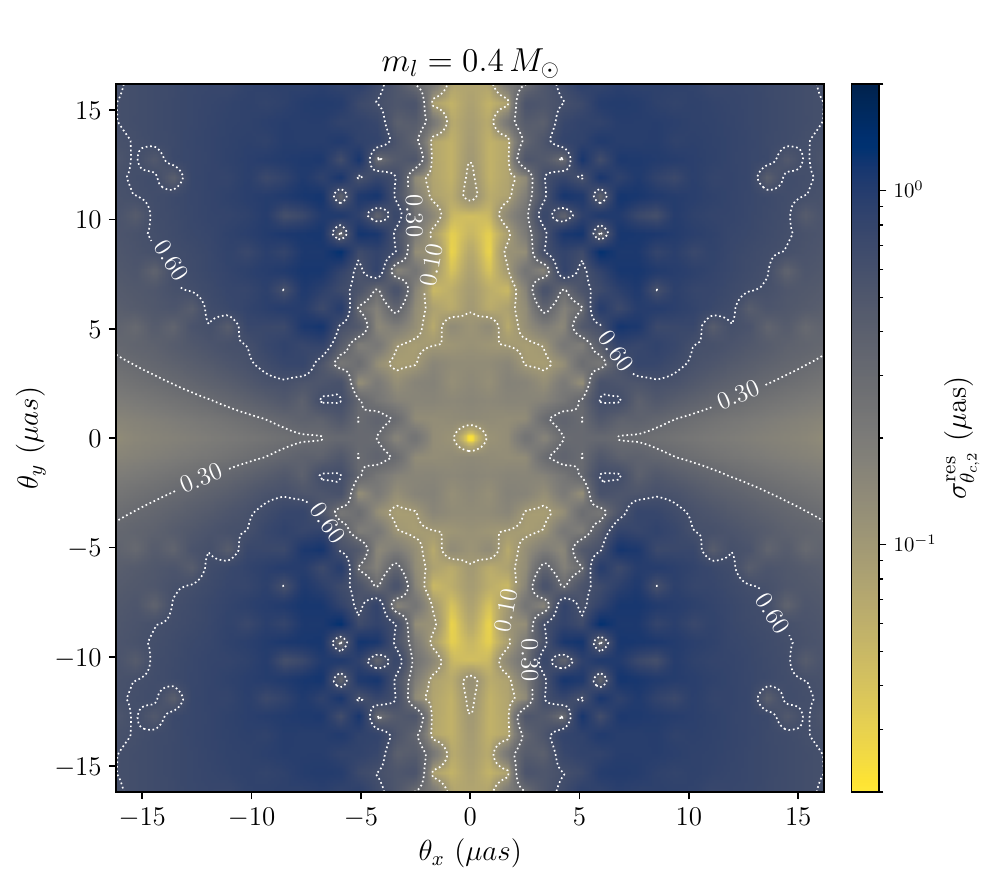}
    \caption{\micronblink{interp_grid04.ipynb} Interpolated uncertainty associated with the residual light centroid position as defined by $\sigma[\theta_{c,1}]$ on the left and $\sigma[\theta_{c,2}]$ in the right panel for $m_l=0.4\,M_\odot$ as a function of lens position $(\theta_x,\theta_y)$. Both colorbars are in $\mu\mathrm{as}$ on a logarithmic scale. The residual error tracks the Cram\'er--Rao precision of Fig.~\ref{fig:lens_grid} where the lens is well measured (small values near the source and along the shear axis), grows toward the single-epoch detection boundary, and is capped by the shrinking microlensing signal $|\Delta\vect{\theta}_c|$ beyond it.}
    \label{fig:thetac_res}
\end{figure*}

The growth of both metrics at large $\theta_\mathrm{sep}$ is physical, but neither of the centroid uncertainties plotted in the bottom two panels of Fig.~\ref{fig:lens_grid} is directly relevant to a centroid-based search for additional weak-lensing perturbers~\cite{DMpaper}. The covariance defined in Eq.~\ref{eq:cov_thetac} is that of the best \emph{unbiased} centroid estimator. Because the lens-parameter covariance $F^{-1}$ diverges faster than the lens Jacobian $J$ shrinks with increasing impact parameter, the centroid uncertainty $\Sigma_{\vect{\theta}_c}=JF^{-1}J^T$ grows even as the perturbation weakens (physically). An unbiased estimator is the wrong target in this regime. The microlensing contribution to the centroid is itself bounded by its shift relative to the macrolensing-only image,
\begin{equation}\label{eq:dthetac}
\Delta\vect{\theta}_c(\vect{p}_0) = \vect{\theta}_c(\vect{p}_0) - \vect{\theta}_c^{\,\mathrm{macro}},
\end{equation}
where $\vect{\theta}_c^{\mathrm{macro}}$ (the origin in our coordinates) is the centroid of the macrolensing-only reference model relative to which the $\chi^2$ and magnification are defined. As $\theta_\mathrm{sep}\to\infty$, $|\Delta\vect{\theta}_c| \simeq |\vect{\alpha}|$ vanishes with the deflection.
If the lens is undetectable one applies no correction and incurs an error of at most $|\Delta\vect{\theta}_c|$; if it is well measured the error is $\sqrt{\Sigma_{\vect{\theta}_c}}$. For forecasting, we regularize the residual covariance by combining these two limiting scales:
\begin{equation}\label{eq:cov_res}
\Sigma_{\vect{\theta}_c}^{\mathrm{res}} = \left(\Sigma_{\vect{\theta}_c}^{-1} + |\Delta\vect{\theta}_c|^{-2}\,\mathbf{I}\right)^{-1},
\end{equation}
with the $2\times2$ identity matrix  $\mathbf{I}$.  
This prescription approaches the Cram\'er--Rao value where the lens is measured and is capped by the shrinking signal where it is not; the crossover is expected near the single-epoch detection threshold, $\chi^2\sim6$. 
One should view $\Sigma_{\vect{\theta}_c}^{\mathrm{res}}$ as a forecast-level interpolation between resolved and unresolved limits, relevant for quantifying stellar microlensing as a systematic background for the DM signal of Ref.~\cite{DMpaper}.
Applying Eqs.~\ref{eq:theta_C1} and \ref{eq:theta_C2} to $\Sigma_{\vect{\theta}_c}^{\mathrm{res}}$ gives the residual centroid precision used below.
We plot this in Fig.~\ref{fig:thetac_res}.

\subsection{Time-Domain Results}
We summarize the results of our time domain analyses for all four lens masses, considering a fiducial observing campagin of $8$ years, with 4 observations spaced evenly throughout each year.

\begin{figure*}
\begin{tabular}{@{}c@{\hskip 0.5em}c@{}}
\includegraphics[width=0.482\textwidth]{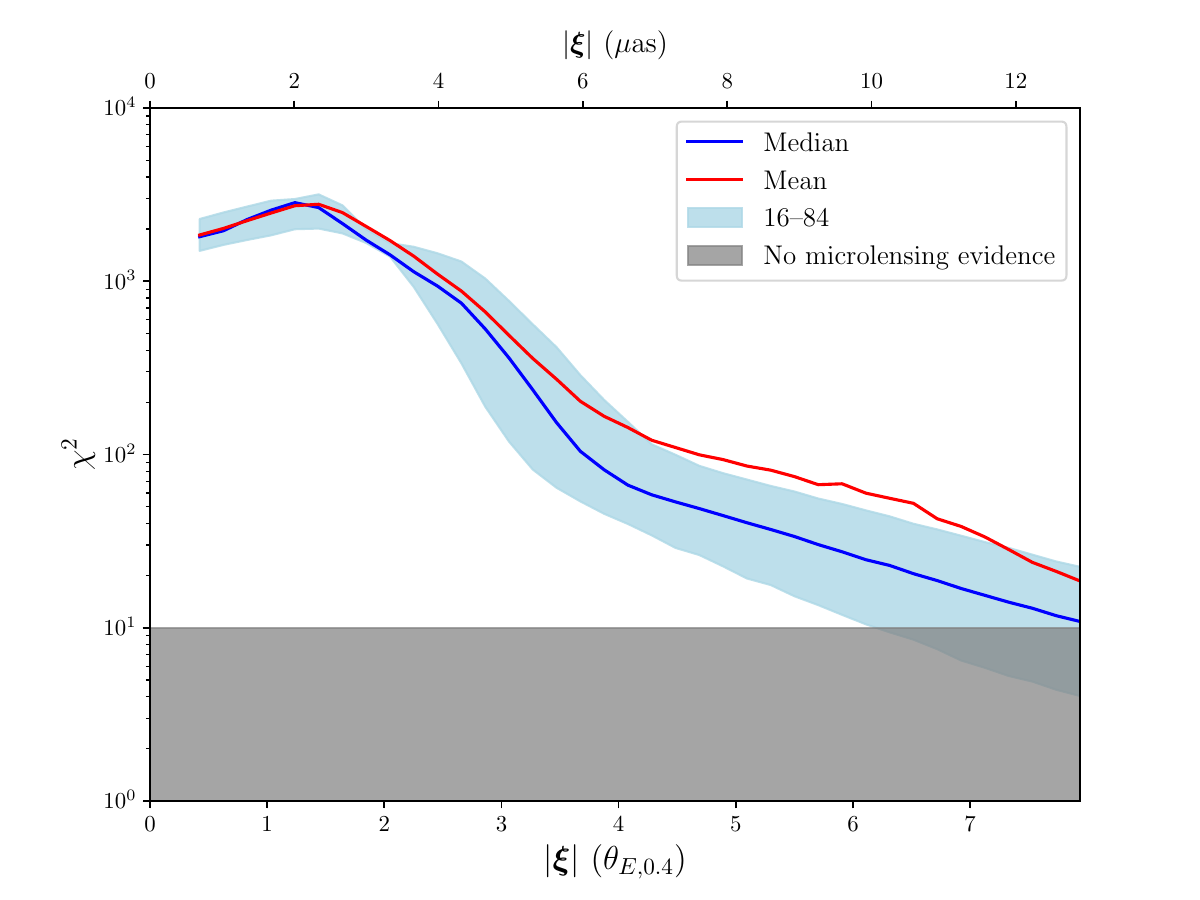} &
\includegraphics[width=0.475\textwidth]{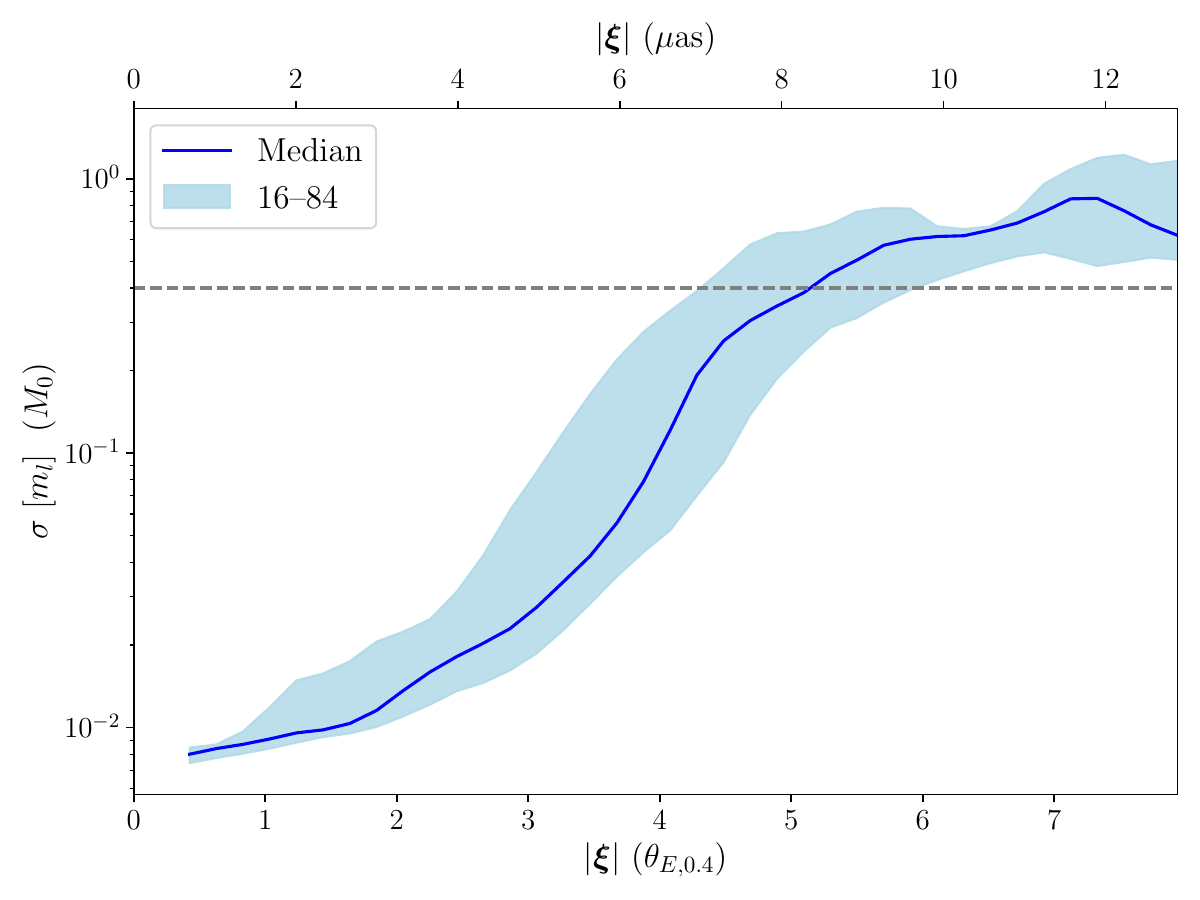} \\[1em]
\hspace{5pt}
\includegraphics[width=0.485\textwidth]{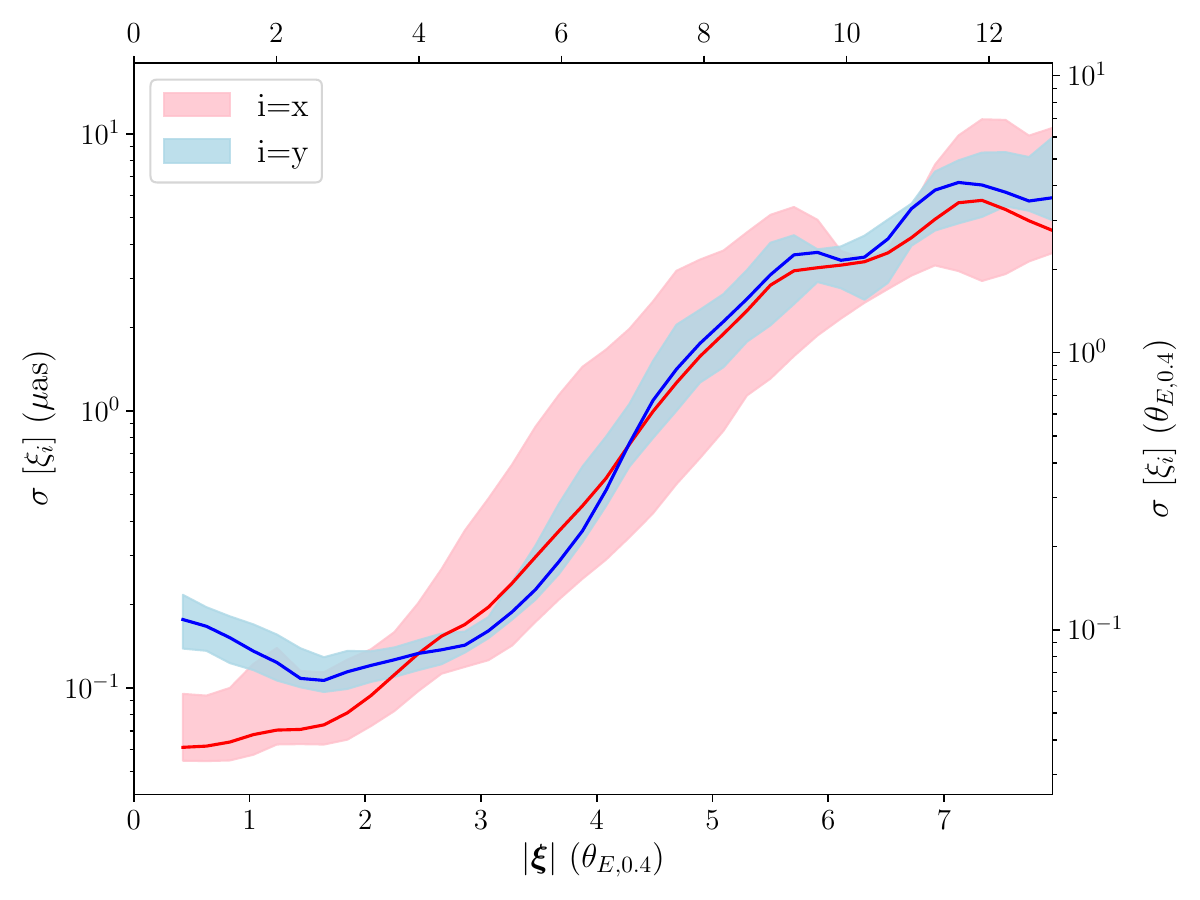} &
\includegraphics[width=0.475\textwidth]{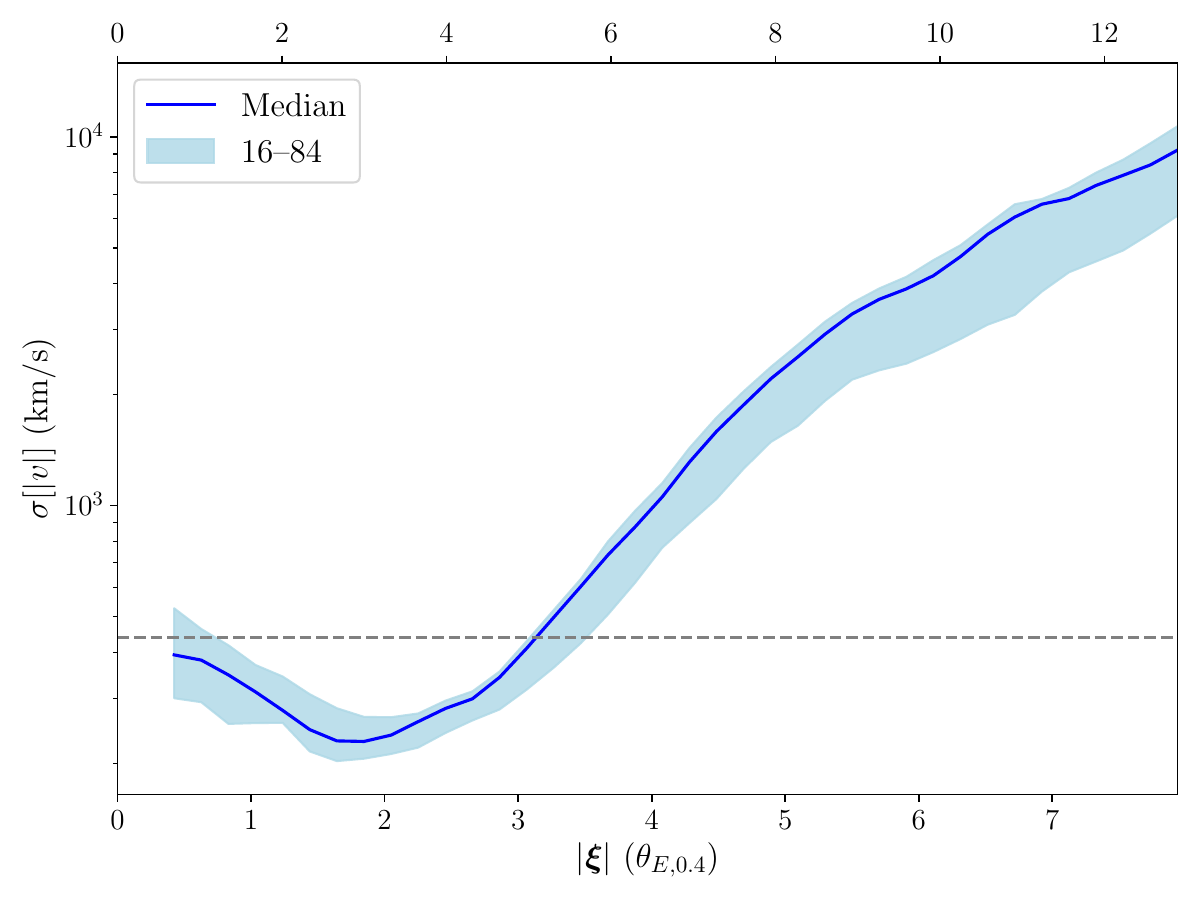}
\\[0.8em]
\hspace{5pt}
\includegraphics[width=0.475\textwidth]{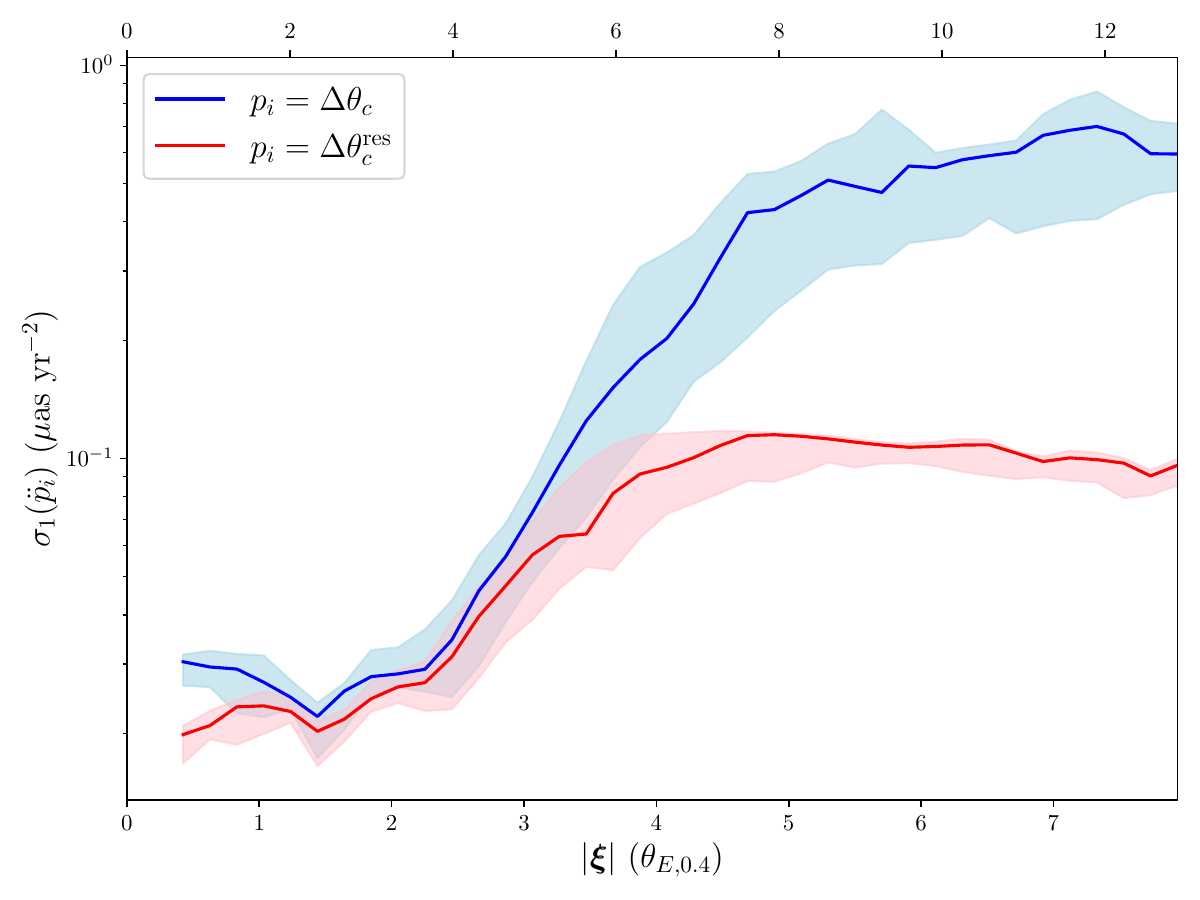}&
\includegraphics[width=0.475\textwidth]{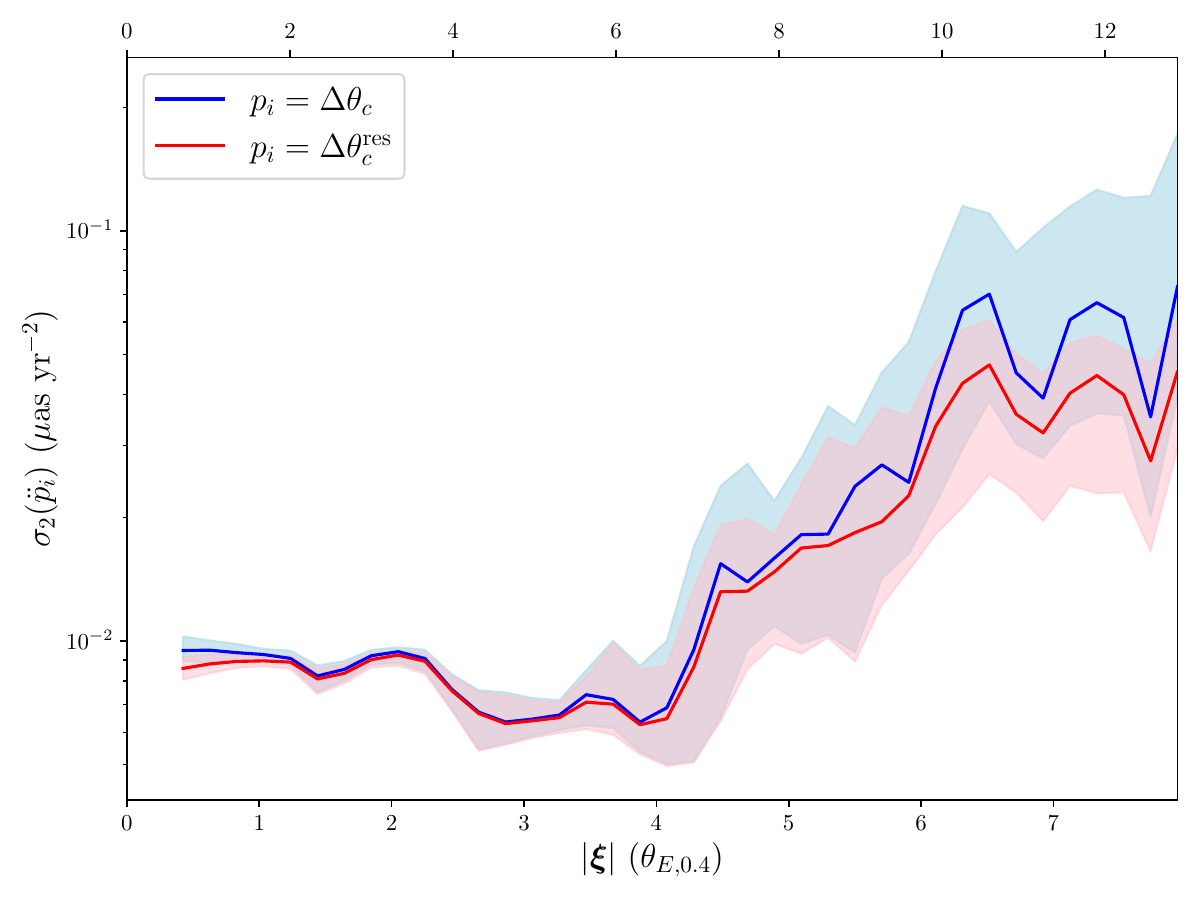}
\\
\end{tabular}
\caption{\micronblink{interp_grid04.ipynb} Results for angle-averaged trajectories as a function of minimum impact parameter $|\vect{\xi}|$, for $m_l=0.4\,M_\odot$. Besides the cumulative $\chi^2$ (top left; dimensionless), we show the median and $16$th--$84$th percentile range of the derived uncertainties on the lens mass $m_l$ (top right; in $M_\odot$), the two components of $\vect{\xi}$ (middle left; in $\mu\mathrm{as}$, with a secondary axis in $\theta_{E,0.4}$), and the lens transverse speed $|\vect{v}|$ (middle right; in $\mathrm{km\,s^{-1}}$). The bottom left panel shows $\sigma_1$ for the direct and residual treatments of the light-centroid acceleration. The bottom right panel shows the corresponding $\sigma_2$ curves, which exhibit similar behavior across the considered range of $|\vect{\xi}|$. 
We use the median rather than the mean because outliers skew the latter for several quantities, making it unrepresentative of a typical lens. The bottom horizontal axis of each panel gives $|\vect{\xi}|$ in units of $\theta_{E,0.4}$ and the top horizontal axis in $\mu\mathrm{as}$; \emph{all vertical axes are logarithmic}. The dashed lines mark the fiducial lens speed (middle right) and $\sigma[m_l]$ equal to the true mass (top right).}
\label{fig:traj_res}
\end{figure*}

We generate trajectories over a range of $|\vect{\xi}|$ for each lens mass, computing for each a combined $\chi^2$ value, all six parameter uncertainties, and the centroid precision, and averaging over approach angle with $\geq10^5$ trajectories per $|\vect{\xi}|$. Figure~\ref{fig:traj_res} shows the resulting $\chi^2$ for $m_l=0.4\,M_\odot$. Because a few outliers skew the mea, especially at large separations, we use the median $\chi^2$ to define when microlensing is typically detectable, and restrict the discussion of derived parameters to this detectable regime.
Fig.~\ref{fig:traj_res} also shows the median parameter uncertainties versus $|\vect{\xi}|$ for $m_l=0.4\,M_\odot$. We resolve the lens mass when $|\vect{\xi}| \lesssim 5.7\,\theta_{E,0.4}$, constrain the microlens impact parameter to within $\sim10\,\mu\mathrm{as}$ across all $|\vect{\xi}|$ considered, and resolve the proper motion for $|\vect{\xi}| \in [1.4, 3.1] \,\theta_{E,0.4}$.

In the same figure, we show $\sigma\left[\frac{\partial^2 \vect{\theta}_c}{\partial t^2}\right]$, as defined in Eq.~\ref{eq:centroid_acc}, quantifying the precision with which the time-dependent centroid perturbation can be inferred from the trajectory. At small and intermediate separations, where the microlens parameters can be inferred with good precision, this uncertainty directly describes the achievable acceleration measurement. The lens parameters become increasingly weakly constrained at large separations, causing the formal acceleration uncertainty to increase even as the physical microlensing-induced centroid perturbation decreases.

To better capture these effects, we extend the same residual treatment defined in Eq.~\ref{eq:cov_res} to our time-domain analysis. Replacing the centroid covariance in the trajectory Fisher matrix with $\Sigma_{\vect{\theta}_c}^{\mathrm{res}}$ gives the corresponding overall localization acceleration precision,
\begin{equation}
\sigma_1\left[\frac{\partial^2 \Delta\vect{\theta}_c^{\text{res}}}{\partial t^2}\right]
=\sqrt{\left[\left(F_{\vect{\theta}_c}^{\mathrm{res}}\right)^{-1}\right]_{ii}},
\qquad i\in\{a_x,a_y\}.
\end{equation}
where $F_{\vect{\theta}_c}^{\mathrm{res}}$ is constructed from the \emph{residual} centroid covariance using the same procedure outlined in Eq.~\ref{eq:fisher_theta_c} and Eq.~\ref{eq:cent_dt}. This quantity is analogous to $\sigma_{| \vect{\sigma}_{c},1|}$, as defined in Eq.~\ref{eq:theta_C1}. We also define the analog to $\sigma_{| \vect{\sigma}_{c},2|}$ (see Eq.~\ref{eq:theta_C2}) as:
\begin{equation}
  \sigma_2\left[\frac{\partial^2 \Delta\vect{\theta}_c^{\text{res}}}{\partial t^2}\right]
= \sqrt{\lambda_{\min}\left[\left(F_{\vect{\theta}_c}^{\mathrm{res}}\right)^{-1}_{a_x a_y}\right]}
\end{equation}
where $\left(F_{\vect{\theta}_{c}}^{\mathrm{res}}\right)^{-1}_{a_x a_y}$ is the $ 2 \times 2$ acceleration sub-block of the inverse residual Fisher matrix, and $\lambda_{\min}$ is its minimum eigenvalue.

Both resulting quantities describe the uncertainty on the measurable microlensing-induced centroid acceleration after accounting for the fact that an unresolved lens cannot produce an arbitrarily large correction.

We show these quantities alongside their direct $\vect{\theta}_c$ analogs (calculated with $F_{\vect{\theta}_c}$ in place of  $F_{\vect{\theta}_c}^{\text{res}}$) in the bottom panels of Fig.~\ref{fig:traj_res}. 
In the bottom left panel, the overall localization precision, $\sigma_1$, diverges for the two treatments of the light centroid at $\sim 3 \theta_{E, 0.4}$, coinciding with the point where the lens mass becomes unresolved. In this regime, the microlens is detectable, but its parameters are no longer individually resolvable. In this case, the residual acceleration precision provides the appropriate continuation of the measurable centroid acceleration uncertainty into the weak-signal regime.

When we consider only the best-constrained direction of $\vect{\theta}_c$, the two $\sigma_2$ curves do not diverge over the range of $\vect{|\xi|}$ considered, as shown in the bottom-right panel of Fig.~\ref{fig:traj_res}. As in the single-epoch analysis, the corresponding eigenvectors show that this direction is aligned primarily with the x-axis, perpendicular to the macrolensing shear axis, where the centroid displacement is least affected by the degeneracy introduced by the shear.

Across the range of impact parameters considered, both $\sigma_1$ and $\sigma_2$ reach sub-$\mu\mathrm{as} \ \mathrm{yr}^{-2}$ precision for the light-centroid acceleration.
Such precision may allow microlensing-induced centroid accelerations to be separated from other astrometric perturbations, such as those of DM subhalos~\cite{DMpaper}.

Table~\ref{tab:tab_main} in App.~\ref{app:microlens} lists the parameter uncertainties as a function of $|\vect{\xi}|$ for all four lens masses. Microlensing is detectable (median combined $\chi^2>10$) out to $\sim[3.3, 5.5, 7.9, 11.0]\,\theta_{E,0.4}$ for $m_l = [0.1, 0.2, 0.4, 0.8]\,M_\odot$, respectively. Over this range, the source-radius uncertainty stays roughly constant, while the lens-mass precision improves with increasing lens mass at fixed position (in fractional terms, not always in absolute terms).
The position and velocity uncertainties vary non-monotonically (Fig.~\ref{fig:traj_res}), reflecting both these trends and the differing effective separations when expressed in units of $\theta_{E,m_l}$. For well-resolved configurations, the centroid-acceleration uncertainty grows weakly with lens mass, but this trend reverses near the detection threshold.

\section{Discussion} \label{sec:discussion}

Intensity interferometry of a strongly lensed quasar, combined with flux-ratio photometry, constrains the mass, impact parameter, and transverse velocity of an individual star in the lens galaxy, along with the angular size of the quasar accretion disk. Photometric microlensing is subject to a scaling degeneracy: the light curve depends on the configuration only through lengths in units of $\theta_{E,m_l}$ and times in units of $\theta_{E,m_l}/|\tilde{\vect{\nu}}|$, leaving $\theta_\mathrm{src}$, $|\vect{\xi}|$, and $|\tilde{\vect{\nu}}|$ determined up to a common factor $m_l^{1/2}$. Intensity correlations are proportional to the square modulus of the visibility, and at sufficiently large baselines, provide angularly resolved information on the macro- and micro-lensed image to break these degeneracies. Our forecasts assume a single point-like microlens, a fixed macrolensing model, and instrumental specifications outlined in Sec.~\ref{sec:setup}. Under our assumptions, inference of indivual stellar masses and transverse velocities becomes possible in galaxies at moderate redshifts, some three orders of magnitude in distance beyond the farthest binary systems from which such measurements are currently obtained\cite{binary_masses, binary_masses_and}.

For B1422+231, one epoch constrains the accretion-disk size to $\sim4\%$ in favorable geometries and $\sim10\%$ for a typical detectable configuration. It measures the mass of a lens with $m_l\gtrsim0.1\,M_\odot$ to order unity where the lensing morphology is informative. An eight-year campaign with 33 epochs would reach sub-percent precision on the source size and percent-level stellar mass precision at small impact parameters. It localizes close trajectories to $\mathcal{O}(0.1\,\mu\mathrm{as})$ and retains sub-$\mu\mathrm{as}$ localization over much of the detectable region. Because most information is gained at close approach (limiting the temporal lever arm), favorable trajectories yield only order-unity proper-motion measurements. These statistical uncertainties scale approximately as $n_\mathrm{obs}^{-1/2}$ at fixed campaign duration and per-epoch sensitivity, so a denser cadence would improve this reach. The corresponding centroid-acceleration precision is well below $0.1\,\mu\mathrm{as}\,\mathrm{yr}^{-2}$ throughout the detectable range.

The forecast has three main astrophysical applications. First, object-by-object mass measurements at cosmological distances would provide direct calibration points for extragalactic stellar mass functions and compact-object populations. Second, resolved transverse motions probe the internal velocity structure of distant lens galaxies, which would also help calibrate the DM signals proposed in the companion paper~\cite{DMpaper}. Third, the same intensity-interferometric observations proposed here would also directly constrain the microarcsecond-scale structure of the inner accretion disks of high-redshift quasars~\cite{II_AGN}. Together, these observations would extend resolved stellar astrophysics and astrometry to cosmological lens galaxies.

The (residual) light-centroid acceleration precision forecast here, significantly below $10^{-2}\,\mu\mathrm{as}\,\mathrm{yr}^{-2}$ in well-resolved configurations, is comparable to the astrometric signatures expected from cold dark matter subhalos along the line of sight~\cite{DMpaper}. Unless extremely star-poor lensing environments are identified, realizing that sensitivity would require modeling and subtracting the stellar microlensing background to high precision.
The forecast indicates that event-by-event subtraction is feasible in the well-resolved regime and that bright, low-$\kappa_*$ macro-images such as image~A of B1422+231 are promising targets. Propagating these centroiding uncertainties into a sensitivity to DM (sub)halos is the subject of a companion paper~\cite{DMpaper}.

In terms of the companion's noise budget, our forecasts translate as follows. Stars of mass $0.1$--$0.8\,M_\odot$ can be detected, and their centroid contribution fit and subtracted, out to impact parameters of $5$--$16\,\mu\mathrm{as}$ from the image (median combined $\chi^2 > 10$; Table~\ref{tab:tab_main}), so the fit-out radius $\theta_\mathrm{fit} \approx 10\,\mu\mathrm{as}$ assumed in Ref.~\cite{DMpaper} corresponds to the detection limit of a $\sim0.3\,M_\odot$ star over our eight-year campaign. Within that radius, the best-case residual acceleration uncertainty, $\sigma[\partial_t^2 \Delta\theta_c^\mathrm{res}] \approx 0.014\,\mu\mathrm{as}\,\mathrm{yr}^{-2}$ (a variance of $\approx2\times10^{-4}\,\mu\mathrm{as}^2\,\mathrm{yr}^{-4}$; Table~\ref{tab:tab_main}), scales in variance approximately as $1/(n_\mathrm{obs}\,\tau^4)$ at fixed per-epoch precision. Rescaled to the fiducial differential-astrometry survey of the companion paper ($n_\mathrm{obs}=300$, $\tau=10$\,yr, $\sigma_{\delta\theta}=0.1\,\mu\mathrm{as}$---a per-epoch precision comparable to the $0.04$--$0.5\,\mu\mathrm{as}$ per-epoch centroid precision of our model fits), it becomes $\approx9\times10^{-6}\,\mu\mathrm{as}^2\,\mathrm{yr}^{-4}$. This is below the $\sim4\times10^{-5}\,\mu\mathrm{as}^2\,\mathrm{yr}^{-4}$ bound on the acceleration noise from the dominant \emph{unfitted} star beyond $\theta_\mathrm{fit}$, and within an order of magnitude of that survey's instrumental acceleration floor of $2.4\times10^{-6}\,\mu\mathrm{as}^2\,\mathrm{yr}^{-4}$~\cite{DMpaper}. Under these rescalings, the post-subtraction stellar noise at image~A is dominated by the unfitted population as opposed to the subtraction residual. For differential observables between image pairs, the residuals of the two images add in quadrature, with images of higher stellar convergence (e.g.~image B, $\kappa_*\approx0.09$) contributing proportionally more.

The same observations can also determine the macrolensing distortions themselves, on which the DM substructure weak-lensing signal depends steeply~\cite{DMpaper}. Resolved image shapes fix the element-wise ratios of the inverse Jacobians $\vect{A}^{-1}$ across images, since all macro-images share a common unlensed source shape; combined with the flux ratios and the relative image proper motions, all components of $\vect{A}^{-1}$ can be determined without recourse to a parametric lens model for systems with three or more images~\cite{KVT2023, Galanis23}. The dominant conversion factor between DM substructure and the astrometric observable can therefore be calibrated from the data themselves; a dedicated forecast of this joint macro--micro inference is left to future work.

Our forecast depends on simultaneous improvements in four instrumental capabilities: large effective collecting area, fast single-photon timing, broad spectral multiplexing, and high photodetection efficiency. Together, these advances would expand the scope of  intensity interferometry beyond morphological measurement on bright stars, to  faint extragalactic sources such as quasars~\cite{II_AGN}.. The compilation of Ref.~\cite{Mosquera} contains 26 lensed quasars brighter than magnitude 17, providing a target pool for instruments that reach the assumed sensitivity; future surveys may enlarge this sample. End-to-end simulations should incorporate realistic source morphologies, multiple microlenses, uncertainties in the macrolensing model, realistic II array geometry, observing cadence and conditions, and calibration systematics. Such simulations will determine whether our forecasted object-by-object measurements remain feasible once the idealized assumptions adopted here are relaxed.

\begin{acknowledgments}
We thank A.~Bracamonte, M.~Galanis, N.~Khusid, A.~Lam, E.~Pinetti, M.~Shara, and R.~Walter for useful conversations and comments on this work.
This material is based upon work supported by the National Science Foundation under Grant No.~PHY-2622106.
AM is supported by NSF Grant No.~DGE-2437839.
\end{acknowledgments}

\appendix

\section{Notation}\label{app:Notation}
For convenience, we provide Table~\ref{tab:notation}, which summarizes select variable names and notation used throughout this work. We show the most important variables here, as well as variables which have similar names to clearly differentiate them from one another.

\begin{table*}[t]
\centering
\small
\caption{Summary of selected notation used throughout the paper, with symbols grouped by role.}
\label{tab:notation}
\begin{tabular}{@{}p{0.17\textwidth} p{0.62\textwidth} p{0.14\textwidth}@{}}
\toprule
\textbf{Symbol} & \textbf{Meaning} & \textbf{First used} \\
\hline
\multicolumn{3}{@{}l}{\textit{Coordinates and angular positions}} \\
$\vect{\beta}$ & Source-plane angular coordinate & Sec.~\ref{subsec:QEM} \\
$\vect{\theta}$ & Image-plane angular coordinate & Sec.~\ref{subsec:QEM} \\
$\theta_\mathrm{src}$ & Angular scale of the (unlensed) emitting region & Eq.~\ref{Toftheta} \\
$\theta_\mathrm{src,lensed}$ & Apparent (magnified) angular scale, $=\sqrt{\mu}\,\theta_\mathrm{src}$ & Eq.~\ref{eq:theta500} \\
$\theta_{E,m_l}$ & Einstein radius of a point lens of mass $m_l$ & Eq.~\ref{eq:thetaE} \\
$\vect{\theta}_c$ & Light centroid of an image & Sec.~\ref{sec:microlensing} \\
$\theta_\mathrm{max}$ & Angular extent of the simulated image region & Sec.~\ref{subsec:sing} \\
\hline
\multicolumn{3}{@{}l}{\textit{Single-Epoch Analysis}} \\
$\sigma_C$ & Per-measurement uncertainty on $C$ & Sec.~\ref{subsec:QEM} \\
$m_l$ & Mass of a single microlens & Sec.~\ref{sec:microlensing} \\
$\vect{\theta}_\mathrm{sep}$ & Angular position of a microlens relative to the quasar & Sec.~\ref{sec:microlensing} \\
$\vect{p}=(\theta_\mathrm{src}, m_l, \vect{\theta}_\mathrm{sep})$ & Single-epoch (4D) microlensing parameter vector & Sec.~\ref{subsec:sing} \\
$F^{\vect{p}}$ & Single-epoch Fisher matrix (4D, over $\vect{p}$) & Eq.~\ref{eq:Fisher} \\
$\Sigma_{\vect{\theta}_c}$ & Centroid covariance, $=JF^{-1}J^T$ & Eq.~\ref{eq:cov_thetac} \\
$\sigma_{|\vect{\theta}_c,1|}$ & Overall centroid localization precision & Eq.~\ref{eq:theta_C1} \\
$\sigma_{|\vect{\theta}_c,2|}$ & Best-direction centroid precision & Eq.~\ref{eq:theta_C2} \\
$\chi^2$ & Detection statistic, $=-2\ln\mathcal{L}$ (relative to macro-only) & Sec.~\ref{subsec:sing} \\
\hline
\multicolumn{3}{@{}l}{\textit{Time-Domain Analysis}} \\
$\vect{\nu}$ & \emph{Unlensed} relative proper motion (source frame) & Eq.~\ref{eq:proper_motion_lens} \\
$\tilde{\vect{\nu}}$ & \emph{Apparent} proper motion in the image-fixed frame, $=A^{-1}\vect{\nu}$ & Eq.~\ref{eq:proper_motion_lens} \\
$\vect{\Theta}=(\theta_\mathrm{src}, m_l, \vect{\xi}, \tilde{\vect{\nu}})$ & Trajectory-level (6D) parameter vector & Sec.~\ref{subsec:traj} \\
$F^{\vect{\Theta}}$ & Trajectory-level Fisher matrix (6D, over $\vect{\Theta}$) & Eq.~\ref{eq:ftheta} \\
$F_{\vect{\theta}_c}$ & Fisher matrix for centroid kinematic parameters & Eq.~\ref{eq:fisher_theta_c} \\
$\Sigma_{\vect{\theta}_c}^\mathrm{res}$ & Regularized ("residual") centroid covariance & Eq.~\ref{eq:cov_res} \\
\hline
\hline
\end{tabular}
\end{table*}

\section{Numerical Methods}\label{app:pipeline}

This appendix describes the numerical methods used to generate microlensed images and forecast parameter uncertainties. We first describe the discretization of the surface-brightness distributions and the calculation of the real- and Fourier-space observables, followed by the numerical evaluation of the Fisher matrix and the convergence tests used to ensure that the results are insensitive to the image resolution.

We compute the magnification and light centroid observables in Eq.~\ref{eq:mag_def} and Eq.~\ref{eq:cent_det} from the discretized surface-brightness distributions generated in Sec.~\ref{subsec:sing}. The simulated surface-brightness distributions are sampled on a two-dimensional grid, with pixel area $\Delta\Omega$. Image-plane integrals can be evaluated as pixel sums. The magnification ratio becomes
\begin{equation}\label{eq:mag}
\mu_{\mathrm{i,ii}}\approx\frac{\sum_\Omega B_\mathrm{(i)}}{\sum_\Omega B_\mathrm{(ii)}}.
\end{equation}
where the common pixel-area factor cancels between numerator and denominator. Similarly, the light centroid is computed as the flux-weighted sum over pixel positions,
\begin{equation}
\vect{\theta}_c \approx \frac{\sum_\Omega B_i \vect{\theta}_i}{\sum_\Omega B_i}.
\end{equation}

We next compute the visibility by evaluating the Fourier transform of the discretized surface-brightness distribution. For each spectral channel centered on $\bar{k}$, the surface brightness $B_\lambda(\vect{\theta})$ is generated on a finite angular region of extent $\theta_\mathrm{max}$ and sampled on a two-dimensional pixel grid. The finite image is then Fourier transformed using a discrete Fourier transform (DFT),
\begin{equation}
\tilde{B}_{u,v} = \mathcal{F}[B_\lambda(\vect{\theta})],
\end{equation}
where $(u,v)$ denote the discrete spatial frequencies sampled by the DFT. For an image with angular extent $\theta_\mathrm{max}$, these modes are given by
\[
(u,v) = (n_u,n_v)\frac{2\pi}{\theta_\mathrm{max}},
\]
where $n_u$ and $n_v$ are integer Fourier modes. The Fourier-space intensity correlation at each mode is then
\begin{equation}\label{C_real}
C_{u,v} = \tilde{B}_{u,v}\tilde{B}^{*}_{u,v}
= |\tilde{B}_{u,v}|^2,
\end{equation}
which is real and non-negative by construction. 
The angular extent of the image, $\theta_{\text{max}}$, determines the density of Fourier-mode sampling, and the pixel scale sets the maximum accessible spatial frequency.

Using these real and Fourier-space images, we calculate the elements of the Fisher matrix separately for the II and magnification terms in Eq.~\ref{eq:LL_full}. The Fisher matrix can be written as the negative expectation value of the log-likelihood Hessian or, for Gaussian data with parameter-independent covariance, as a Gram matrix of observable derivatives. In our case, however, the likelihood is generally shallow as a function of the model parameters $p_i$, leading to very small second derivatives. This effect is more significant in the magnification term, but also becomes significant for the II term in the case of distant and/or low-mass lenses. Under these conditions, floating-point error can dominate the numerical second derivatives and drive the estimated curvature to the wrong sign. This in turn makes inversion of the Fisher information matrix, which should theoretically be positive semidefinite, numerically unstable or impossible.

We therefore use the first-derivative (Gram) form of the Fisher matrix given in Eq.~\ref{eq:Fisher}, $F = J^{T}J/\sigma^2$ with $J$ the Jacobian of the model observables, which is positive semidefinite by construction. The required derivatives are evaluated with central differences for the II correlation and the magnification, and forward differences for the centroid Jacobian, with step sizes of $1\%$ in $\theta_\mathrm{src}$, $5\%$ in $m_l$, and half an image pixel in the lens position. This approach is numerically stable across a wide range of perturbation choices $\delta$, provided that $\delta$ is chosen such that it resolves the local variation in $\log \mathcal{L}(\vect{p})$ while remaining above floating-point precision.

We also encounter potential numerical noise issues when resolving the simulated lensed images.
To ensure that the simulations accurately represent a well-resolved image, we must choose a total angular image size $\theta_\mathrm{max}$ (which sets the pixel spacing in Fourier space) and number of pixels $N_\text{pix}$ such that the image is resolved in both real and Fourier space. In practice, this requires that the real-space pixel size $\Delta x
< \theta_\mathrm{src}$ and the Fourier-space pixel size $\Delta u < \theta_\mathrm{src}^{-1}$. In this regime, the forecast uncertainties are insensitive to numerical discretization.
We determine suitable values by progressively increasing $\theta_\mathrm{max}$ and $N_\text{pix}$ until further increases produce no change in the derived uncertainties. These tests indicate that $\theta_\mathrm{max} = 4 \times 10^{-10}$ radians with $N_\text{pix} = 8000$ pixels is sufficient. This corresponds to
\[\Delta x  = 5\times 10^{-14} \ \text{rad} < \theta_\mathrm{src} = 2.0 \times 10^{-13}\ \text{rad}\]
and
\[
\Delta u = \frac{2\pi}{\theta_\mathrm{max}} = 1.6\times 10^{10}\  \text{rad}^{-1} <  \theta_\mathrm{src}^{-1} = 5.1 \times 10^{12} \ \text{rad}^{-1},\]
where $\theta_\mathrm{src}$ here is the (unlensed) fiducial source scale of our simulations.

These inequalities are satisfied, so we adopt these values for the II and centroid analyses; the magnification-only images use a smaller region of $3\times10^{-10}$ radians at the same pixel count. Finally, the central temperature divergence of Eq.~\ref{Toftheta} is regularized on a fixed angular scale by evaluating $T\propto[(|\vect{\beta}|+\theta_\mathrm{src, lensed})/\theta_\mathrm{src}]^{-3/4}$, which caps the central temperature at $\approx0.49\,T_{500}$ for the fiducial radius. The derived uncertainties are insensitive to this numerical regularization over the tested range.

\section{Additional Microlensing Configurations}\label{app:microlens}

\begin{figure*}
    \centering
    \includegraphics[width=0.95\linewidth]{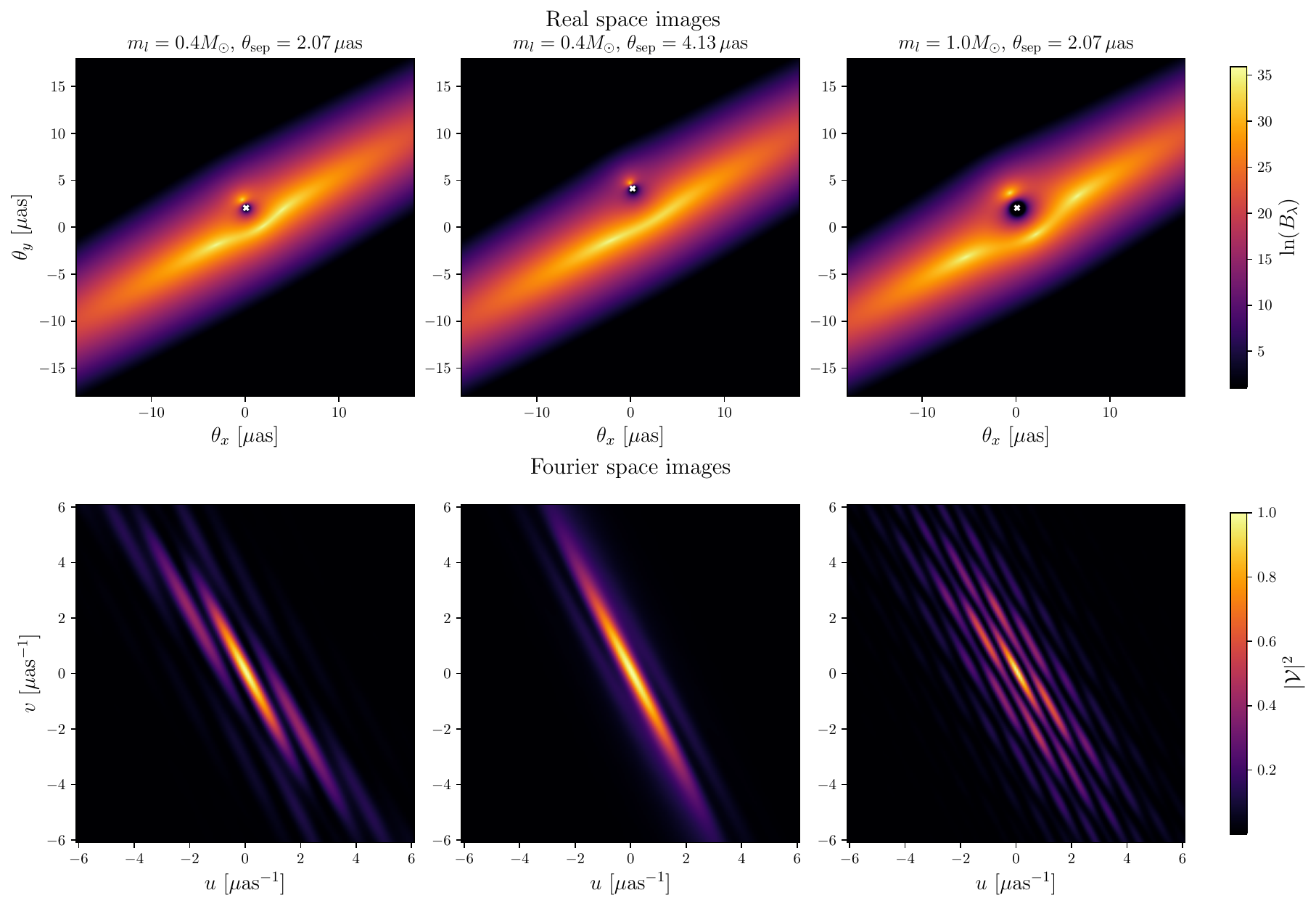}
    \caption{\micronblink{Plots.ipynb} Several microlensing scenarios for image A of B1422+231. As in Fig.~\ref{fig:basic_lensing}, the real-space panels (axes in $\mu\mathrm{as}$) show the logarithmic surface brightness $\ln B_\lambda$ (arbitrary units) and the Fourier-space panels (axes in $\mu\mathrm{as}^{-1}$) show the square visibility modulus $C = |\mathcal{V}|^2$. The microlens is marked in the real-space images with a white ``x.'' The microlens parameters are selected arbitrarily for illustrative purposes. }
    \label{fig:lens_comp}
\end{figure*}

Figure~\ref{fig:lens_comp} illustrates how varying the microlens mass and position changes the lensed image structure in real and Fourier space.
Increasing the angular separation $\theta_{\mathrm{sep}}$ reduces the strength of the perturbation to the image. In real space, the induced micro-images shift outward and the primary image is less distorted. In Fourier space, this corresponds to a weaker modulation of the visibility amplitude, suppressing the high-spatial-frequency structure introduced by the microlens.
A similar effect is seen with decreasing lens mass; because the Einstein radius scales as $\theta_E \propto \sqrt{m_l}$, higher masses increase the characteristic separation between lensed components in real space and generally enhance the total magnification. In Fourier space, this larger angular scale redistributes power to smaller spatial scales (larger $|\vect{u}|$), producing oscillatory structure in the visibility pattern at high angular wavenumbers.

\subsection{Single-Epoch Analysis}
 
\begin{figure*}
    \centering
    \includegraphics[width=0.48\linewidth]{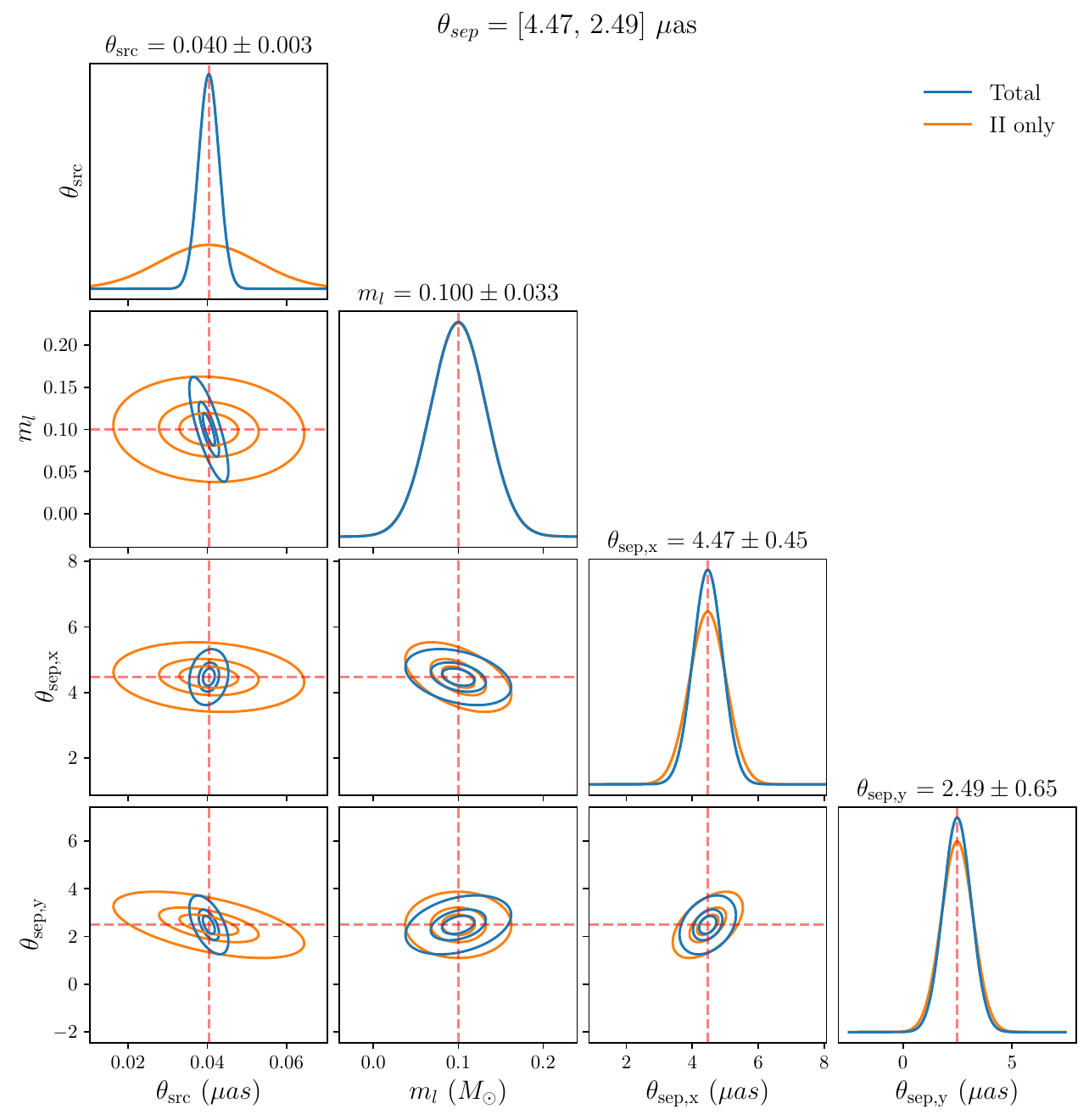}
    \includegraphics[width=0.48\linewidth]{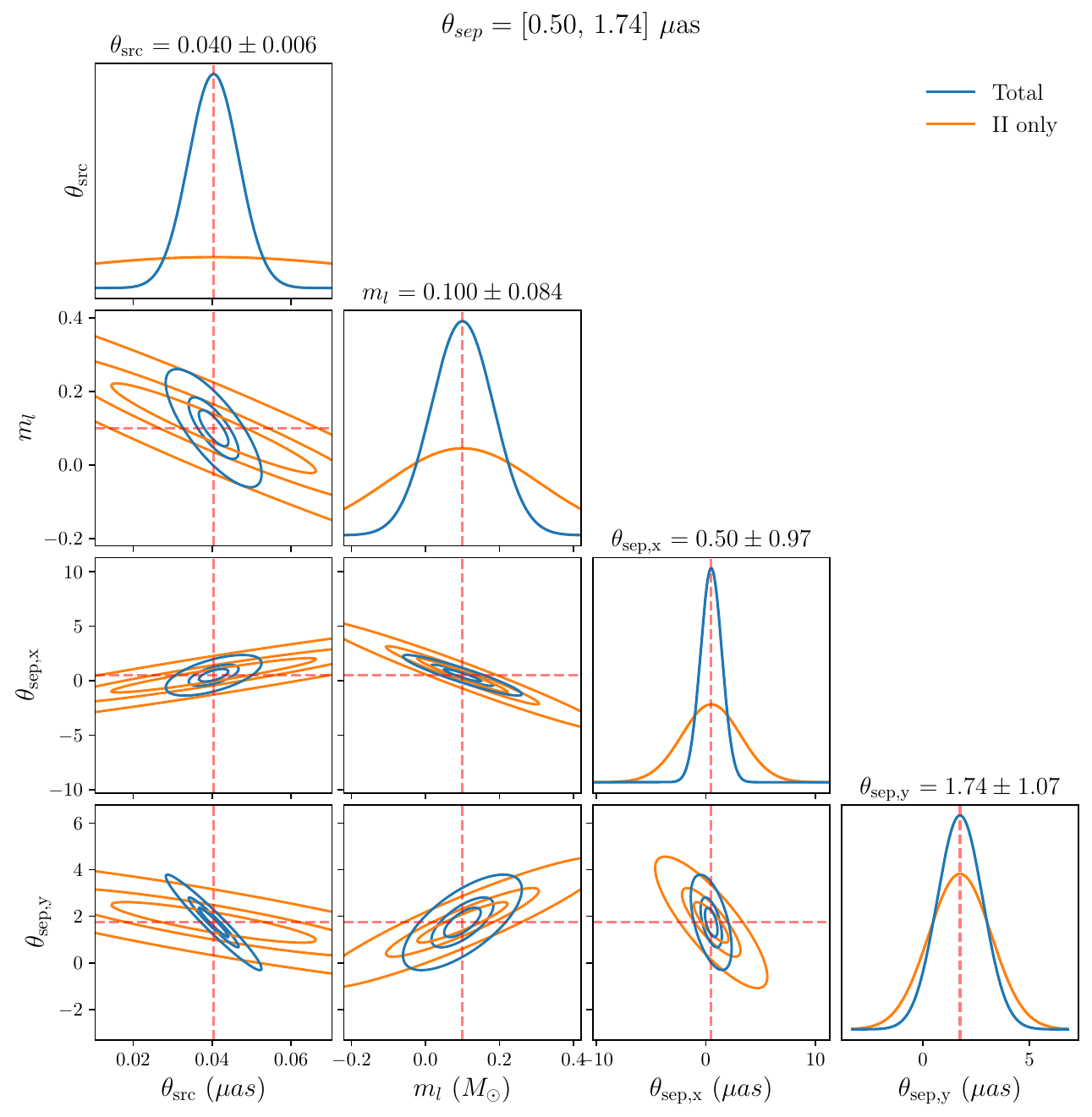}\\
    \includegraphics[width=0.48\linewidth]{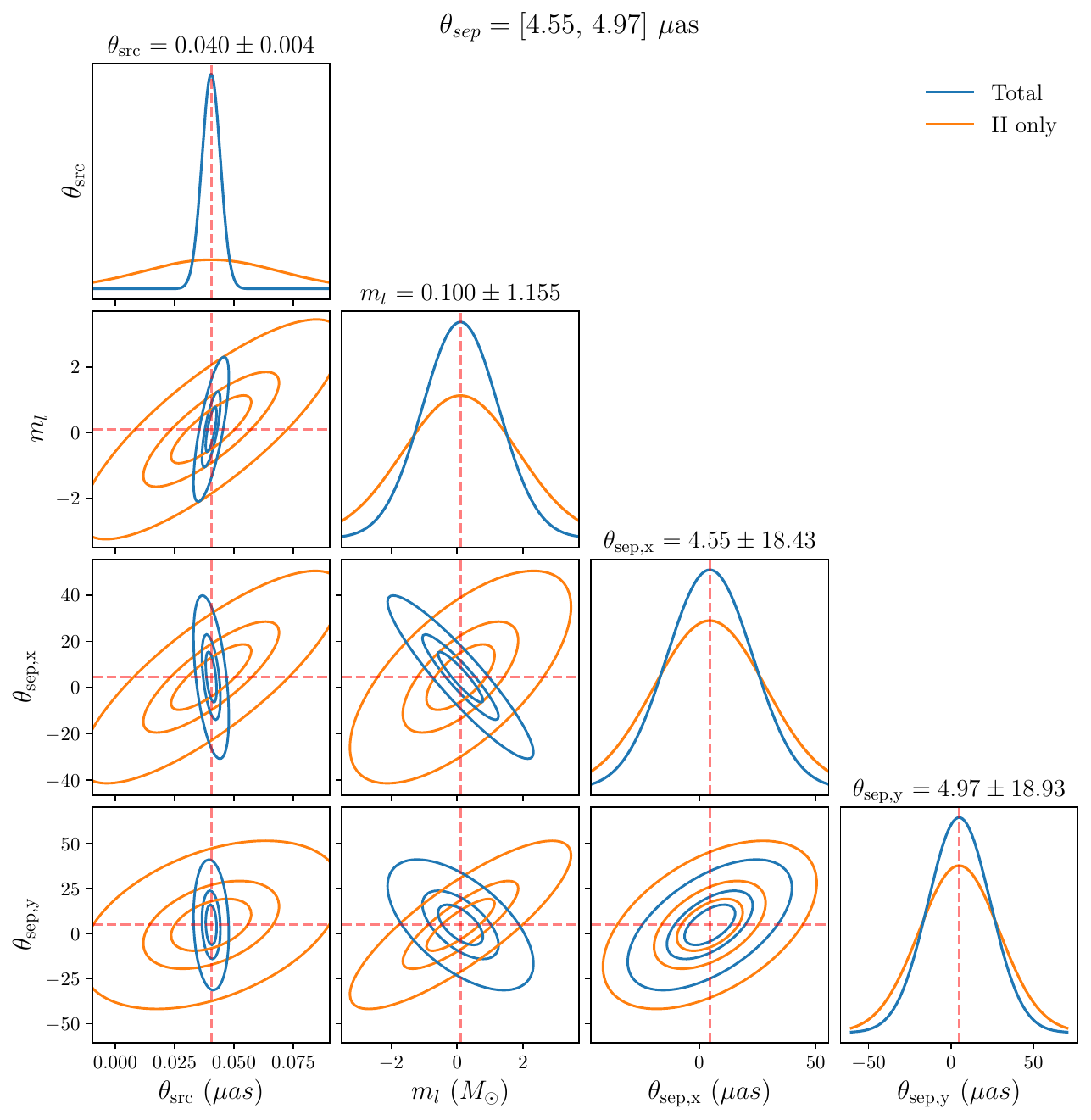}
    \caption{\micronblink{interp_grid01.ipynb} Corner plots for $m_l=0.1\,M_\odot$ for a well-resolved microlens (top left), a barely resolved lens (top right) and an unresolved lens (bottom row). The red dotted lines show the true parameter values.}
    \label{fig:m1_corners}
\end{figure*}

\begin{figure*}
    \centering
    \includegraphics[width=0.45\linewidth]{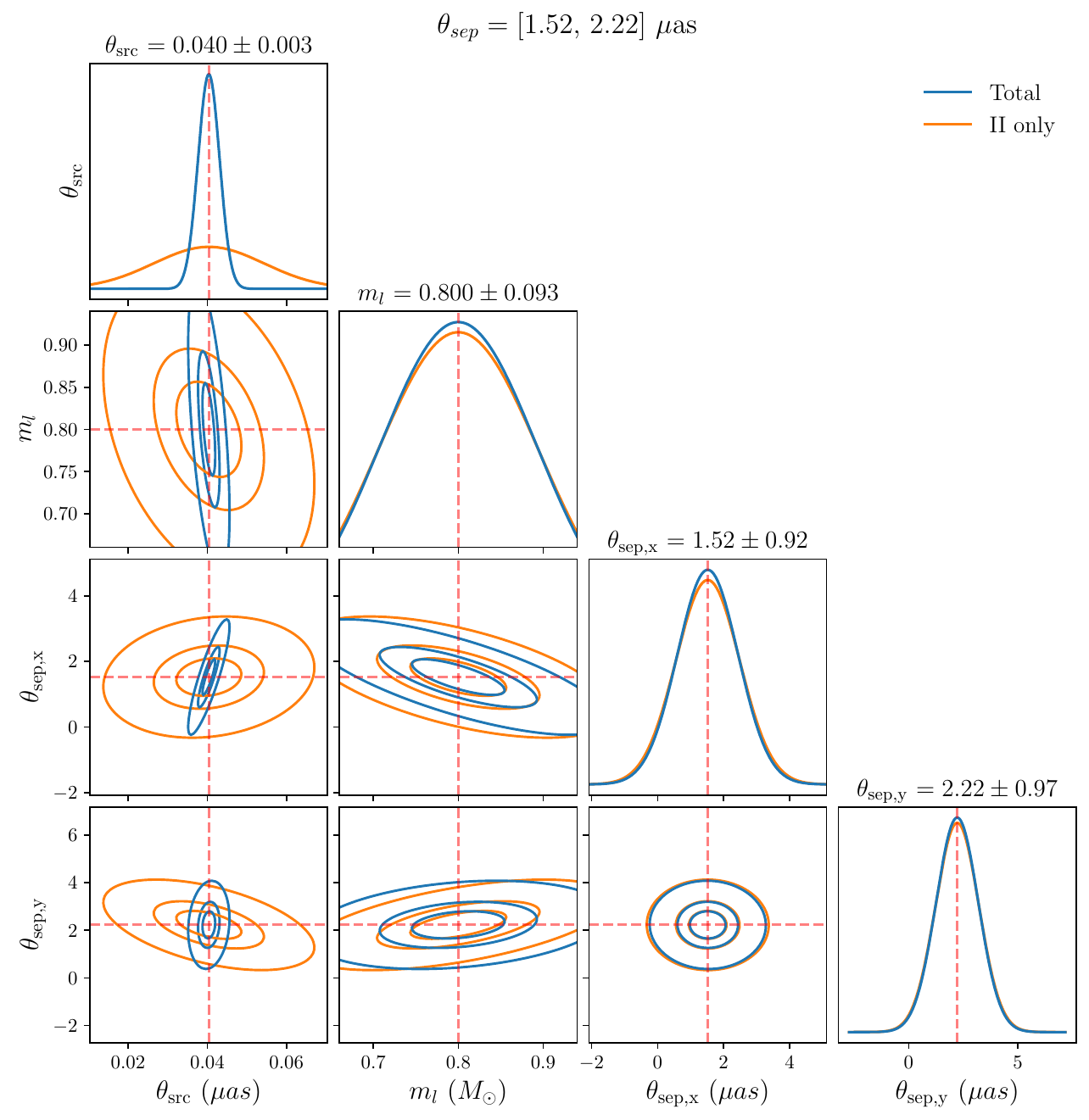}
    \includegraphics[width=0.45\linewidth]{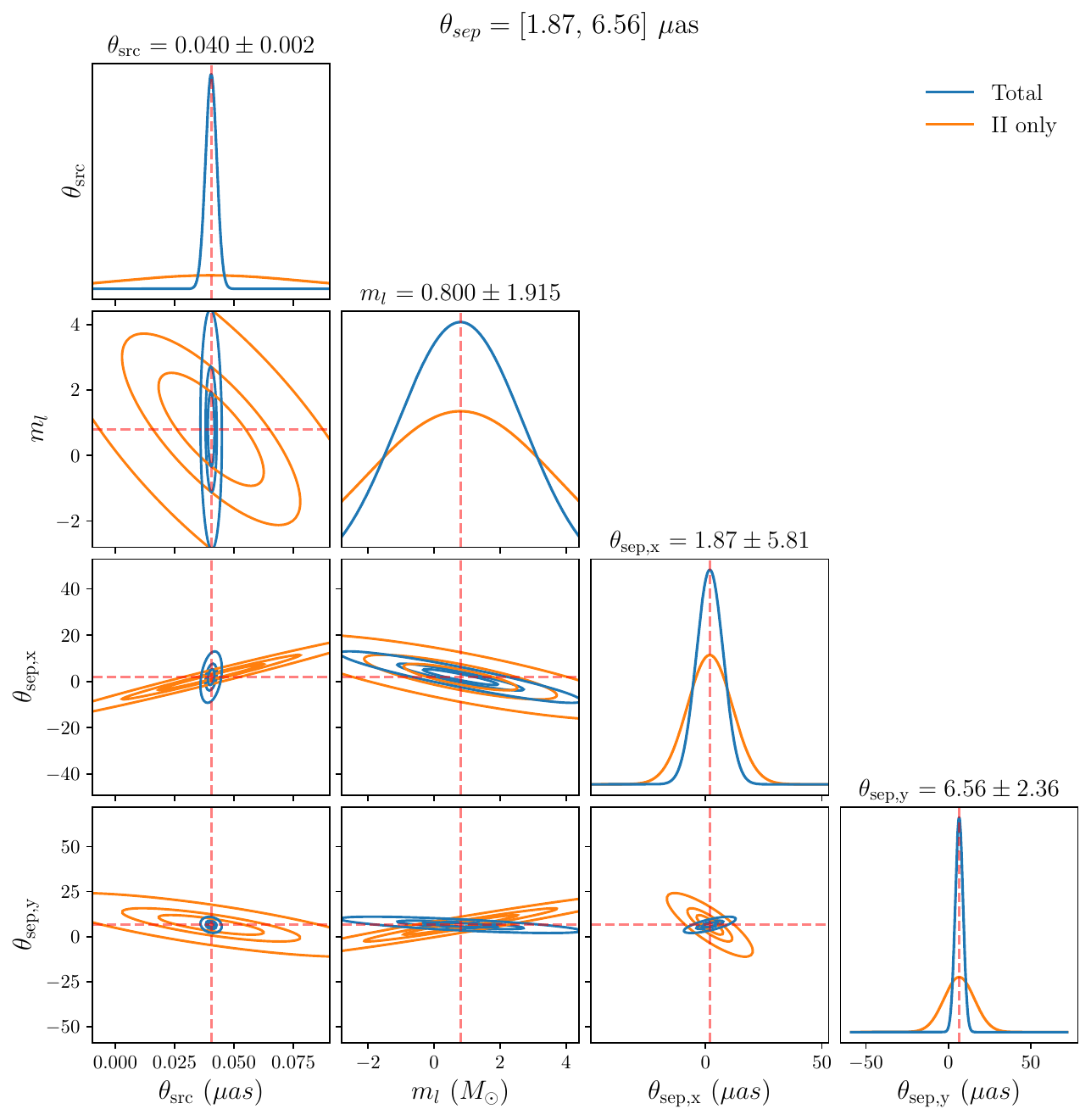}\\
    \caption{\micronblink{interp_grid08.ipynb} Corner plots of parameter constraints for $m_l=0.8\,M_\odot$ for a well-resolved microlens (left) and an unresolved microlens (right). The true values are shown in red. The illustrated positions are representative examples of the resolved and unresolved regimes.}

    \label{fig:m8_corners}
\end{figure*}

We first illustrate the parameter degeneracies in the single-epoch inference using representative posterior distributions for $m_l=[0.1, 0.8]M_\odot$. Figure~\ref{fig:m1_corners} and Fig.~\ref{fig:m8_corners} show corner plots for low- and high-mass lenses across a range of source-lens separations $\vect{\theta}_{\rm sep}$.
The magnification term contributes most when the II likelihood leaves a broad degeneracy. This behavior is most clearly illustrated in the $m_l=0.1\,M_\odot$ case in Fig.~\ref{fig:m1_corners}, where the magnification term has little visible impact on the posterior constraints for the lens mass or position, both in the unresolved and resolved configurations. For the higher-mass case, $m_l=0.8\,M_\odot$ in Fig.~\ref{fig:m8_corners}, the magnification term contributes a more noticeable amount of information even in the unresolved configuration, reflecting the larger overall scale of the lensing signal. Across all configurations the magnification term contributes significantly to the source radius constraint, and dominates the inference in some cases.

This behavior arises from the structure of the covariances introduced by the magnification term. The magnification likelihood constrains parameters along a single direction in 4D space, which all preserve the same image brightness within the small magnification uncertainty. Within this range of total intensities, the source radius is tightly constrained because very small radial changes move the model rapidly off a given brightness contour. In contrast, remaining on the same brightness contour through changes in the lens parameters requires large shifts in lens mass and position. These degenerate directions correspond to substantially different lens configurations, which are already ruled out by the II term once it provides even modest spatial information. As a result, the magnification term primarily sharpens the source radius constraints; the other parameter constraints see only minor improvements with the inclusion of the magnification term in the unresolved cases, and see essentially no improvements when parameters are resolved by the II term alone.

\begin{figure*}[htbp]
    \centering
    \def\arraystretch{1}
    \setlength\tabcolsep{0pt}
    \begin{tabular}{cc}

        \begin{subfigure}[b]{0.32\textwidth}
            \includegraphics[width=\textwidth]{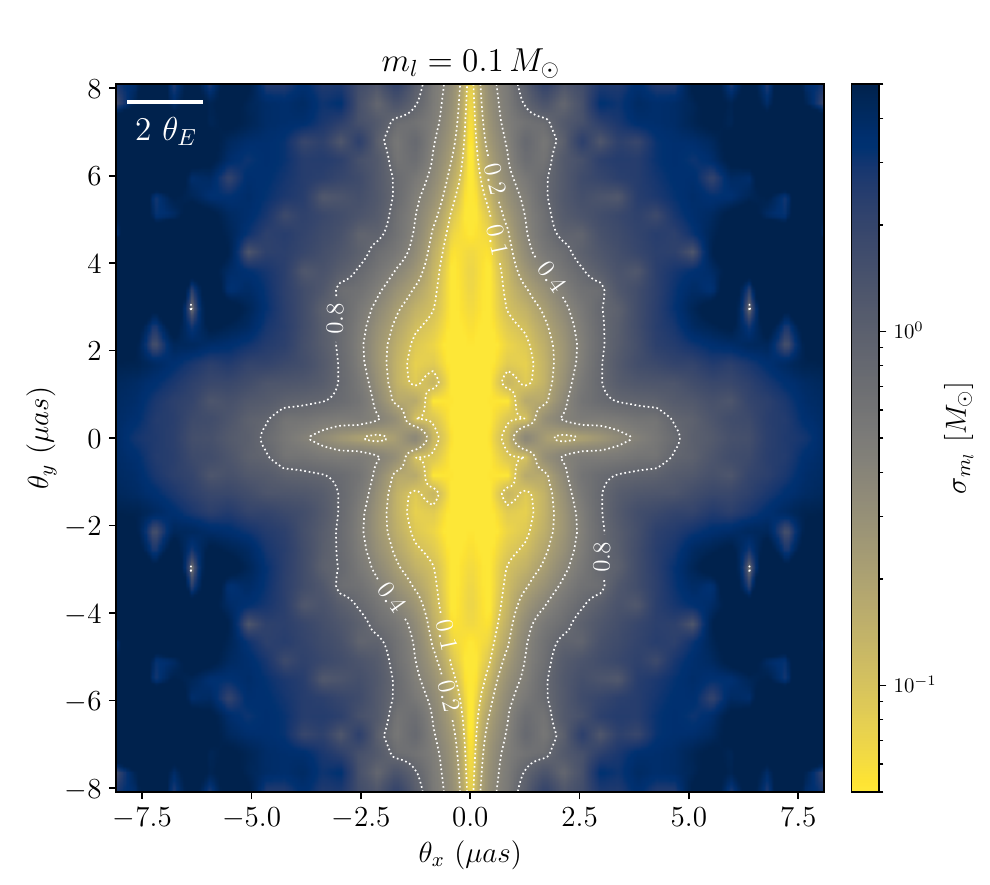}
        \end{subfigure} &
        \begin{subfigure}[b]{0.32\textwidth}
            \includegraphics[width=\textwidth]{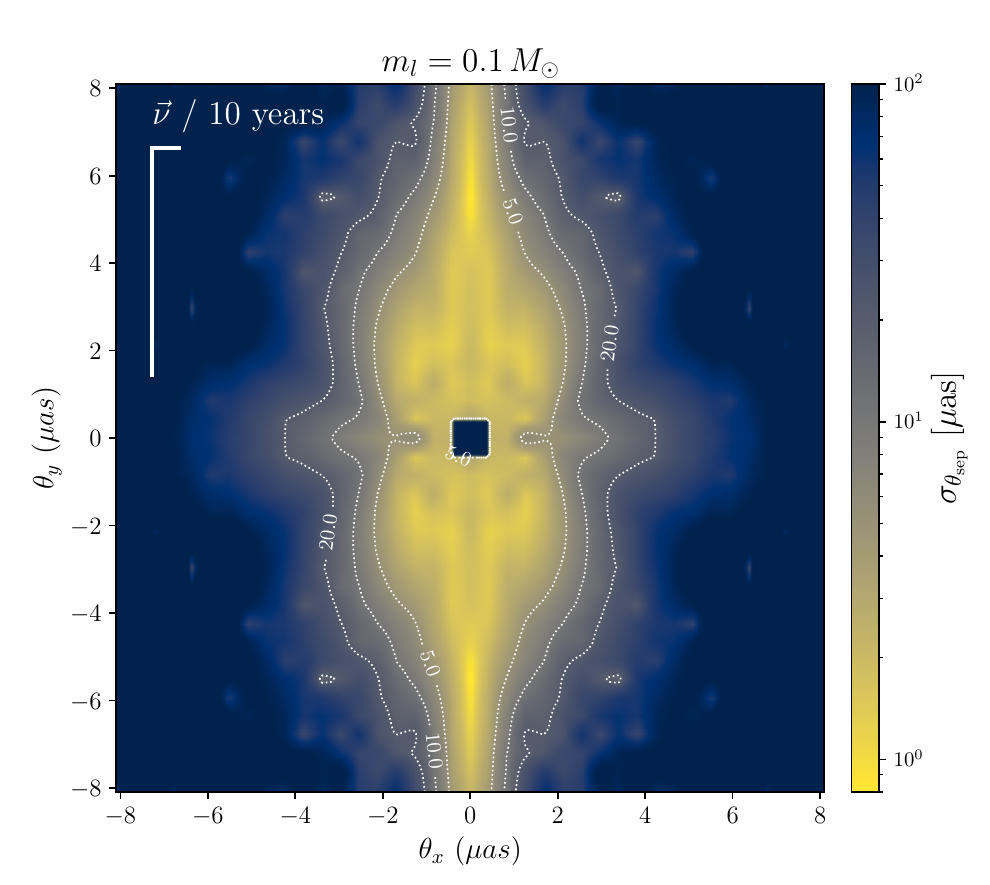}
        \end{subfigure} \\

        \begin{subfigure}[b]{0.32\textwidth}
            \includegraphics[width=\textwidth]{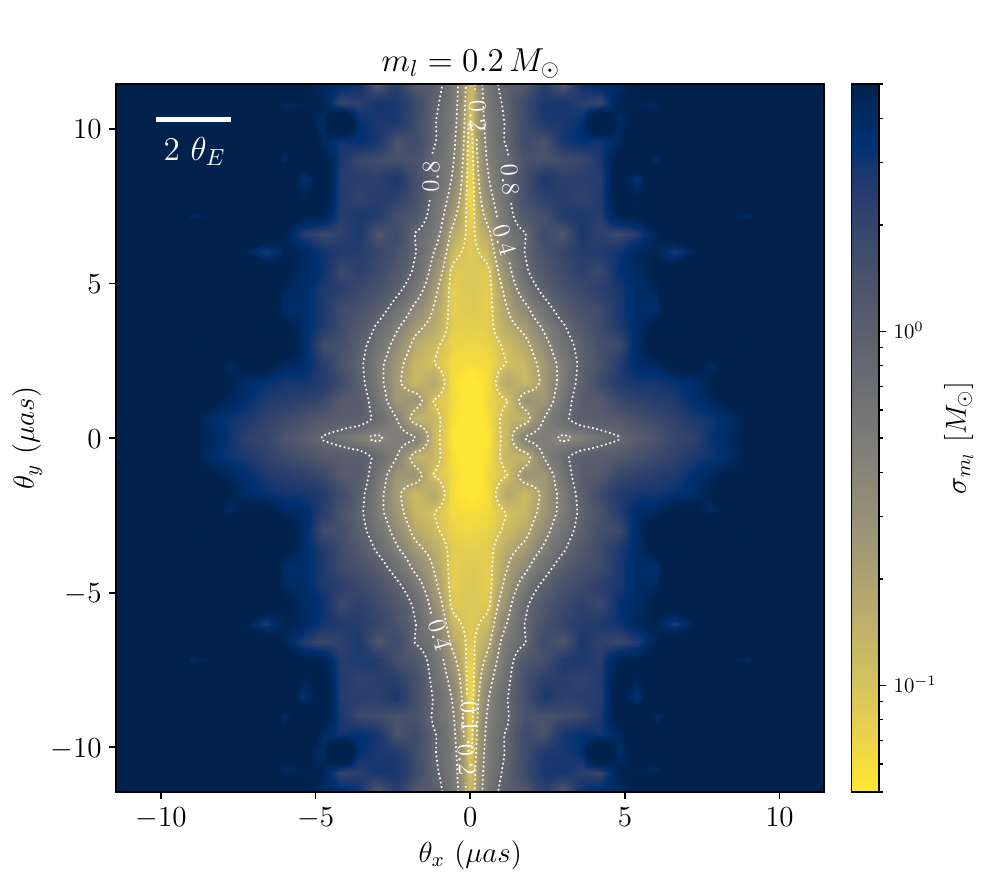}
        \end{subfigure} &
        \begin{subfigure}[b]{0.32\textwidth}
            \includegraphics[width=\textwidth]{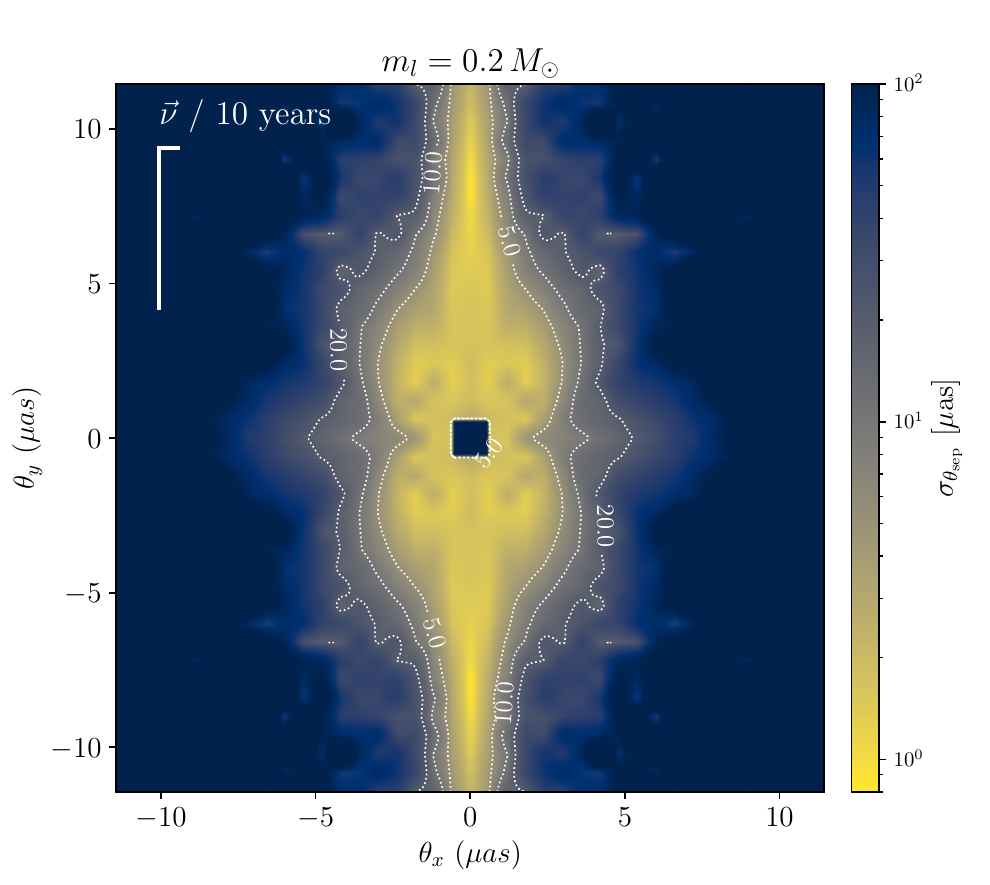}
        \end{subfigure} \\

        \begin{subfigure}[b]{0.32\textwidth}
            \includegraphics[width=\textwidth]{mass_04.pdf}
        \end{subfigure} &
        \begin{subfigure}[b]{0.32\textwidth}
            \includegraphics[width=\textwidth]{pos_04.pdf}
        \end{subfigure} \\

        \begin{subfigure}[b]{0.32\textwidth}
            \includegraphics[width=\textwidth]{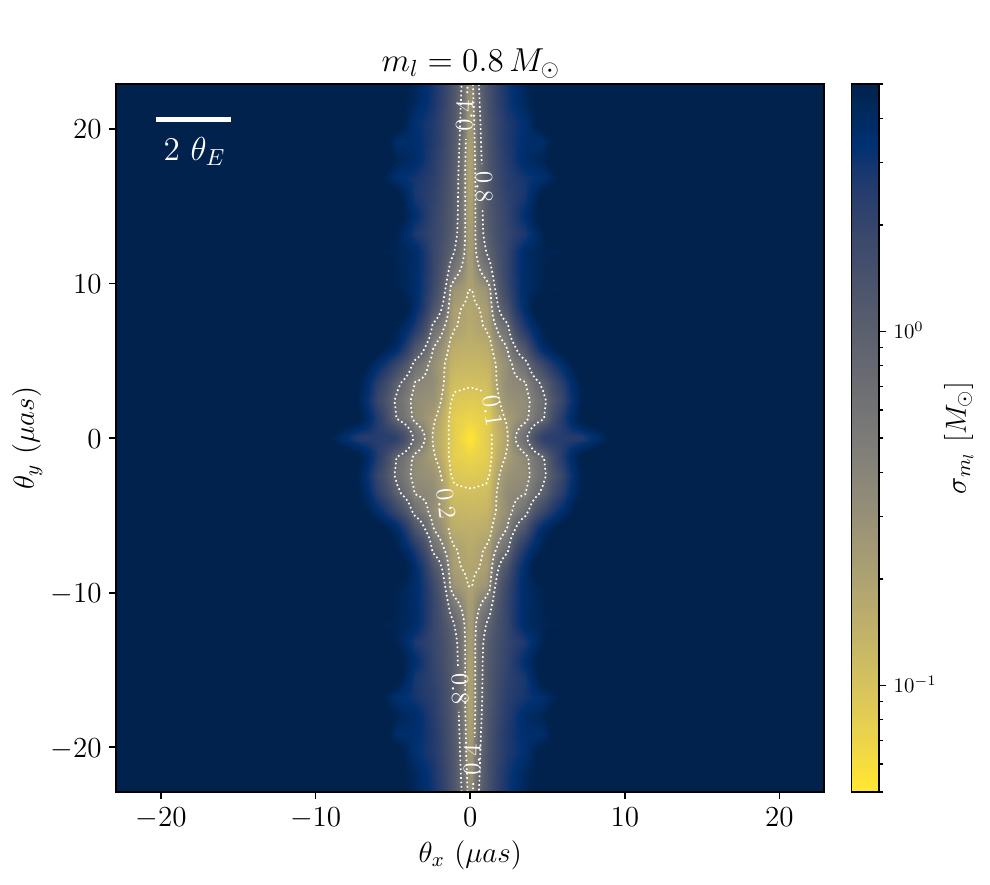}
        \end{subfigure} &
        \begin{subfigure}[b]{0.32\textwidth}
            \includegraphics[width=\textwidth]{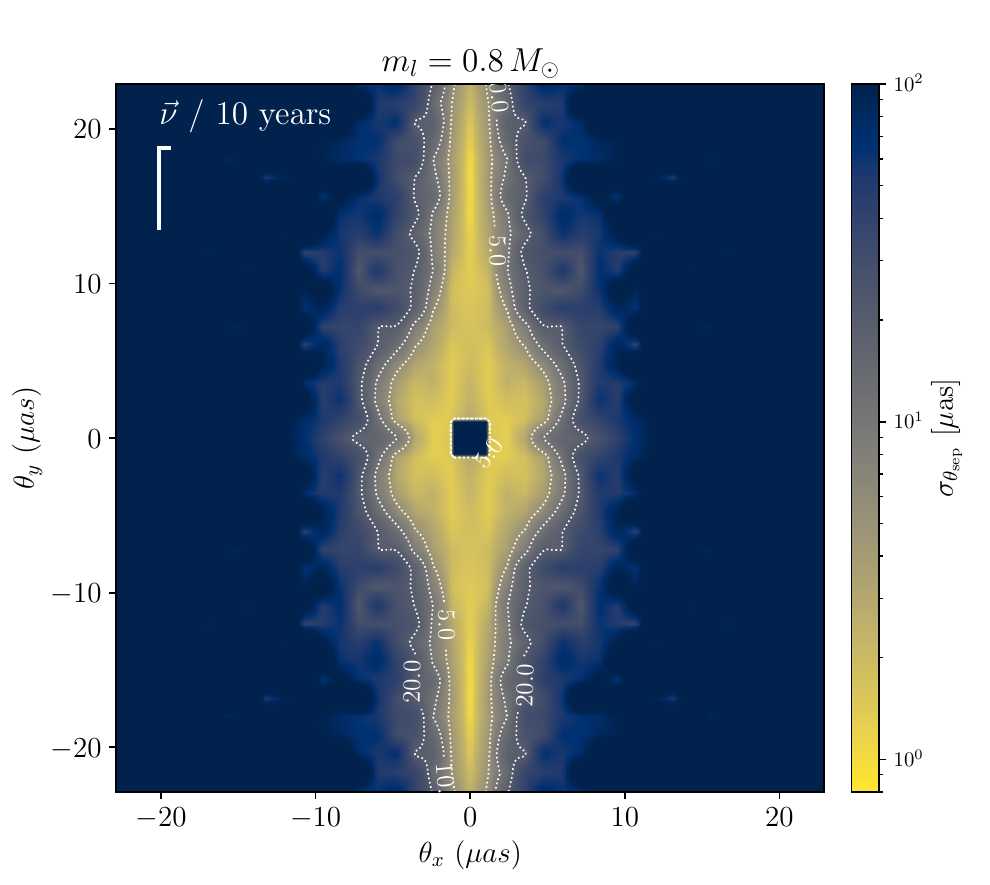}
        \end{subfigure} \\
    \end{tabular}
    \caption{\micronblink{interp_grid01.ipynb}\micronblink{interp_grid02.ipynb}\micronblink{interp_grid04.ipynb}\micronblink{interp_grid08.ipynb} Interpolated single-epoch precisions (including II and magnification information) as a function of lens position $\vect{\theta}_\mathrm{sep} = (\theta_x, \theta_y)$ (in $\mu\mathrm{as}$), for the lens mass $m_l$ (left) and lens position $\theta_\mathrm{sep}$ (right). The colorbars show $\sigma[m_l]$ (left column, in $M_\odot$) and $\sigma[\theta_\mathrm{sep}]$ (right column, in $\mu\mathrm{as}$), as indicated by the colorbar labels, on a logarithmic scale that is identical across lens masses. Each row is a different lens mass, with twice the Einstein radius $2\theta_{E,m_l}$ marked for reference. The reference velocity vectors $\vect{\nu}$ are drawn along the $x$ and $y$ axes of the macrolensing shear frame, corresponding to the minimally and maximally macrolensed trajectories, respectively.
    }
    \label{fig:other_mass_grid}
\end{figure*}

\begin{figure*}[htbp]
    \centering
    \def\arraystretch{1}
    \setlength\tabcolsep{0pt}
    \begin{tabular}{cc}

        \begin{subfigure}[b]{0.32\textwidth}
            \includegraphics[width=\textwidth]{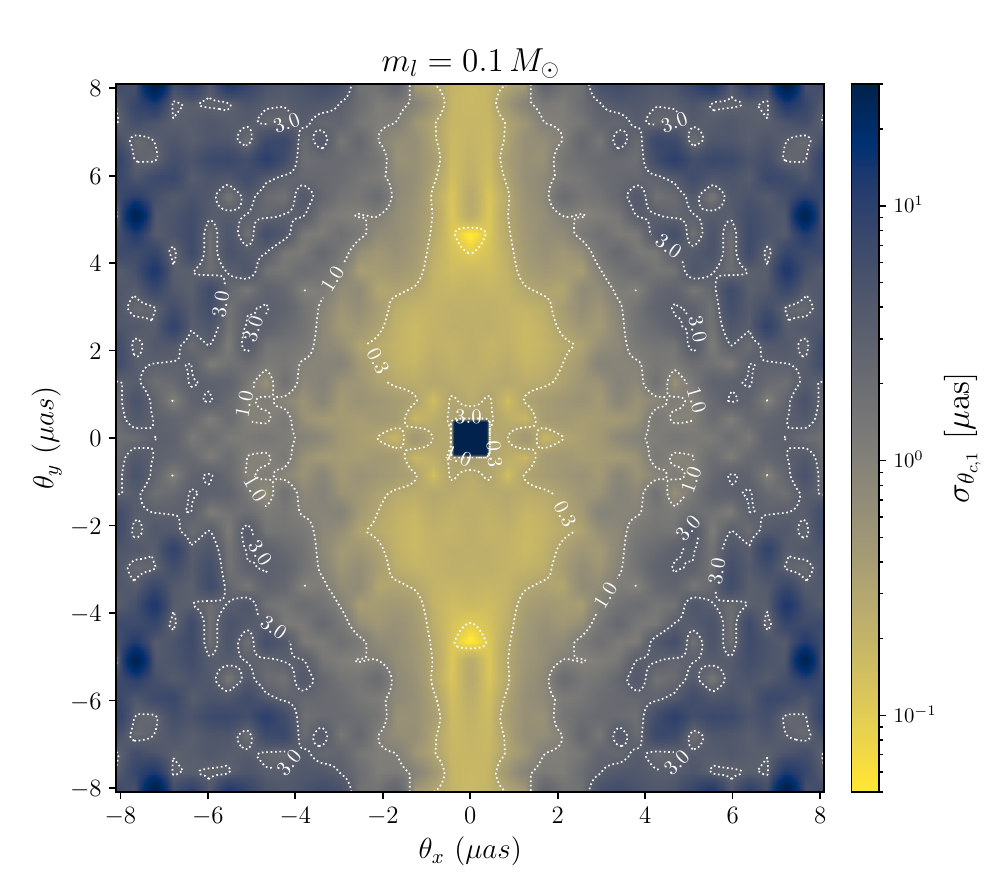}
        \end{subfigure} &
        \begin{subfigure}[b]{0.32\textwidth}
            \includegraphics[width=\textwidth]{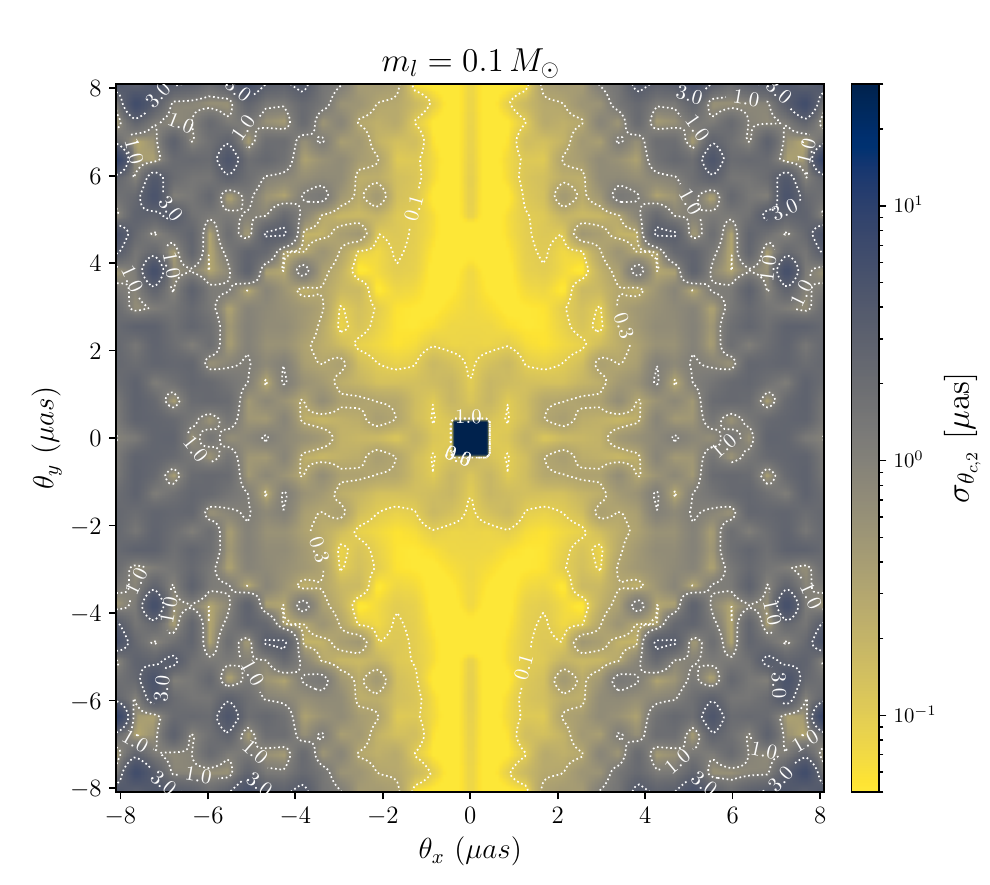}
        \end{subfigure} \\

        \begin{subfigure}[b]{0.32\textwidth}
            \includegraphics[width=\textwidth]{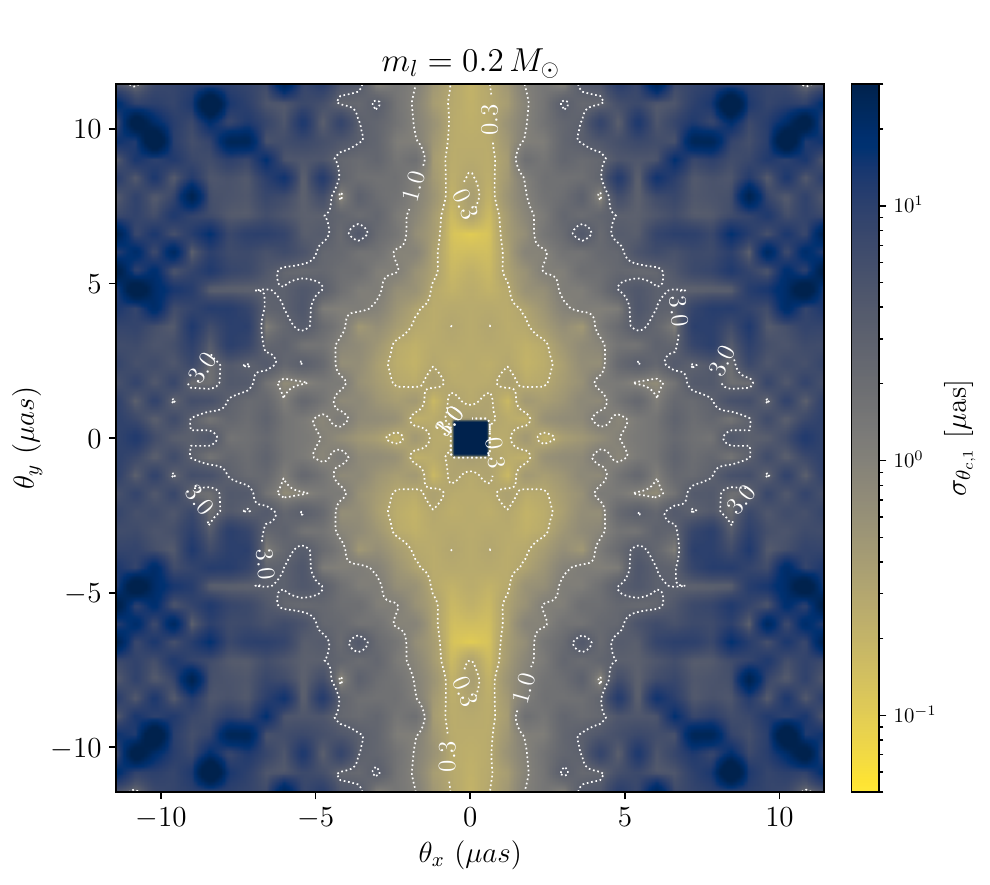}
        \end{subfigure} &
        \begin{subfigure}[b]{0.32\textwidth}
            \includegraphics[width=\textwidth]{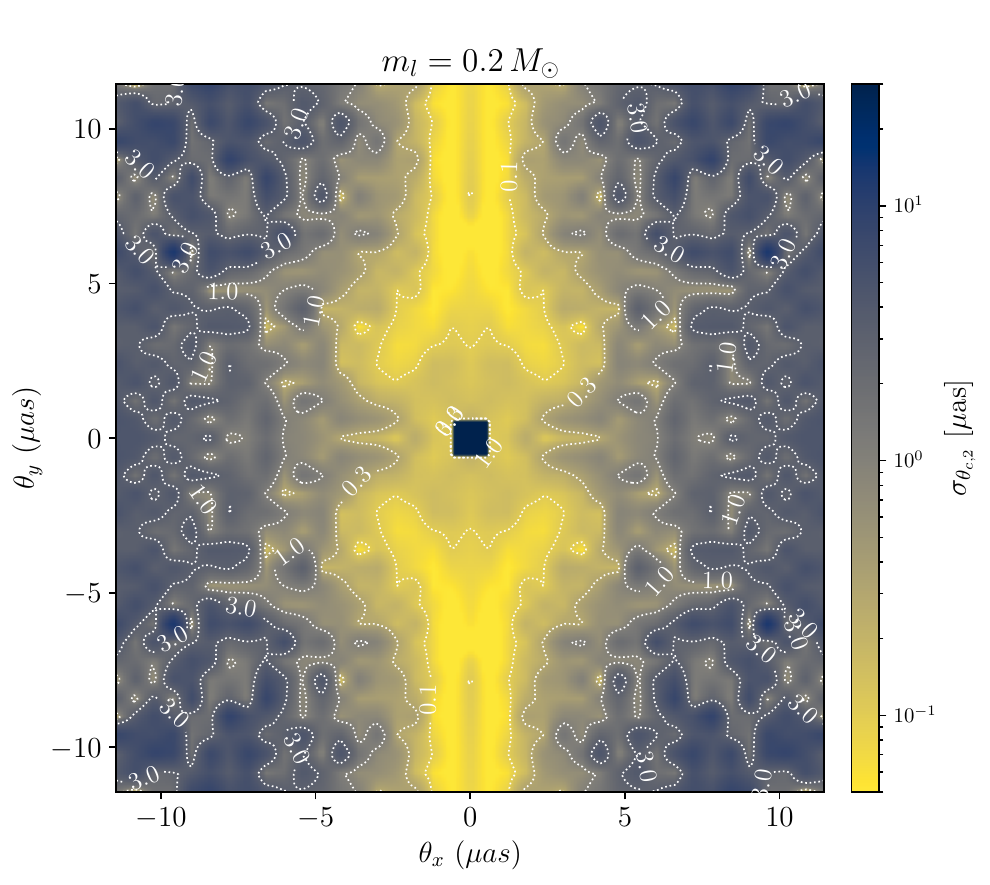}
        \end{subfigure} \\

        \begin{subfigure}[b]{0.32\textwidth}
            \includegraphics[width=\textwidth]{thetac1_04.pdf}
        \end{subfigure} &
        \begin{subfigure}[b]{0.32\textwidth}
            \includegraphics[width=\textwidth]{thetac2_04.pdf}
        \end{subfigure} \\

        \begin{subfigure}[b]{0.32\textwidth}
            \includegraphics[width=\textwidth]{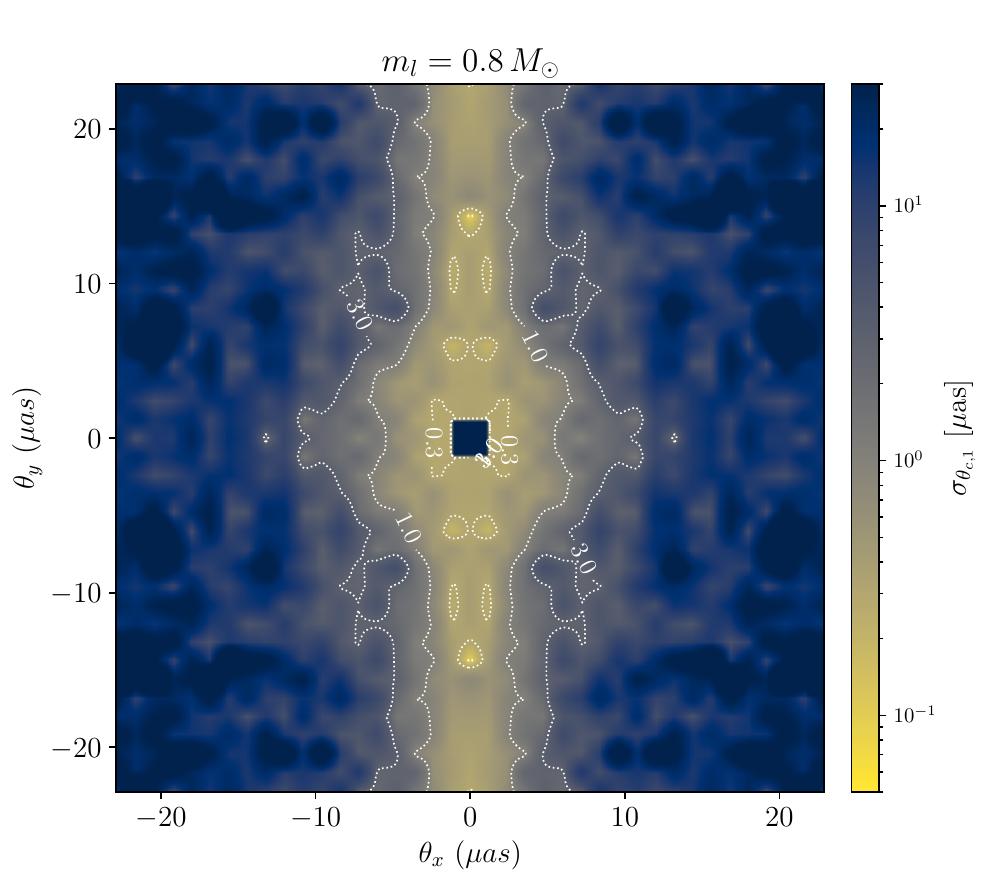}
        \end{subfigure} &
        \begin{subfigure}[b]{0.32\textwidth}
            \includegraphics[width=\textwidth]{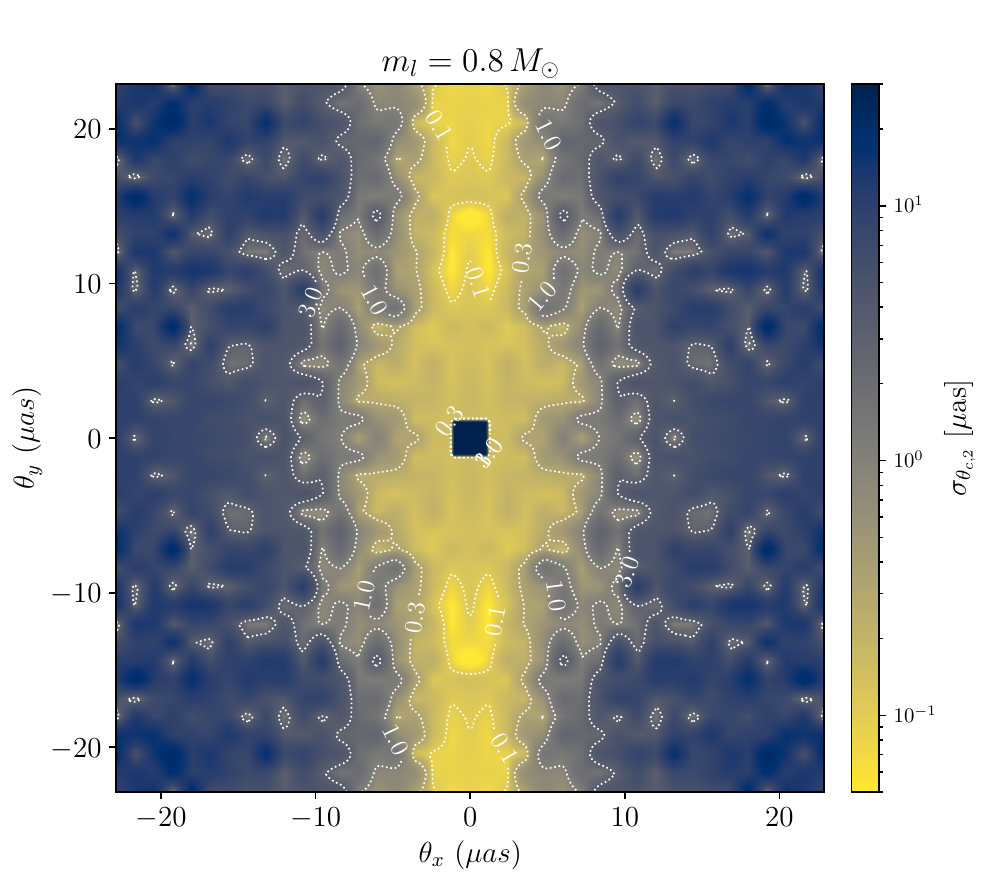}
        \end{subfigure} \\
    \end{tabular}
    \caption{\micronblink{interp_grid01.ipynb}\micronblink{interp_grid02.ipynb}\micronblink{interp_grid04.ipynb}\micronblink{interp_grid08.ipynb} Interpolated single-epoch light-centroid precision, $\sigma[\theta_{c,1}]$ defined in Eq.~\ref{eq:theta_C1} (left) and $\sigma[\theta_{c,2}]$ defined in Eq.~\ref{eq:theta_C2} (right), as a function of lens position $\vect{\theta}_\mathrm{sep} = (\theta_x, \theta_y)$ (in $\mu\mathrm{as}$). Both colorbars are in $\mu\mathrm{as}$ on a logarithmic scale. Results combine II and magnification constraints, with the same layout and image scales as Fig.~\ref{fig:lens_grid}. }
    \label{fig:other_centroids}
\end{figure*}

\begin{figure*}[htbp]
    \centering
    \def\arraystretch{1}
    \setlength\tabcolsep{0pt}
    \begin{tabular}{cc}

        \begin{subfigure}[b]{0.32\textwidth}
            \includegraphics[width=\textwidth]{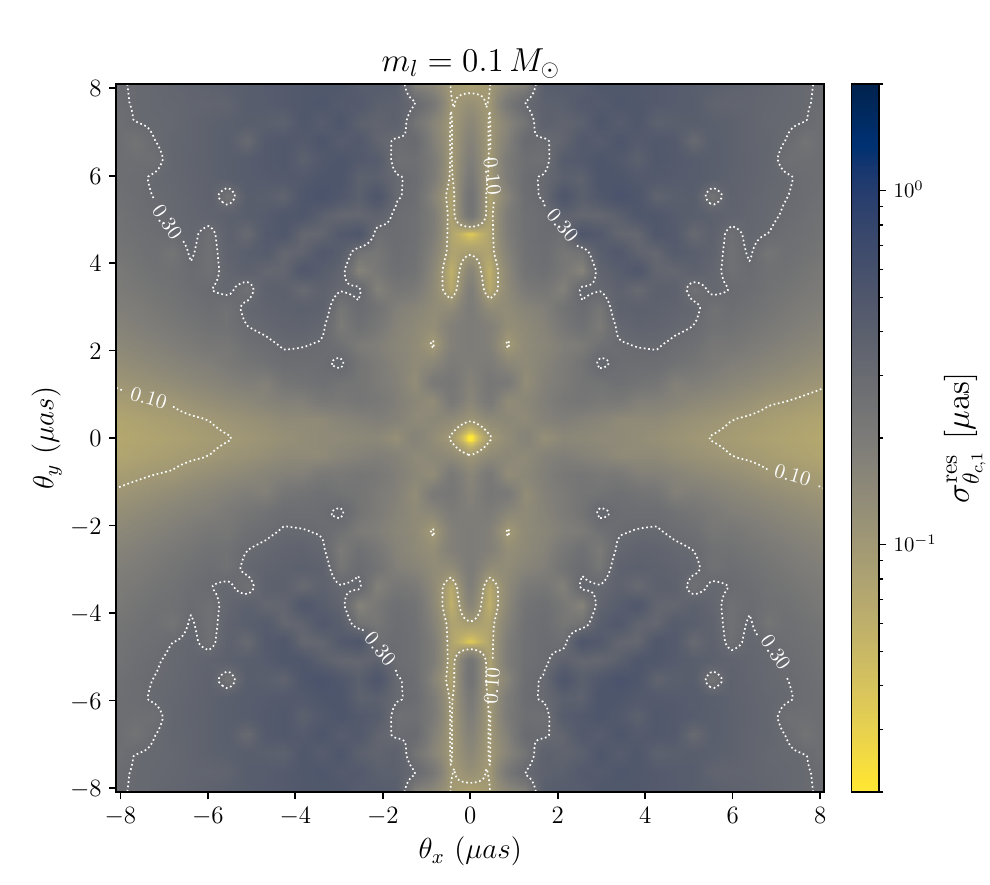}
        \end{subfigure} &
        \begin{subfigure}[b]{0.32\textwidth}
            \includegraphics[width=\textwidth]{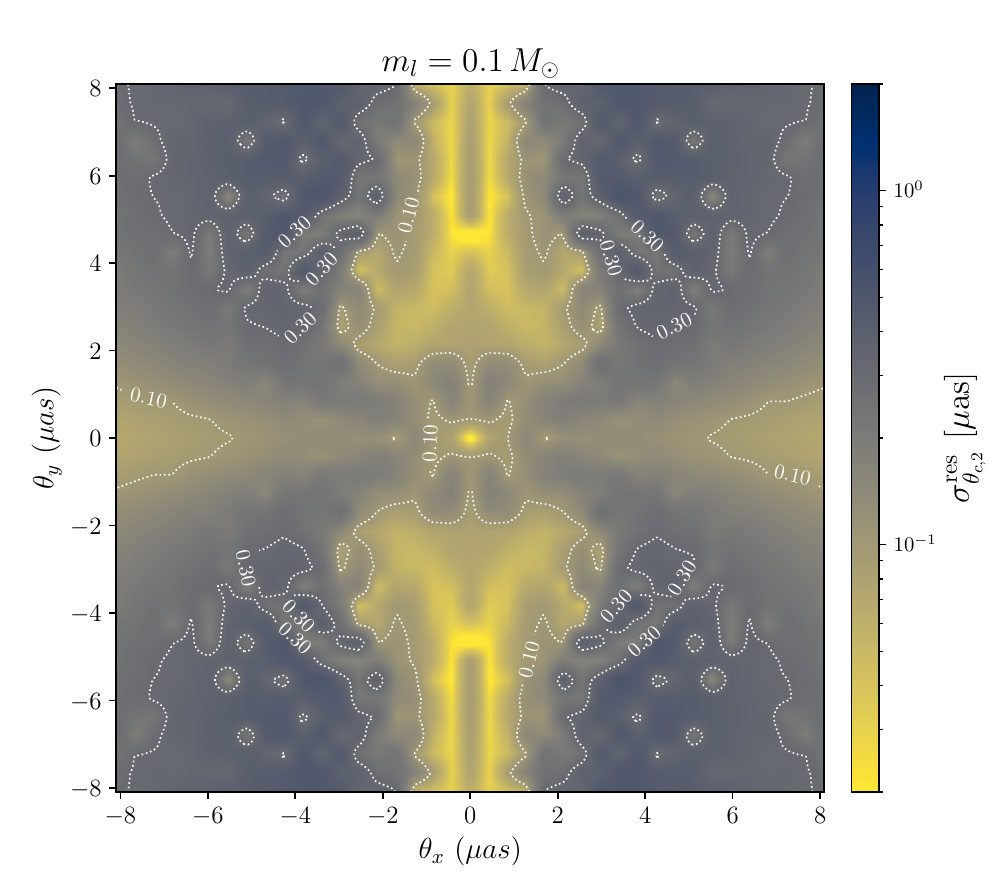}
        \end{subfigure} \\

        \begin{subfigure}[b]{0.32\textwidth}
            \includegraphics[width=\textwidth]{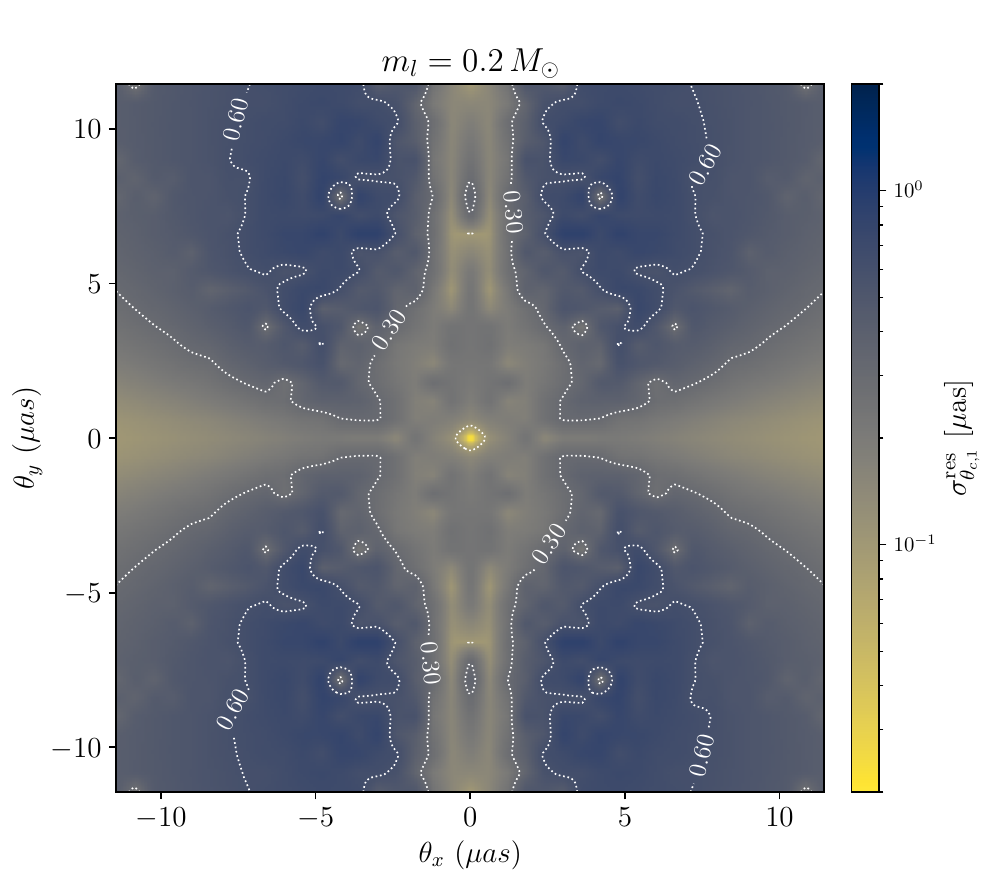}
        \end{subfigure} &
        \begin{subfigure}[b]{0.32\textwidth}
            \includegraphics[width=\textwidth]{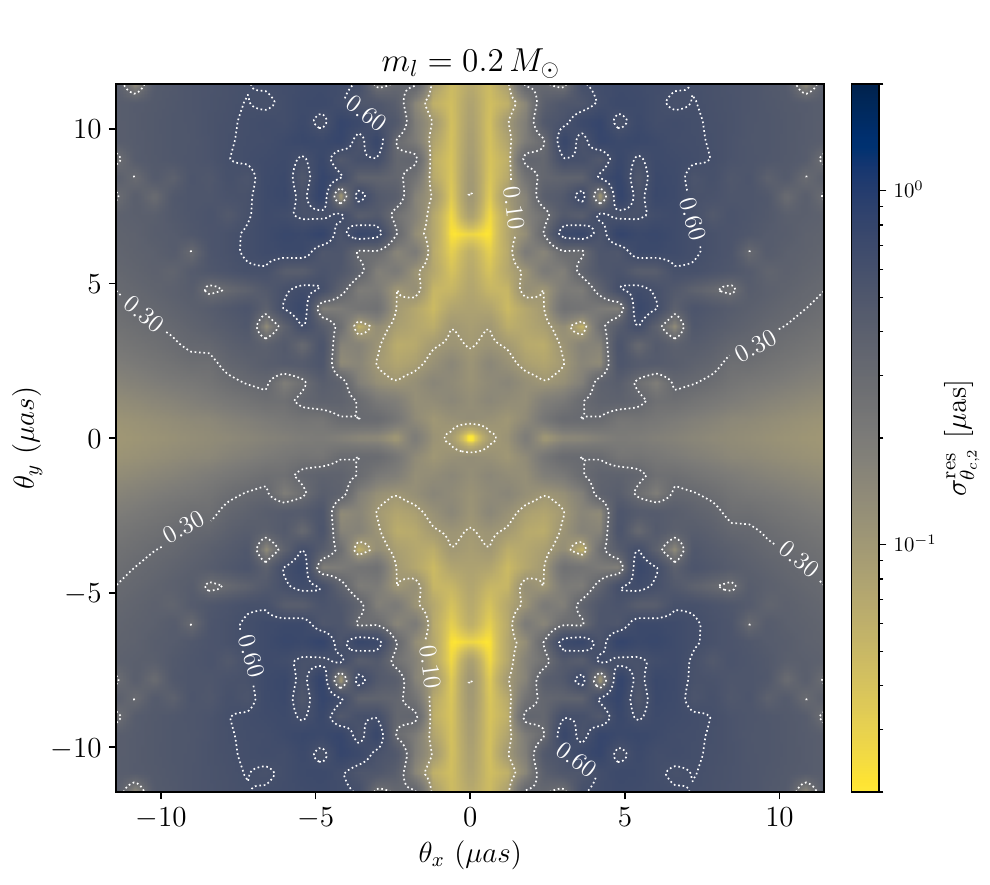}
        \end{subfigure} \\

        \begin{subfigure}[b]{0.32\textwidth}
            \includegraphics[width=\textwidth]{thetac1res_04.pdf}
        \end{subfigure} &
        \begin{subfigure}[b]{0.32\textwidth}
            \includegraphics[width=\textwidth]{thetac2res_04.pdf}
        \end{subfigure} \\

        \begin{subfigure}[b]{0.32\textwidth}
            \includegraphics[width=\textwidth]{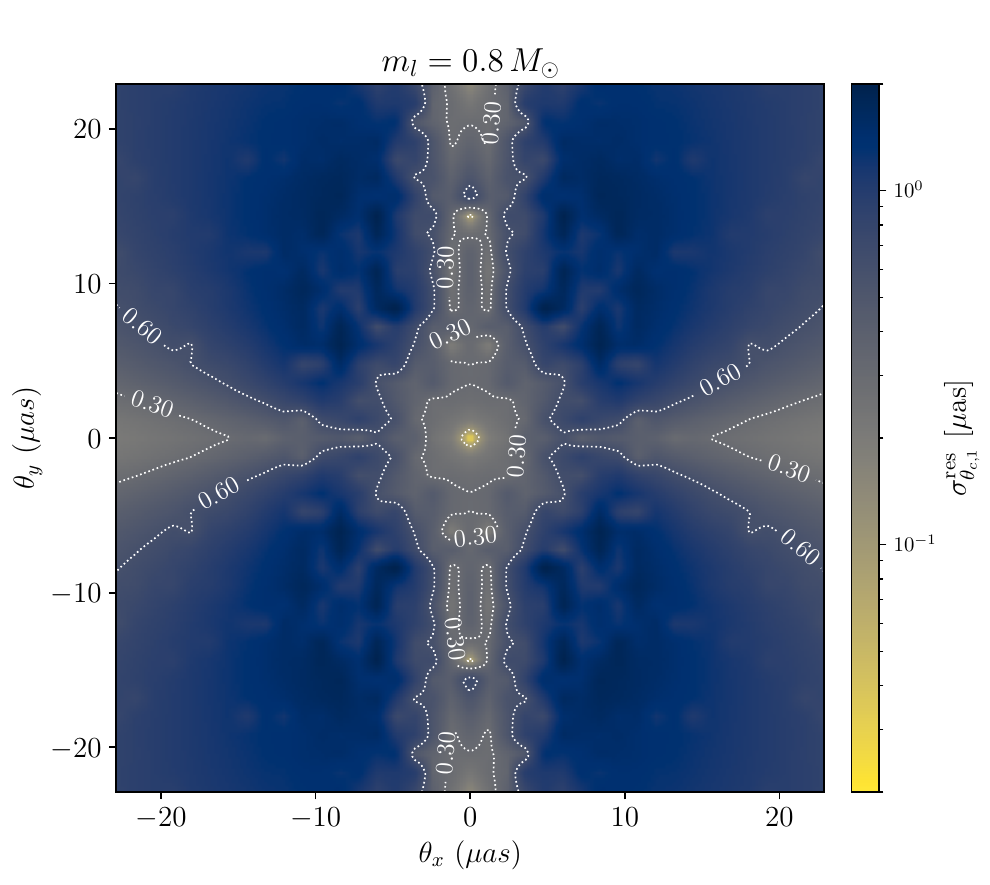}
        \end{subfigure} &
        \begin{subfigure}[b]{0.32\textwidth}
            \includegraphics[width=\textwidth]{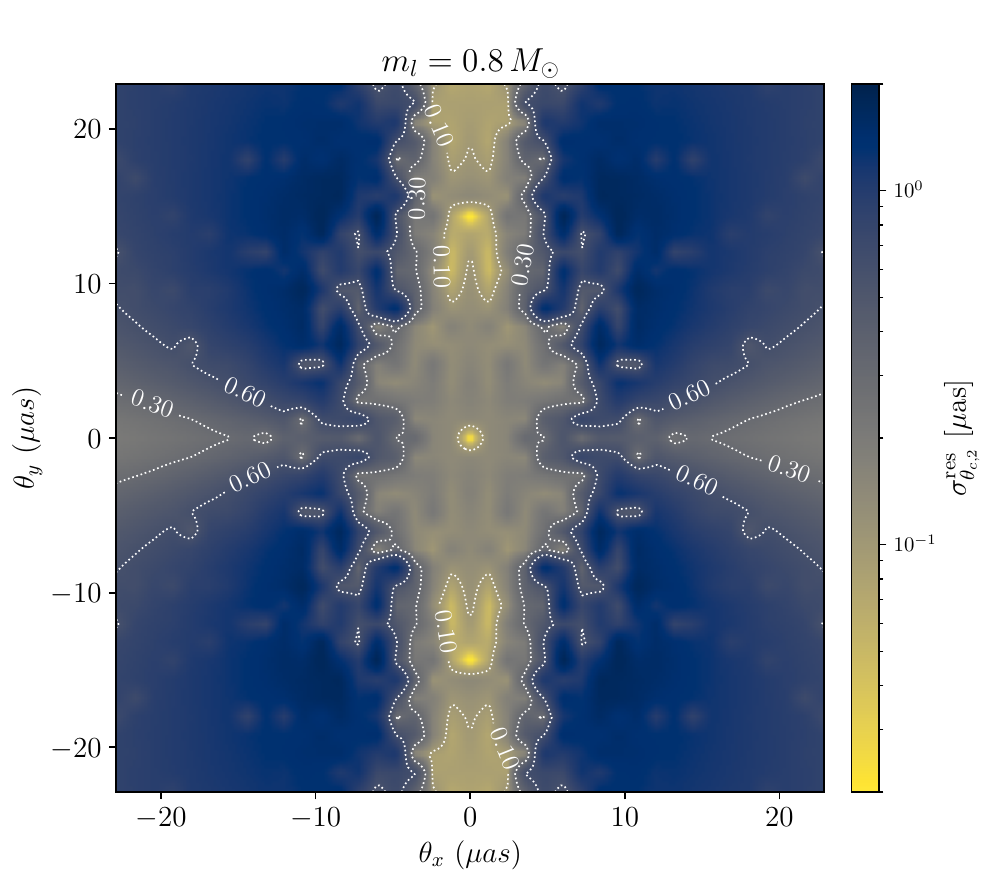}
        \end{subfigure} \\
    \end{tabular}
    \caption{\micronblink{interp_grid01.ipynb}\micronblink{interp_grid02.ipynb}\micronblink{interp_grid04.ipynb}\micronblink{interp_grid08.ipynb} Interpolated residual light-centroid uncertainty (Eq.~\ref{eq:cov_res} evaluated with the metrics of Eq.~\ref{eq:theta_C1}, left, and Eq.~\ref{eq:theta_C2}, right) as a function of lens position $(\theta_x,\theta_y)$ (in $\mu\mathrm{as}$), for $m_l = 0.1, 0.2, 0.4$, and $0.8\,M_\odot$. Both colorbars are in $\mu\mathrm{as}$ on a logarithmic scale.}
    \label{fig:other_centres}
\end{figure*}

The posterior examples illustrate the qualitative behavior of the single-epoch constraints; we now show the corresponding uncertainties across the full lens-position parameter space for all considered lens masses, $m_l=[0.1, 0.2, 0.4, 0.8],M_\odot$. The interpolated lens mass and position constraints are shown in Fig.~\ref{fig:other_mass_grid}, repeating the fiducial $m_l=0.4\,M_\odot$ case of the main text for ease of comparison. We show the derived light centroid uncertainties in Fig.~\ref{fig:other_centroids} and residual light centroid position uncertainties in Fig.~\ref{fig:other_centres}. Within each figure, the color scale of a given quantity is identical across lens masses. In general, higher-mass lens parameters are resolved with greater precision and out to greater separations.

\subsection{Time-Domain Analysis}

For each lens mass, we perform the time-domain analysis as a function of minimum impact parameter $|\vect{\xi}|$ as described in Sec.~\ref{subsec:traj}. We show these results in Table~\ref{tab:tab_main}.

\begin{turnpage}
\begin{table*}[ht]
\centering
\caption{Median parameter uncertainties as a function of minimum trajectory impact parameter $|\vect{\xi}|$ and lens mass after averaging over all lens approach angles.}
\label{tab:tab_main}
\begin{ruledtabular}
\begin{tabular}{c c c c c c c c c c}
$|\vect{\xi}|$ ($\mu\mathrm{as}$)
& $m_l\,(M_\odot)$
& $\left[ \sigma[\theta_\mathrm{src}]\times 10^4 \right]\;(\mu\mathrm{as})$
& $\sigma[m_l]\;(M_\odot)$
& $\sigma[\xi_x]\;(\mu\mathrm{as})$
& $\sigma[\xi_y]\;(\mu\mathrm{as})$
& $\sigma[v_x]\;(\mathrm{km\,s^{-1}})$
& $\sigma[v_y]\;(\mathrm{km\,s^{-1}})$
& $\sigma_1 [\frac{\partial^2 \Delta \theta_c}{\partial t^2}]$ ($\mu\mathrm{as}\,\mathrm{yr}^{-2}$)
& $\sigma_2 [\frac{\partial^2 \Delta \theta_c^{\mathrm{res}}}{\partial t^2}]$ ($\mu\mathrm{as}\,\mathrm{yr}^{-2}$) \\
\hline

\multirow{4}{*}{0.69}
 & 0.1 & 3.01 & 0.005 & 0.051 & 0.089 & 87.6 & 159.8 & 0.020 & 0.007 \\
 & 0.2 & 2.95 & 0.006 & 0.050 & 0.129 & 131.2 & 254.5 & 0.023 & 0.007 \\
 & 0.4 & 2.90 & 0.008 & 0.061 & 0.177 & 174.4 & 352.3 & 0.030 & 0.009 \\
 & 0.8 & 2.63 & 0.010 & 0.092 & 0.206 & 219.2 & 434.6 & 0.032 & 0.009 \\
\hline

\multirow{4}{*}{2.00}
 & 0.1 & 3.72 & 0.006 & 0.086 & 0.100 & 97.9 & 174.6 & 0.026 & 0.007 \\
 & 0.2 & 3.11 & 0.007 & 0.064 & 0.096 & 97.1 & 179.5 & 0.020 & 0.008 \\
 & 0.4 & 2.75 & 0.010 & 0.069 & 0.123 & 125.0 & 248.5 & 0.025 & 0.009 \\
 & 0.8 & 2.75 & 0.012 & 0.093 & 0.172 & 177.9 & 357.0 & 0.032 & 0.010 \\
\hline

\multirow{4}{*}{3.33}
 & 0.1 & 2.99 & 0.010 & 0.210 & 0.154 & 180.9 & 327.6 & 0.041 & 0.005 \\
 & 0.2 & 3.10 & 0.010 & 0.118 & 0.110 & 124.3 & 217.8 & 0.035 & 0.006 \\
 & 0.4 & 3.19 & 0.014 & 0.093 & 0.121 & 114.1 & 209.3 & 0.028 & 0.009 \\
 & 0.8 & 2.85 & 0.016 & 0.100 & 0.133 & 142.1 & 270.8 & 0.022 & 0.009 \\
\hline

\multirow{4}{*}{4.64}
 & 0.1 & 3.61 & 0.042 & 0.590 & 0.515 & 466.7 & 833.7 & 0.104 & 0.006 \\
 & 0.2 & 2.88 & 0.026 & 0.280 & 0.201 & 245.1 & 454.4 & 0.092 & 0.006 \\
 & 0.4 & 2.95 & 0.023 & 0.171 & 0.143 & 158.8 & 299.4 & 0.056 & 0.006 \\
 & 0.8 & 3.52 & 0.024 & 0.118 & 0.146 & 130.8 & 237.7 & 0.034 & 0.010 \\
\hline

\multirow{3}{*}{5.97}
 & 0.2 & 4.56 & 0.121 & 0.822 & 0.840 & 596.0 & 1154.1 & 0.215 & 0.008 \\
 & 0.4 & 2.74 & 0.055 & 0.364 & 0.284 & 328.2 & 656.8 & 0.151 & 0.007 \\
 & 0.8 & 2.86 & 0.038 & 0.195 & 0.189 & 182.6 & 338.0 & 0.051 & 0.008 \\
\hline

\multirow{3}{*}{7.28}
 & 0.2 & 5.89 & 0.256 & 2.212 & 2.508 & 1338.1 & 2563.9 & 0.321 & 0.014 \\
 & 0.4 & 4.93 & 0.258 & 0.999 & 1.091 & 720.5 & 1414.4 & 0.328 & 0.013 \\
 & 0.8 & 2.63 & 0.078 & 0.360 & 0.326 & 373.1 & 755.0 & 0.159 & 0.009 \\
\hline

\multirow{3}{*}{8.61}
 & 0.2 & 6.35 & 0.372 & 3.476 & 4.379 & 2265.2 & 4319.1 & 0.453 & 0.023 \\
 & 0.4 & 5.94 & 0.457 & 2.335 & 2.571 & 1292.2 & 2597.9 & 0.514 & 0.017 \\
 & 0.8 & 4.20 & 0.389 & 0.915 & 0.901 & 752.6 & 1554.9 & 0.359 & 0.013 \\
\hline

\multirow{2}{*}{9.92}
 & 0.4 & 6.22 & 0.622 & 3.383 & 3.493 & 1898.1 & 3755.1 & 0.550 & 0.033 \\
 & 0.8 & 5.84 & 0.689 & 2.116 & 2.308 & 1226.9 & 2571.5 & 0.609 & 0.023 \\
\hline

\multirow{2}{*}{11.23}
 & 0.4 & 6.37 & 0.761 & 4.929 & 6.264 & 2974.2 & 5876.0 & 0.666 & 0.032 \\
 & 0.8 & 6.08 & 0.758 & 2.182 & 2.633 & 1665.5 & 3460.5 & 0.626 & 0.026 \\
\hline

\multirow{2}{*}{12.56}
 & 0.4 & 6.46 & 0.681 & 4.876 & 5.729 & 3731.1 & 7487.6 & 0.595 & 0.028 \\
 & 0.8 & 6.17 & 0.842 & 2.760 & 3.156 & 2274.6 & 4639.4 & 0.679 & 0.034 \\
\hline

\multirow{1}{*}{13.87}
 & 0.8 & 6.28 & 0.834 & 2.739 & 3.515 & 3040.6 & 6182.4 & 0.730 & 0.041 \\
\hline

\multirow{1}{*}{15.20}
 & 0.8 & 6.32 & 1.188 & 4.939 & 5.712 & 3852.4 & 8051.4 & 0.768 & 0.053 \\
 
 \hline
\multirow{1}{*}{ 16.50}
 & 0.8 & 6.38 & 1.378 & 6.207 & 6.680 & 4592.6 & 9646.0 & 0.930 & 0.069 \\
 \hline

 \multirow{1}{*}{17.82}
 & 0.8 &  6.33 & 1.351 & 5.592 & 7.292 & 5664.3 & 12034 & 1.02 & 0.079\\

\end{tabular}
\end{ruledtabular}
\end{table*}
\end{turnpage}

\clearpage 
\bibliography{bib1}

\end{document}